\documentclass[11pt,leqno]{article}
\usepackage[left=1in,right=1in,top=1in,bottom=1in]{geometry}
\usepackage{amssymb,amsmath,amsthm}
\usepackage{amsfonts,bbm}
\usepackage{blindtext}
\usepackage{booktabs}
\usepackage{bm}
\usepackage{caption}
\usepackage{comment}
\usepackage[shortlabels]{enumitem}
\usepackage{float}
\usepackage{graphicx}
\usepackage{bbm}
\graphicspath{ {../figures/} }
\usepackage{indentfirst}
\usepackage{mathtools}
\usepackage{mathrsfs}
\usepackage{multirow}
\usepackage{upgreek}
\usepackage[normalem]{ulem}
\usepackage[utf8]{inputenc}
\usepackage[bibstyle=numeric, citestyle=authoryear, natbib=true,doi=false, url=true, backend=biber, maxbibnames=6, maxcitenames=4, uniquelist=false, uniquename=false, sorting=nyt]{biblatex}
\renewbibmacro{in:}{}
\DeclareNameAlias{default}{last-first/first-last}
\DeclareMathOperator*{\argmin}{arg\,min}

\DeclareMathOperator*{\arginf}{arg\,inf}
\DeclareMathOperator*{\plim}{plim}
\DeclareMathOperator*{\sgn}{sgn}
\newcommand{\indicator}[1]{\mathbbm{1} \{#1 \}}
\newcommand\independent{\protect\mathpalette{\protect\independenT}{\perp}}
\def\independenT#1#2{\mathrel{\rlap{$#1#2$}\mkern2mu{#1#2}}}
\usepackage{array}
\newcolumntype{L}[1]{>{\raggedright\arraybackslash}m{#1}}
\usepackage{rotating}
\usepackage{setspace}
\usepackage{subcaption}
\usepackage{tikz}
\usepackage{ctable}
\usepackage[colorlinks,citecolor=blue,urlcolor=blue,linkcolor=blue]{hyperref}
\usepackage{cleveref}
\crefname{enumi}{}{} 
\usepackage{dcolumn}
\usepackage{float}
\usepackage{pdflscape}
\usepackage{longtable}
\usepackage{threeparttable}

\makeatletter

\newcommand{\Rmnum}[1]{\expandafter\@slowromancap\romannumeral #1@}
\newcommand{\E}{\mathbb{E}}
\newcommand{\Prob}{\mathbb{P}}

\makeatother

\newtheorem{theorem}{Theorem}

\newtheorem{lemma}{Lemma}

\newtheorem{definition}{Definition}
\newtheorem{assumption}{Assumption}

\newtheorem{remark}{Remark}
\newtheorem{example}{Example}
\newtheorem{proposition}{Proposition}

\newtheorem*{property*}{Properties of dCov}
\newtheorem{inneruassumption}{Assumption}
\newenvironment{namedassumption}[1]
  {\renewcommand\theinneruassumption{#1}\inneruassumption}
  {\endinneruassumption}

\crefname{figure}{Figure}{Figures}
\crefname{assumption}{assumption}{assumptions}
\crefname{lemma}{Lemma}{Lemmata}

\numberwithin{lemma}{section}
\numberwithin{result}{section}
\numberwithin{definition}{section}
\numberwithin{remark}{section}
\numberwithin{example}{section}
\numberwithin{table}{section}
\numberwithin{proposition}{section}
\numberwithin{condition}{section}
\numberwithin{equation}{section}

\newfloat{sfigure}{tbp}{sfig}
\begin{document}
	\title{A Distance Covariance-based Estimator\thanks{This paper benefitted from valuable feedback from Brantly Callaway, Feiyu Jiang, Weige Huang, Julius Owusu, Oleg Rytchkov, Rosnel Sessinou, Dai Zusai, and participants at the 2021 AFES, 2021 EWMES, 2022 AMES, 2023 MEG, and the 2025 University of Georgia Econometrics Reading Group Meetings.}}
	\date{\today}
	\author{Emmanuel Selorm Tsyawo\footnote{email: estsyawo@gmail.com, Department of Economics, Finance and Legal Studies, Culverhouse College of Business, University of Alabama} \and  Abdul-Nasah Soale\footnote{ abdul-nasah.soale@case.edu, Department of Mathematics, Applied Mathematics, and Statistics, Case Western
Reserve University.}}
	\maketitle
\begin{refsection}
\begin{abstract}

\noindent 
This paper proposes an estimator that relaxes the conventional relevance condition in instrumental variable (IV) analyses. The method allows endogenous covariates to be weakly correlated, uncorrelated, or even mean-independent---though not independent---of the instruments, enabling the use of the maximal set of relevant instruments in a given application. Identification is attainable without exclusion restrictions and without finite-moment assumptions on the disturbance term. Under either of two non-nested exogeneity conditions, combined with mild regularity conditions, the parameter of interest is identified. The estimator is shown to be consistent and asymptotically normal, and the relaxed relevance condition required for identification is testable.
		
		\noindent \textit{Keywords:} distance covariance, dependence, weak instrument, endogeneity, $ U $-statistics

		\noindent \textit{JEL classification: C13, C14, C26} 
		
\end{abstract}
	
	\newpage

\section{Introduction}

Empirical work in economics often relies on instrumental variable (IV) methods. However, when instruments are weakly correlated with endogenous covariates, conventional IV methods such as two-stage least squares (TSLS), the control function (CF) method, and the generalised method of moments (GMM) become unreliable, leading to biased estimates and hypothesis tests with significant size distortions. Furthermore, conventional IV methods are infeasible when excluded instruments are unavailable or uncorrelated with the endogenous variables. These conventional methods are also highly sensitive to outliers or non-existent moments of the disturbance term $U$. While much of the econometric literature on the weak instrument problem is focused on detection and weak-instrument-robust inference, theoretical progress on estimation is scant \citep{andrews2019weak}. This paper introduces a new single-step estimator that minimises a scalar-valued measure of stochastic dependence between a parametrised disturbance $U(\theta)$ and a set of instruments $Z$ using the distance covariance measure (dCov) proposed by \textcite{szekely2007measuring}. The proposed Minimum Dependence estimator (MDep) substantially relaxes the instrument relevance requirement, allows for instruments $Z$ that are not independent of covariates $X$, and remains robust even when the disturbance term $U$ lacks finite moments.

The MDep has remarkable features that render it \emph{fundamentally} different from existing IV methods. (1) The non-independence identifying variation means the MDep can exploit the maximum number of instruments available in any given empirical setting.\footnote{In a class of single-index models, for example, MDep relevance requires that no non-trivial linear combination of $X$ be independent of $Z$.} (2) In the absence of excluded instruments, identification in the MDep framework continues to hold as long as covariates $X$ are not independent of instruments $Z$. (3) Although the MDep does not estimate a quantile model, it shares the ``robustness'' property of quantile estimators—see, e.g., \textcite{powell1991estimation,oberhofer2016asymptotic}—in that its asymptotic properties do not rely on the existence of moments of~$U$. By replacing $Z$ with a bounded one-to-one mapping such that $ Z $ and the mapping generate the same Euclidean Borel field, one obviates moment existence conditions on $Z$ as well.\footnote{An example of such a mapping is $ z \mapsto \mathrm{atan}(z) $.} This third feature is important as economic theory can go as far as justifying the exogeneity of instruments, but typically \emph{cannot} go far enough to justify the existence of moments of $U$. This paper appears to be the first to introduce an IV estimator that exploits identifying variation from arbitrary stochastic dependence---of unknown and unspecified form---between \( X \) and \( Z \), in a broad class of models.

As the form of identifying variation needs to be neither known nor specified, the MDep framework effectively eliminates the sensitivity of estimates to first-stage model specification.\footnote{\textcite{dieterle2016simple}, for example, uncovers substantial sensitivity of conclusions to specification (linear versus quadratic) of the first stage.} Thus, often-imposed linearity or monotonicity restrictions on first-stage relationships, e.g., \citet{wooldridge2010econometric,dhaultfoeuille-fevrier-2015-identification,torgovitsky2017minimum}, are unnecessary in the MDep framework. Although this property is also shared by integrated conditional moment estimators (ICM hereafter), e.g., \textcite{dominguez2004consistent,escanciano2006consistent,antoine2014conditional,escanciano2018simple,tsyawo2023feasible}, it is worth emphasising that the MDep relevance condition is more general. For example, $ \Prob\big(\E[X\mid Z] \neq \E[X]\big) > 0 $ neither implies nor is implied by $ \Prob\big(\E[Z\mid X]\neq \E[Z] \big) > 0 $. For identification, the MDep exploits both forms of dependence, while ICM estimators can only exploit the former. The MDep can achieve identification without excludability; this is more general than similar identification highlighted in \citet{tsyawo2023feasible,gao-wang-2023linIV} for the ICM and IV classes of estimators, respectively.

The rest of the paper is organised as follows. \Cref{Sect:Literature} discusses strands of related literature, while \Cref{Sect:The dCov_Estimator} describes the dCov measure and presents the MDep estimator. \Cref{Sect:Asymp_Theory} derives theoretical results viz. identification, consistency, asymptotic normality, consistency of the covariance matrix estimator, and testability of the MDep relevance condition. \Cref{Sect:Sim} examines the small sample performance of the MDep via simulations, and \Cref{Sect:Conclusion} concludes. All proofs are relegated to the Appendix. Additional theoretical and simulation results are available in the Supplemental Appendix. 

\paragraph*{Notation:} Define $ \E_n[\xi_i] := \frac{1}{n}\sum_{i=1}^{n}\xi_i $ and $ \E_n [\xi_{ij}] := \frac{1}{n(n-1)}\sum_{i=1}^{n}\sum_{j\neq i}^{n}\xi_{ij} $. For a random variable $\xi$, let $\xi^\dagger$ denote its independent and identically distributed ($i.i.d.$) copy, and define its symmetrised version as $\widetilde{\xi}:= \xi - \xi^\dagger $. Similarly for observations $ i \neq j$, define $\widetilde{\xi}_{ij} := \xi_i - \xi_j $. Independence between random variables is denoted by $\xi_1 \independent \xi_2 $. Let $p_\xi$ denote the dimension of $\xi$, and define $[p]:= \{1,\ldots,p\} $ for $ p\in \mathbb{N} $. The symbol $ || \cdot || $ denotes the usual Euclidean norm; $ a \vee b:= \max\{a,b\} $; and $ a \wedge b := \min\{a,b\} $. Finally, let $\widetilde{\sigma}\big(\xi\big)$ denote the sigma-algebra generated by $ [\xi,\xi^\dagger] $, and define the sign function as \( \sgn( \xi ) := \big(1-2\indicator{ \xi \leq 0 }\big) \).

\section{Related Literature}\label{Sect:Literature}

The MDep minimises a scalar-valued criterion of stochastic dependence between a parametrised error $U(\theta)$ and a set of instruments $Z$. This approach builds on the tradition of Minimum Distance from Independence (MDI) estimators initiated by \citet{manski1983closest} and further developed by \citet{brown-wegkamp-2002weighted,komunjer-santos-2010semi,gao-galvao-2014-minimum,dhaultfoeuille-fevrier-2015-identification,torgovitsky2017minimum,poirier-2017-efficient}. Of the foregoing, only \citet{torgovitsky2017minimum} explicitly considers identification cum estimation under endogeneity, as does this paper. \citet{torgovitsky2017minimum} specifies and models a first-stage infinite-dimensional nuisance parameter (the conditional distribution $X\mid Z$). \citet{komunjer-santos-2010semi,dhaultfoeuille-fevrier-2015-identification,torgovitsky2017minimum} require that covariates be continuously distributed---a substantive restriction, e.g., in settings with endogenous binary treatment. This paper imposes no support restrictions on \([X, Z]\), thereby accommodating a broader class of models, covariates and instruments, allowing for potentially non-monotonic first-stage relationships, and obviating continuity assumptions in the first stage. Moreover, the current paper appears to be the first to provide a tractable IV relevance condition in the class of MDI estimators.

The MDep estimator is related to ICM estimators, e.g., \textcite{dominguez2004consistent,escanciano2006consistent,antoine2014conditional,escanciano2018simple,wang2018consistent,antoine2022partially,tsyawo2023feasible,song-jiang-ke-2024estimation}. This class of estimators minimises the mean dependence of $U(\theta)$ on $Z$. Continuum Moment (CM) estimators---a related class of estimators---convert mean-independence restrictions into a continuum of unconditional moment conditions indexed by a nuisance parameter on an index set, and are typically estimated using IV methods such as Two-Stage Least Squares (TSLS) or the Generalized Method of Moments (GMM) (see, e.g., \textcite{carrasco2000generalization,donald2003empirical,hsu2011estimation,carrasco2015regularized}). Despite the advantages of both the ICM and Continuum Moment (CM) classes of estimators, two key differences set the MDep apart. First, endogenous covariates in the MDep framework can be mean-independent but stochastically dependent on instruments, e.g., at some quantile(s) that need not be known or determined. Thus, ICM/CM-relevant instruments are MDep-relevant by construction, whereas the converse does not hold. Second, unlike ICM/CM estimators, which require the existence of at least the first two moments of the disturbance for consistency and asymptotic inference, the MDep obviates the existence of any moment of the disturbance. Mean independence assumptions apply to the ICM, CM, and conventional IV classes and are often imposed as replacements for distributional exogeneity conditions. While mean independence is implied by distributional exogeneity, this holds under the \textit{implicit} assumption that the mean exists.

Some existing works consider IV estimation without excludability by exploiting and modelling non-linear forms of dependence between endogenous and exogenous covariates, e.g.,  \textcite{cragg1997using,dagenais1997higher,lewbel1997constructing,erickson2002two,rigobon2003identification,klein2010estimating,gao-wang-2023linIV}. Unlike the foregoing, the MDep does not require the practitioner to construct moments or model first-stage relationships. It suffices that there be dependence between covariates and instruments that ought not to be known, modelled, or estimated. To enhance the practicality of this important feature, this paper demonstrates the testability of the MDep relevance condition.
	
The econometric literature on weak instruments largely focuses on detection and weak-instrument-robust inference (e.g., \textcite{staiger1997instrumental,andrews2006optimal,kleibergen2006generalized,andrews2016condInference,sanderson2016weak,andrews2017unbiased})---see \textcite{andrews2019weak} for a review. Normal distributions of conventional IV estimates can be poor and hypothesis tests based on them can be unreliable when instruments are weak \citep{nelson1990distribution,nelson1990somefurther,bound1995problems}. The MDep gives a new perspective to handling weak IVs in empirical practice; IV- or ICM/CM-irrelevant instruments can be MDep-strong, and this condition is testable.
	
By extracting non-linear identifying variation in instruments in order to boost instrument strength, some works employ flexible methods such as the non-parametric IV, e.g., \textcite{donald2001choosing,newey2003instrumental,donald2003empirical,kitamura2004empirical,das2005instrumental}, machine learning techniques, e.g., \textcite{chen2020mostly}, and regularisation or moment selection schemes, e.g., \textcite{ng2009selecting,darolles2011nonparametric,belloni2012sparse,hansen2014instrumental,carrasco2015regularized}. While it is conceivable to take transformations of instruments to extract more identifying variation, this approach may be limited, for example, when available instruments are non-monotone in endogenous covariates.\footnote{E.g., $ X = X^* + U $, $ Z = |X^*| $, $ U \independent Z $, and $ X^* $ is symmetrically distributed with mean zero. $ \mathrm{cov}[X,Z]=0 $, and no measurable (feasible) transformation of $ Z $ can induce correlation with $ X $.} Further, the aforementioned approach usually results in high dimensionality, unlike the MDep, which remains parsimonious in $Z$.
	
The dCov measure is primarily used in tests of independence that are consistent against all forms of dependence, including linear, non-linear, monotone, and non-monotone alternatives. This feature of the dCov measure accounts for the weak relevance condition in the MDep framework. Several applications of the dCov measure have emerged since the seminal work \textcite{szekely2007measuring}---see, e.g., \textcite{sheng2013direction,szekely2014partial,shao2014martingale,park2015partial,su2017martingale,davis2018applications,xu2020martingale}. The current paper departs from this literature by leveraging the dCov for estimation and inference under possible endogeneity.

\section{The MDep Estimator}\label{Sect:The dCov_Estimator}
This section presents (1) motivating illustrative examples highlighting the MDep's key features, (2) the dCov measure, (3) an interesting class of applicable models, and (4) the MDep estimator.

\subsection{Motivating examples}\label{Sub_Sect:Illus_Eg}
The MDep estimator has unique strengths relative to existing IV estimators. To explore these, consider the linear model
\begin{equation*}
  Y = X_1\theta_1 + X_2\theta_2 + U  
\end{equation*}
in the following examples. $Z$ is MDep-relevant as long as it is \emph{not} independent of any non-trivial linear combination of $X$.

\begin{example}[Non-monotone first stage]\label{ex:illus_transform} Suppose $X_1$ and $Z_1$ are such that 
\[
    X_1 = Z^* + U \quad \text{ and } \quad Z_1 = \indicator{ |Z^*| < -\Phi^{-1}(0.25) } ,
\]
where $ [Z^*, U] \sim \mathcal{N}(0,\mathrm{I}_2)$ and $\Phi^{-1}(\cdot)$ is quantile function of the standard normal distribution. Clearly, $X_1$ and $Z_1$ are related through $Z^*$. However, there is no possible transformation of $Z_1$, without extra information, that induces mean dependence or correlation between $X_1$ and $Z_1$.
\end{example}

\begin{example}[Identification without excludability I - non-monotone first-stage]\label{ex:illus_no_excl_nmono}
Consider a slight modification of Example 3.1, where
\[
    X_2 = Z = 0.2Z^* + Z^{*2}.
\]
$Z$ is MDep-relevant without being excluded, as no non-trivial linear combination of $X_1$ and $X_2$ is independent of $Z$.
\end{example}

\begin{example}[Identification without excludability II - skedastic function]\label{ex:illus_no_excl_skedastic}
Consider the setting where 
\[
    X_1 = U\sqrt{1+Z^2}, \quad X_2 = Z, \quad \text{and} \quad \E[U] = 0.
\]
$Z$ is not IV-relevant. Moreover, $X_1$ is mean-independent of $Z$. However, relevance in the MDep framework holds as any non-trivial linear combination of $X_1$ and $X_2$ is dependent on $Z$.
\end{example}

 \begin{example}[Non-existent first moment of $U$]\label{ex:illus_moment}
     The disturbance, $U$, follows the Cauchy distribution $U \sim \mathcal{C}\big(0,\ 0.1 + |Z_1| \big)$ with conditional scale heterogeneity. Existing IV methods, such as Conventional IV, non-parametric IV, ICM, and CM estimators, are inconsistent when the first moment of $U$ does not exist. The MDep, in contrast, is consistent.
 \end{example} 

 \noindent The MDep explores identifying variation in all the examples given above while conventional IV, non-parametric IV, ICM, and CM methods fail. The above scenarios serve to highlight the remarkable features of the MDep relative to existing conventional methods. 

\subsection{The dCov measure}
It is instructive to briefly review the distance covariance (dCov) measure introduced by \textcite{szekely2007measuring}, which underpins the MDep objective function.
\begin{definition}\label{Def:Dist_Cov_Indep}
The square of the distance covariance between random variables $ \Upsilon $ and $ Z $ with finite first moments is defined by \textcite{szekely2007measuring} as
\begin{equation}\label{eqn:dCov1}
 \begin{split}
 \mathcal{V}^2(\Upsilon,Z)
 &= \int\big| \varphi_{\Upsilon,Z}(t,s) - \varphi_{\Upsilon}(t)\varphi_{Z}(s) \big|^2 w(t,s)dtds\\
 & = \int \Big| \E\big[\exp(\iota(t'\Upsilon+s'Z))\big] - \E\big[\exp(\iota t'\Upsilon)\big]\E\big[\exp(\iota s'Z)\big] \Big|^2w(t,s)dtds
 \end{split}
\end{equation}
where $ \varphi_{\xi}(.)$ denotes the characteristic function of $\xi$,  $\iota = \sqrt{-1} $, and the integrating measure $ w(t,s) $ is an arbitrary positive function for which the integral exists. The modulus is defined as $ |\zeta|^2 = \zeta\bar{\zeta} $, where $ \bar{\zeta} $ is the complex conjugate of $\zeta$. 
\end{definition} 
\noindent Using the integrating measure $ w(t,s) = (c_{p_\Upsilon}c_{p_Z}||t||^{1+p_\Upsilon}||s||^{1+p_Z})^{-1} $ where $ c_p = \frac{\pi^{(1+p)/2}}{\Gamma((1+p)/2)}, \ p \geq 1 $, and $ \Gamma(\cdot) $ is the complete gamma function, \textcite{szekely2007measuring} obtains a distance covariance measure, which is shown in \Cref{Prop:dCov_eqn} to have the representation 
\[
\mathcal{V}^2(\Upsilon,Z) = \E[\mathcal{Z}||\Upsilon-\Upsilon^\dagger||]
\] 
where $ \mathcal{Z} := h(Z,Z^\dagger) $ and $h(z_a,z_b) := ||z_a - z_b|| - \E\big[||z_a - Z||+||Z - z_b||\big] + \E\big[||Z - Z^\dagger||\big]$. From \eqref{eqn:dCov1}, one observes that $ |\varphi_{U,Z}(t,s) - \varphi_{U}(t)\varphi_{Z}(s)|_w^2 \geq 0 $. 

This paper follows \textcite{szekely2014partial} in using the following algebraically equivalent form of the unbiased estimator of the distance covariance measure
 	\begin{equation}\label{eqn:dCov_Measure}
 	\mathcal{V}_n^2(\Upsilon,Z)  := \frac{1}{n(n-3)}\sum_{i = 1}^n \sum_{j \neq i}^n\mathcal{Z}_{ij,n} ||\widetilde{\Upsilon}_{ij}||
 	\end{equation}
 	
\noindent where $ \mathcal{Z}_{ij,n} := h_n(Z_i,Z_j)$ such that
 	\begin{equation}\label{eqn:dcov_sample}
 		h_n(Z_i,Z_j) = 
 		\begin{cases} 
 			||\widetilde{Z}_{ij}|| - \frac{1}{n-2}\sum_{l=1}^{n}(||\widetilde{Z}_{il}||+||\widetilde{Z}_{lj}||) + \frac{1}{(n-1)(n-2)}\sum_{k=1}^n \sum_{l\neq k}^n||\widetilde{Z}_{kl}||, & i\neq j \\
 			0, & i=j 
 		\end{cases}.
 	\end{equation}
 	
\citeauthor{szekely2007measuring}'s \citeyearpar{szekely2007measuring} integrating measure $ w(t,s) = (c_{p_\Upsilon}c_{p_Z}||t||^{1+p_\Upsilon}||s||^{1+p_Z})^{-1} $, besides yielding a reliable measure of dependence, results in a computationally tractable measure, which does not require numerical integration, obviates the choice of smoothing parameters (e.g., bandwidth or number of approximating terms in non-parametric approaches), and admits multiple instruments. The simplified formulation \eqref{eqn:dCov_Measure} offers two key advantages for the proposed estimator: the permutation symmetry of $ \mathcal{Z}_{ij,n} = \mathcal{Z}_{ji,n} $ facilitates the use of U-statistic theory in establishing asymptotic normality and reduces the computational burden in evaluating \eqref{eqn:dCov_Measure}.    

For ease of reference, the properties of the dCov measure in \textcite{szekely2007measuring,szekely2009brownian} are stated below. 
 \begin{property*} The following properties hold for the distance covariance measure under the condition $ \E\big[||\Upsilon||^2 + ||Z||^2\big]< \infty $: 
		\begin{enumerate}[label=(\alph*)]
			\item\label{Property:dCov_geq0} $ \mathcal{V}^2(\Upsilon,Z) \geq 0 $;
			\item \label{Property:dCov_indep0} $ \mathcal{V}^2(\Upsilon,Z) = 0 \text{ if and only if } \Upsilon $ and $ Z $ are independent;
			\item \label{Property:dCov_expression} $ \mathcal{V}^2(\Upsilon,Z) = \E\big[\mathcal{Z}||\Upsilon - \Upsilon^\dagger||\big] $; and
			\item\label{Property:dCov_unbiased} $ \E[\mathcal{V}_n^2(\Upsilon,Z)] = \mathcal{V}^2(\Upsilon,Z) $ for $ n>3 $ and $i.i.d.$ samples $ \big\{ [\Upsilon_i,Z_i] \, : \, i \in [n]  \big\} $.
		\end{enumerate}
	\end{property*} 
	\noindent The properties are proved in the following: Property \ref{Property:dCov_geq0} in \textcite[Theorem 4 (i)]{szekely2009brownian}, Property \Cref{Property:dCov_indep0} in \textcite{szekely2007measuring}, Property \Cref{Property:dCov_expression} in \Cref{Prop:dCov_eqn} of this paper, and Property \Cref{Property:dCov_unbiased} in \textcite[Proposition 1]{szekely2014partial}.
	
\subsection{Model specification}\label{SubSect_Model}
For a tractable characterisation and statistical testing of the MDep relevance identification condition, consider regression models in which the outcome \( Y_i \) is generated as
\begin{equation}\label{eqn:DGP}
    Y_i = G\big(\theta_{o,c} + g(X_i\theta_o) + U_i\big),
\end{equation}
where
$ G(\cdot) $ is a known invertible function and $ g(\cdot) $ is a known differentiable function with unknown parameter vector $ \theta_o \in \mathbb{R}^{p_\theta} $. $ \theta_{o,c} $ is the location parameter of $U(\theta_o)$, where $ U(\theta) := G^{-1}(Y) - g(X\theta) $ denotes the parametrised disturbance function. $X_i$ contains a constant term. The dependence of $ U_i(\theta) $ on $ X_i $ is suppressed for notational ease. 

The class of models under consideration includes interesting examples such as the linear model $ U_i(\theta) = Y_i - X_i\theta $ (where the location parameter coincides with the intercept), non-linear parametric models, e.g., $ U_i(\theta) = Y_i - \exp(X_i\theta) $, fractional response models, e.g., $ U_i(\theta) = \log( Y_i/(1-Y_i)) - X_i\theta $, and special cases of Box-Cox models e.g., $ U_i(\theta) = \log(Y_i) - X_i\theta $. See \Cref{Rem:mod_class} for a more general class of applicable models. 

\subsection{Estimation}

Let \( \big\{W_i = [Y_i, X_i, Z_i]: i \in [n] \big\} \) be a random sample of $W:= [Y, X, Z] $ defined on a probability space \( (\mathcal{W}, \mathscr{W}, \mathbb{P}) \). The MDep estimator is the minimiser of \( \mathcal{V}_n^2\big(U(\theta),Z\big) \), namely
	\begin{equation}\label{eqn:dCov_Estimator}
	\widehat{\theta}_n = \argmin_{\theta \in \Theta } \frac{1}{n(n-3)}\sum_{i = 1}^n \sum_{j \neq i}^n\mathcal{Z}_{ij,n} |\widetilde{U}_{ij}(\theta)|,
	\end{equation}
where $\mathcal{Z}_{ij,n} := h_n(Z_i,Z_j)$ as defined in \eqref{eqn:dcov_sample} and $\widetilde{U}_{ij}(\theta) = U_i(\theta) - U_j(\theta)$. It may be of interest to estimate a location parameter for $U$, e.g., the median: $ \displaystyle \widehat{\theta}_{n,c} = \argmin_{t} \sum_{i=1}^n | U_i(\widehat{\theta}_n) - t | $.\footnote{The asymptotic properties of $\widehat{\theta}_{n,c}$ are omitted since they can be derived straightforwardly from those of $\widehat{\theta}_n$.} 

Following \textcite{huber1967behavior}, the minimand in (\ref{eqn:dCov_Estimator}) is normalised as
	\begin{equation}\label{eqn:obj_fun_gen_model}
		Q_n(\theta) := \frac{1}{n(n-3)}\sum_{i = 1}^n \sum_{j \neq i}^n \mathcal{Z}_{ij,n}\big(|\widetilde{U}_{ij}(\theta)|-|\widetilde{U}_{ij}|\big)
	\end{equation}
\noindent in order to avoid unnecessary moment conditions on $ U $---e.g.,  \citep{powell1991estimation,oberhofer2016asymptotic}. This holds even though the dCov measure itself requires the existence of $\E[|U|]$---cf. \citet{szekely2007measuring}.
		
\section{Asymptotic Theory}\label{Sect:Asymp_Theory}

It follows from \eqref{eqn:obj_fun_gen_model} that the asymptotic theory for the MDep estimator belongs to the broader class of estimators based on \( U \)-statistics--type objective functions (e.g., \textcite{honore1994pairwise,honore-powell-2005-pairwise,jochmans2013pairwise}), as well as those involving non-smooth objective functions such as quantile regression (QR) (e.g., \textcite{koenker1978regression,powell1991estimation,oberhofer2016asymptotic}), instrumental-variable QR methods (e.g., \textcite{chernozhukov2006instrumental,chernozhukov2008instrumental}), and the control-function approach to QR of \textcite{lee2007endogeneity}. Let the Jacobian and its symmetrised version be defined by  
\[
    X^g(\theta) := -\frac{\partial U(\theta)}{\partial \theta'} \text{ and } \widetilde{X^g}(\theta) := X^g(\theta)-X^{g\dagger}(\theta),
\]respectively. Also, define \( \displaystyle \widetilde{X^{gg}}(\theta):= \frac{\partial \widetilde{X^g}(\theta)}{\partial \theta}. \) The parameter vector $ \theta_o$ is the MDep estimand, and $U:= U(\theta_o) - \theta_{o,c} $.

\subsection{Regularity conditions}
Two sets of regularity conditions imposed in the paper guarantee the consistency of the MDep estimator $ \widehat{\theta}_n $. The first set outlined in the following comprises smoothing and dominance conditions, ensuring that the difference between the normalised minimand and its expectation converges to zero uniformly in $ \theta \in \Theta $.
	
	\begin{assumption}[Regularity]\label{Ass_dC:Reg}
		\quad
		\begin{enumerate}[label=(\alph*)]
			\item\label{Ass_dC:Reg_UdiffXMeas} $ U(\theta) $ is measurable in $ [U,X] $ for all $ \theta $ and is twice continuously differentiable in $ \theta $ for all $ [U,X] $ on the support of $ [U_i,X_i] $. $ X^g(\theta) = g'(X\theta)X $ is measurable in $ X $ and $ \Prob\big( g'(X\theta) = 0 \big) < 1 $ for all $ \theta \in \Theta $.
			\item\label{Ass_dC:Reg_Dominance} For some constant $ C \in (0,\infty) $, \quad $\displaystyle \E\Big[ \Big( \{|\mathcal{Z}|\vee 1\} \cdot \{ \sup_{\theta \in \Theta }||\widetilde{X^g}(\theta)|| \vee 1 \} \Big)^4\Big] \leq C $, \quad $ \displaystyle \E\Big[ \sup_{\theta \in \Theta }\Big\lVert \{|\mathcal{Z}|\vee 1\} \cdot \widetilde{X^{gg}}(\theta) \Big\rVert^2 \Big] \leq C $, \quad and \quad $\displaystyle \E\Big[ \big\lVert \widetilde{Z} \big\rVert^4\Big] <\infty. $
			\item\label{Ass_dC:Reg_CompactTheta} $ \Theta $ is a compact parameter space.
		\end{enumerate}
	\end{assumption}

\eqref{eqn:DGP} and the differentiability requirement in \Cref{Ass_dC:Reg}\Cref{Ass_dC:Reg_UdiffXMeas} characterise an interesting class of models considered in this paper, e.g., the linear model. $ \widetilde{U}(\theta) = \widetilde{U} - \widetilde{X^g}(\bar{\theta})(\theta-\theta_o) $ for some $ \bar{\theta} $ lying on the line segment between $ \theta $ and $ \theta_o $ is a useful expression for subsequent analyses thanks to \Cref{Ass_dC:Reg}\Cref{Ass_dC:Reg_UdiffXMeas} and the Mean-Value Theorem (MVT). The technical requirement $ \Prob\big(g'(X\theta) = 0 \big) < 1 $ is important for identification as the expression $ \widetilde{U}(\theta) = \widetilde{U} - \widetilde{X^g}(\bar{\theta})(\theta-\theta_o) $ with $ \widetilde{X^g}(\theta) = \big(g'(X\theta)X - g'(X^\dagger\theta)X^\dagger\big) $ shows that $ \widetilde{U}(\theta) $ can equal $ \widetilde{U} $ \emph{almost surely} ($a.s.$) for some $ \theta\neq \theta_o $ if it is violated.
	
\Cref{Ass_dC:Reg}\Cref{Ass_dC:Reg_Dominance} is an MDep analogue of uniform moment bounds; it implies 
\( \displaystyle \E\big[\sup_{\theta \in \Theta }||\mathcal{Z}\widetilde{X^g}(\theta)||^4\big] \leq C \) and \( \displaystyle \E\big[\sup_{\theta \in \Theta }||\widetilde{X^g}(\theta)||^4\big] \leq C.\) \Cref{Ass_dC:Reg}\ref{Ass_dC:Reg_Dominance} can be further weakened by replacing $Z$ with bounded one-to-one mappings such that $ Z $ and the mapping generate the same Euclidean Borel field, e.g., $ \mathrm{atan}(Z) $ ---see \textcite[p. 108]{bierens1982consistent} and \citet[Remark 1]{szekely2007measuring}---thereby allowing $Z$ (in addition to $U$) to have no finite moments. In that case, $\mathcal{Z}$ can be dropped from \Cref{Ass_dC:Reg}\ref{Ass_dC:Reg_Dominance}. \Cref{Ass_dC:Reg}\Cref{Ass_dC:Reg_CompactTheta} is required since the objective function \eqref{eqn:obj_fun_gen_model} is non-convex.

\subsection{Identification and consistency}

The second set of regularity conditions for consistency (\Cref{Ass_dC:Ident_Relv}, \Cref{Ass_dC:Ident_Exog_I}, and \Cref{Ass_dC:Ident_Exog_II}) are identification conditions that ensure that \( Q(\theta) := \E\big[ \mathcal{Z}(|\widetilde{U}(\theta)| - |\widetilde{U}|) \big] \) is uniquely minimised at $\theta_o$. The first identification assumption concerns the relevance condition in the MDep framework.

\begin{assumption}[Relevance]\label{Ass_dC:Ident_Relv} 
    $ X\uptau \not\independent Z $ for all $ \uptau \neq 0$.
    \end{assumption}

\Cref{Ass_dC:Ident_Relv} is the condition of non-independence between non-trivial linear combinations of $ X $ and $ Z $; it is the MDep analogue of the relevance condition in the IV setting, e.g., \textcite[Assumption 2SLS.2(b)]{wooldridge2010econometric}, and an MDep analogue of the linear completeness condition in ICM estimators, e.g., \textcite{escanciano2018simple,tsyawo2023feasible}. In the IV setting, the relevance condition requires that no non-zero linear combination of $ X $ be uncorrelated with $ Z $. The ICM relevance condition requires that no non-zero linear combination of $ X $ be mean-independent of $ Z $.  \Cref{Ass_dC:Ident_Relv} requires that no non-zero linear combination $ X $ be independent of $ Z $. As independence implies mean independence, which in turn implies uncorrelatedness, it follows that the MDep relevance condition (\Cref{Ass_dC:Ident_Relv}) is the weakest possible. In a simple case with a univariate $ X $, \Cref{Ass_dC:Ident_Relv} allows $ X $ to be uncorrelated, or mean-independent of $ Z $ as long as $ X $ is \emph{not} independent of $ Z $.  All IV-strong or ICM-strong instruments are therefore MDep-strong by construction. The converse is, however, not true. Like in the case of ICM estimators, \Cref{Ass_dC:Ident_Relv} can hold even if there are fewer instruments than covariates, e.g., \textcite{tsyawo2023feasible}. This feature of the MDep can be explored to attain identification without excludability: \Cref{ex:illus_no_excl_nmono,ex:illus_no_excl_skedastic}.
	
\begin{remark}
The MDep accommodates the broadest possible set of instruments in any empirical setting: it includes all IV- and ICM-relevant instruments, and even those that are IV- or ICM-irrelevant yet dependent on covariates in the sense of \Cref{Ass_dC:Ident_Relv}.
\end{remark}

The single-index structure of the models in \Cref{SubSect_Model} offers the advantage of a tractable characterisation and statistical testing of the relevance condition (\Cref{Ass_dC:Ident_Relv}). General non-single-index structures and non-additively separable disturbance functions can be considered at the cost of a less intuitive relevance identification condition. 

\begin{remark}\label{Rem:mod_class}
A more general class of applicable models accommodates potentially non-single-index structures, non-additive disturbances, or both, taking the form $Y = G(X, U; \theta_o)$, where $G(X, U; \theta)$ is invertible in $U$, such that $U(\theta_o) := G^{-1}(Y, X; \theta_o)$, and $X$ may be endogenous. In this broader setting, the relevance condition becomes $X^g(\theta)\uptau \not\independent Z$ for all $\uptau \neq 0$ and $\theta \in \Theta$.\footnote{See the discussion around \eqref{eqn:dCov_pos}.}
\end{remark}

Two non-nested exogeneity conditions apply under the MDep framework. The first is a standard MDI exogeneity condition of independence between $Z$ and $U$.
\begin{assumption}[Exogeneity I]\label{Ass_dC:Ident_Exog_I} \( U \independent Z \). 
	\end{assumption}
\noindent From a model specification perspective, \Cref{Ass_dC:Ident_Exog_I} is testable using the tests of \textcite{sen2014testing,davis2018applications,xu2021omnibus}. \Cref{Ass_dC:Ident_Exog_I} rules out conditional scale heterogeneity, e.g., heteroskedasticity. However, exploiting the absolute value in the objective function \eqref{eqn:obj_fun_gen_model}, the following exogeneity condition can also be exploited for identification.
    \begin{namedassumption}{\ref{Ass_dC:Ident_Exog_I}$^\prime$}[Exogeneity II]\label{Ass_dC:Ident_Exog_II}
    \( \mathrm{med}\big[ (U - U^\dagger) \mid \widetilde{\sigma}( [X,Z] )\big] = 0 \ a.s. \)
\end{namedassumption}
    \noindent Exogeneity in the MDep framework only requires either \Cref{Ass_dC:Ident_Exog_I} or \Cref{Ass_dC:Ident_Exog_II} to hold. Moreover, both exogeneity conditions are non-nested. Consider two DGPs with $X=Z+V$: (a) $U=\rho V + \xi$, $ \rho \neq 0 $ with $ Z, \, V, \, \xi $ all independent and (b) $U=|X|\xi, \, \xi \sim \mathcal{N}(0,1) $. (a) satisfies \Cref{Ass_dC:Ident_Exog_I} but not \Cref{Ass_dC:Ident_Exog_II}, whereas (b) satisfies \Cref{Ass_dC:Ident_Exog_II} but not \Cref{Ass_dC:Ident_Exog_I}. 
    
    \Cref{Ass_dC:Ident_Exog_II} accommodates some form of \emph{conditional scale heterogeneity} of $U$ in $ [X,Z] $, e.g., conditional heteroskedasticity.\footnote{Heteroskedasticity in the traditional sense does not apply to heavy-tailed distributions such as the Cauchy. However, it is conceivable that the scale parameter of \( (U - U^\dagger) \mid [X, X^\dagger, Z, Z^\dagger] \) is non-degenerate.} In the aforementioned example (b), $(U-U^\dagger)\mid [X,X^\dagger,Z,Z^\dagger] \ \sim \mathcal{N}(0,|X|+|X^\dagger|) $, whence $ \mathrm{med}\big[ (U - U^\dagger) \mid X,X^\dagger,Z,Z^\dagger\big] = 0 \ a.s.$\footnote{This type of characterisation applies to the entire family of symmetric $\alpha$-stable distributions.} Unlike the ICM and conventional IV estimators, the MDep is not robust to arbitrary forms of heteroskedasticity if $\E[U^2]<\infty$.\footnote{If the violation of \Cref{Ass_dC:Ident_Exog_II} arises \emph{solely} from arbitrary scale heterogeneity in \( U \), a potential remedy---left unexplored in this paper---is to estimate the conditional scale function alongside $\theta_o$ and scale-standardise $U(\theta)$ \`a la, e.g., \citet{wooldridge2010econometric,romano2017resurrecting,alejo2024endogenous}.} \Cref{Ass_dC:Ident_Exog_II} requires that the median of $ (U - U^\dagger) $ conditional on $\widetilde{\sigma}( [X,Z] )$ be zero \emph{almost surely}, thereby unifying the median, the mean (if it exists), and the mode (if $ \widetilde{U} $ is unimodal) as a natural point on which to impose exogeneity, thanks to symmetrisation. Unlike \Cref{Ass_dC:Ident_Exog_II}, which is imposed on \emph{pairwise differences} in disturbances, similar exclusion restrictions on conditional quantiles are imposed on the \emph{levels} of disturbances for quantile estimators under (possible) endogeneity, see e.g., \textcite[Assumption A.2]{chernozhukov2006instrumental}, \textcite[Assumption 3.6]{lee2007endogeneity}, and \textcite[Assumption B2]{powell1991estimation}. \Cref{Ass_dC:Ident_Exog_II} can be expressed as $ \E\big[\indicator{\widetilde{U} \leq 0} - 0.5 \mid \widetilde{\sigma}( [X,Z] )\big] = 0 \ \text{a.s.} $; this condition is testable from a model specification perspective using a suitable extension of, for example, ICM tests---see \textcite{bierens1982consistent,dominguez2015simple,su2017martingale,xu2020martingale,jiang-tsyawo-2022consistent}.\footnote{This task, however, is left for future work due to considerations of scope and space. }

    \begin{remark}\label{Remark:Exog_Szekely_w}
    Neither \Cref{Ass_dC:Ident_Exog_I} nor \Cref{Ass_dC:Ident_Exog_II} requires the existence of any moment of \( U \). \Cref{Ass_dC:Ident_Exog_II} is tied to the integrating measure of \citet{szekely2007measuring}, which yields the absolute value function in \eqref{eqn:obj_fun_gen_model}. As a result, the MDep behaves like a specially weighted least absolute deviations (LAD) estimator on pairwise differences in disturbances. In contrast, arbitrary integrating measures in \eqref{eqn:dCov1} do not deliver this extra property.
    \end{remark} 

The MDep objective function \eqref{eqn:obj_fun_gen_model} is non-convex because $ \mathcal{Z}_{ij,n} $ is not non-negative. This renders typical QR identification proof techniques that draw on the convexity of the objective function, e.g., \textcite{koenker1978regression,powell1991estimation,oberhofer2016asymptotic}, inapplicable. In contrast, this paper leverages the non-negativity and ``omnibus" properties of the dCov measure---namely Properties \Cref{Property:dCov_geq0} and \Cref{Property:dCov_indep0}---to establish identification.

\begin{theorem}\label{Theorem:Identification}
Suppose \Cref{Ass_dC:Reg,Ass_dC:Ident_Relv} hold. If, in addition, either Assumption~\ref{Ass_dC:Ident_Exog_I} or ~\ref{Ass_dC:Ident_Exog_II} is satisfied, then for every $\varepsilon > 0$, there exists a constant $\delta_\varepsilon > 0$ such that
    \[
        \inf_{\{\theta \in \Theta : \|\theta - \theta_o\| \ge \varepsilon\}} Q(\theta) > \delta_\varepsilon.
    \]
\end{theorem}
\noindent \Cref{Theorem:Identification} shows that under the given assumptions, the minimand $Q(\theta)$ has a unique minimum.

\begin{figure}[H]
\centering 
\caption{$Q(\theta)$}
\begin{subfigure}{0.32\textwidth}
\centering
\includegraphics[width=1\textwidth]{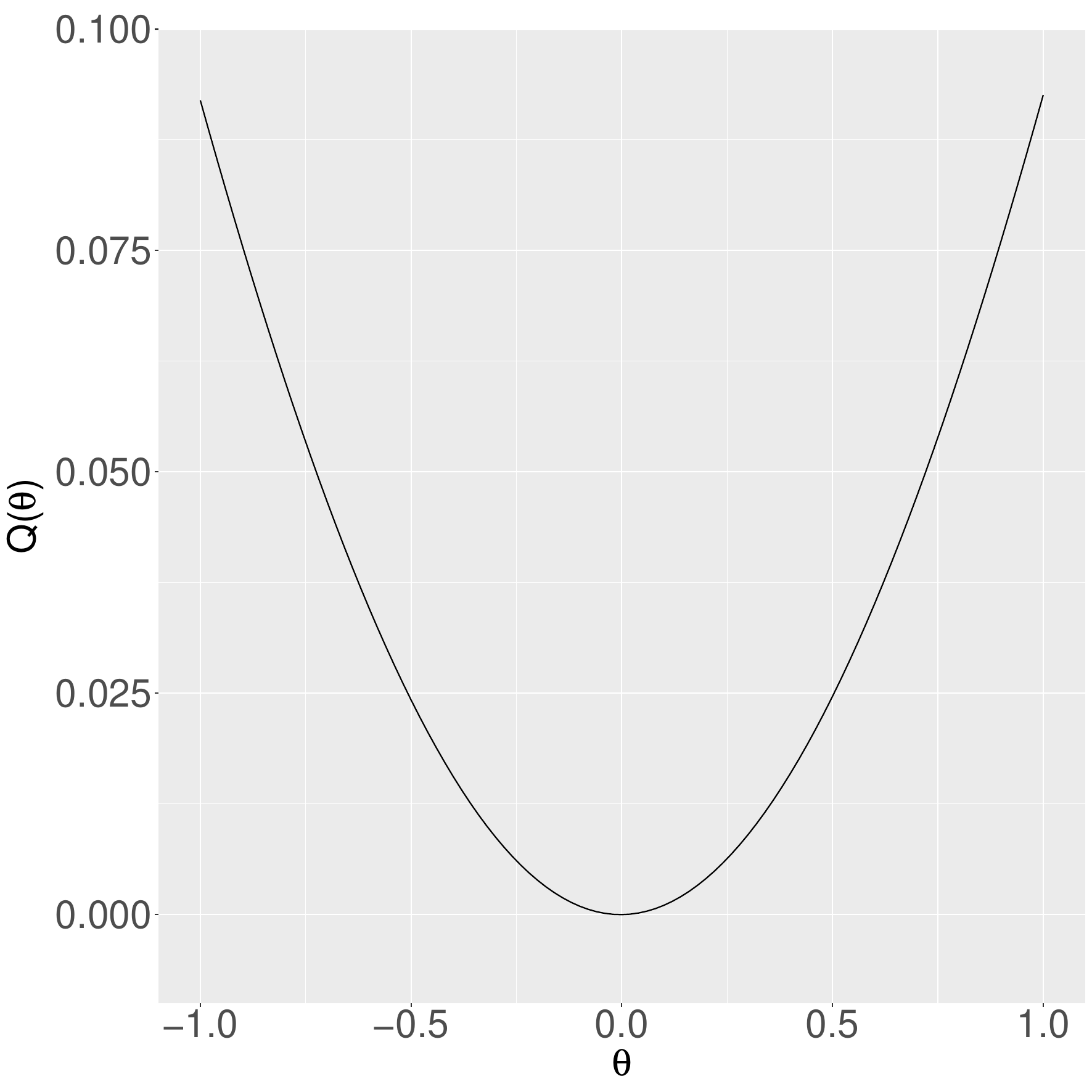}
\caption{ $ U \sim \mathcal{N}(0,0.5)$}
\end{subfigure}
\begin{subfigure}{0.32\textwidth}
\centering
\includegraphics[width=1\textwidth]{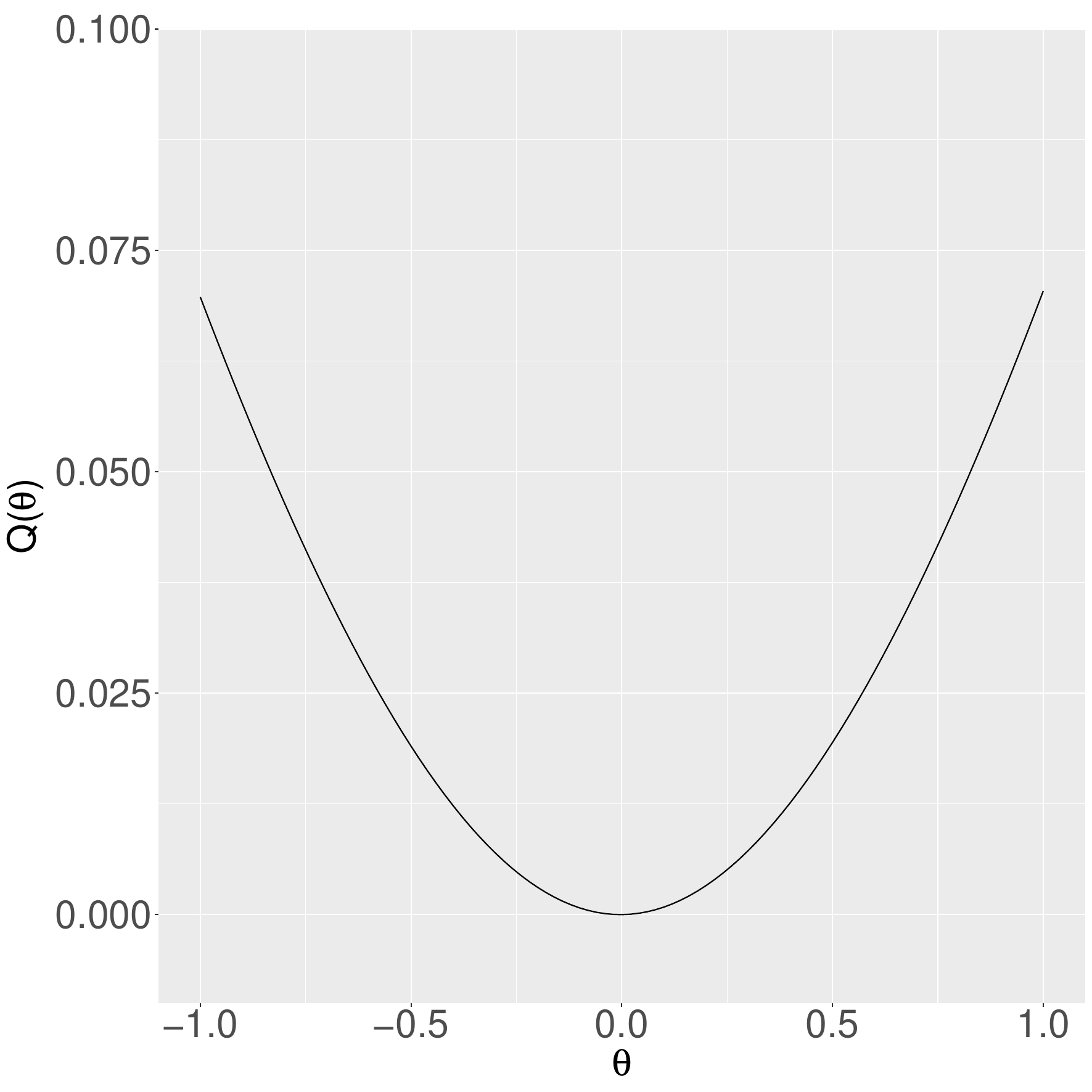}
\caption{ $ U \sim \mathcal{C}(0,0.5)$}
\end{subfigure}
\begin{subfigure}{0.32\textwidth}
\centering
\includegraphics[width=1\textwidth]{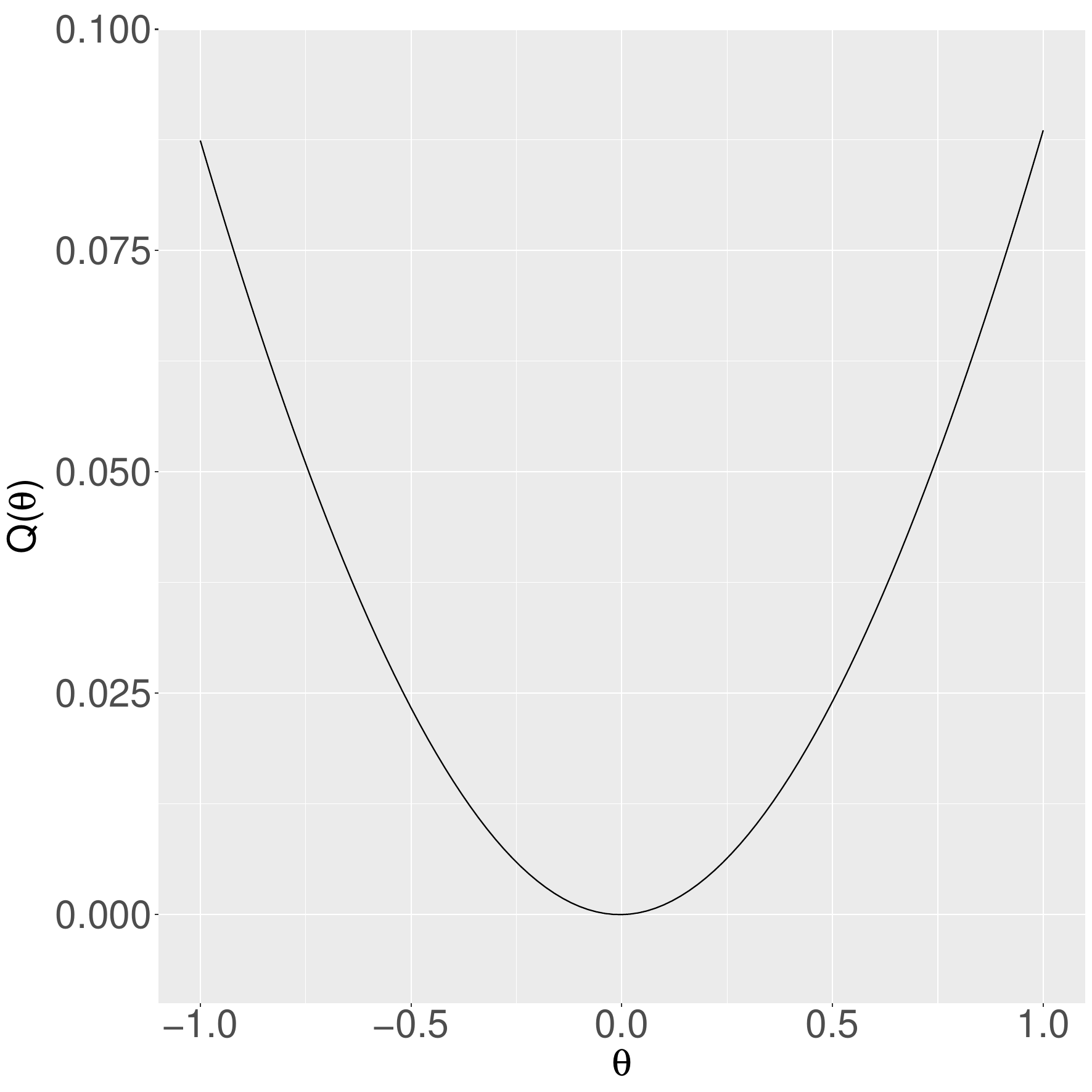}
\caption{ $ U \sim \mathcal{U}[0,\sqrt{6}]$}
\end{subfigure}
\label{fig:Qtheta}
\end{figure}

For illustrative purposes, consider the setting where \( \theta_o = 0 \), \( X \sim \mathrm{Ber}(0.5) \), \( X=Z \independent U \), and \( Y = X\theta_o + U \), under three distributions for \( U \): (a) \( U \sim \mathcal{N}(0,0.5) \), (b) \( U \sim \mathcal{C}(0,0.5) \), and (c) \( U \sim \mathcal{U}[0,\sqrt{6}] \). The corresponding population objective functions \( Q(\theta) := \E[\mathcal{Z}(|\widetilde{U}(\theta)| - |\widetilde{U}|)] \) are plotted in \Cref{fig:Qtheta}. The minima are well defined, \( X \) is discrete, and \( U \), in case (b), lacks a finite first moment.

With the identification result in hand, this subsection concludes with a proof of consistency of the MDep. The following standard sampling scheme is imposed.
\begin{assumption}
    \label{Ass_dC:Sampling} $ \big\{W_i: i \in [n] \big\} $ are independently and identically ($i.i.d.$) distributed random vectors.
\end{assumption}
\begin{theorem}\label{Theorem:Consistency}
    Suppose the conditions of \Cref{Theorem:Identification} hold, then in addition to \Cref{Ass_dC:Sampling}, the MDep $\widehat{\theta}_n$ converges almost surely to $\theta_o$ as $n \to \infty$, i.e., \( \widehat{\theta}_n \xrightarrow{a.s.} \theta_o. \)
\end{theorem}

\subsection{Conditional functionals and parameters of interest}\label{Sub_Sect:Par_Interpret}
Whenever elements of $\theta_o$ are themselves of interest, e.g., in a structural economic model with an economically meaningful $\theta_o$, the interpretation is direct. However, when $\theta_o$ is not of direct interest \emph{per se}, but the partial effects obtained therefrom are, it is essential first to determine the identified conditional functional.

Consider the simple linear model $ Y = X\theta_o + U $ where $ X=Z $ and $p_X=1$. Under \Cref{Ass_dC:Ident_Exog_I}, $ Q_{Y|X}(\tau|x) = x\theta_o $ for all $\tau \in (0,1)$ where $ Q_{Y|X}(\tau|x) $ is the $\tau$'th quantile of $Y$ conditional on $X=x$. When $ \E[|U|]<\infty $, then $ \E[Y|X=x] = x\theta_o $ as well. Under \Cref{Ass_dC:Ident_Exog_II}, $ \mathrm{med}[ (Y - Y^\dagger) \mid (X-X^\dagger) ] = (X-X^\dagger)\theta_o $. Hence, $\theta_o$ is \emph{the median partial effect} of a unit increase in $X$ on the outcome $Y$, \emph{relative to an observationally equivalent agent}. 

Unlike the simple linear example above, the partial effect of $X$ is not constant for non-linear $G(\cdot)$. For example, consider the model $ \log(Y) = X\theta_o + U $ where $G(\cdot)=\exp(\cdot)$. Under \Cref{Ass_dC:Ident_Exog_II}, $ \mathrm{med}[ \log(Y) - \log(Y^\dagger) \mid X,X^\dagger] = \log\big(\mathrm{med}[(Y/Y^\dagger) \mid X,X^\dagger]\big) = (X-X^\dagger)\theta_o $, i.e., $ \displaystyle \mathrm{med}\Big[ \frac{Y-Y^\dagger}{Y^\dagger} \Big| (X-X^\dagger) \Big] = \exp\big((X - X^\dagger)\theta_o\big)-1$, and the partial effects are interpretable as changes in fractions or percentages. As the resulting partial effect is a function of $[X,X^\dagger]$, interesting summaries of this heterogeneity can be reported, such as the average partial effect or the partial effect at the average.

\subsection{Asymptotic normality}
Define the score function $ \mathcal{S}_n(\theta):= \E_n[\psi(W_i,W_j;\theta)]$ where $ \psi(W_i,W_j;\theta) := \mathcal{Z}_{ij}\sgn\big( \widetilde{U}_{ij}(\theta) \big) \widetilde{X^g}_{ij}(\theta)' $, with $ \psi(W_i,W_j) := \psi(W_i,W_j;\theta_o) $. $ h(Z_i,Z_j) =: \mathcal{Z}_{ij} = \mathcal{Z}_{ji} $ and $ \sgn( \widetilde{U}_{ij} ) \widetilde{X^g}_{ij} = \sgn( \widetilde{U}_{ji} ) \widetilde{X}_{ji}^g $ with $ \widetilde{X^g}_{ij} := \widetilde{X^g}_{ij}(\theta_o) $ hence $ \psi(\cdot,\cdot) $ is permutation symmetric. Denote the cumulative distribution function and the probability density functions of $ \widetilde{U} $ conditional on $\widetilde{\sigma}( [X,Z] )$ by $ F_{\widetilde{U} \mid \widetilde{\sigma}( [X,Z] )}(\cdot) $ and $ f_{\widetilde{U} \mid \widetilde{\sigma}( [X,Z] )}(\cdot) $, respectively. Further, define  $ v_n(\theta):= \sqrt{n}\big( \mathcal{S}_n(\theta) - \mathcal{S}(\theta) \big)  $ where $ \mathcal{S}(\theta) := \E[\mathcal{S}_n(\theta)] = \E\big[\psi(W,W^\dagger;\theta)\big] $, $ \psi^{(1)}(W_i) := \E\big[\psi(W_i,W_j)|W_i\big] $, and the Hessian 
\[ 
\mathcal{H} := 2\E\big[f_{\widetilde{U} \mid \widetilde{\sigma}( [X,Z] )}(0)\mathcal{Z}\widetilde{X^g}'\widetilde{X^g}\big] + \E\big[\sgn(\widetilde{U}) \mathcal{Z}\widetilde{X^{gg}}\big]. 
\] Finally, let $ \partial^{-}|\hat{q}| $ and $ \partial^{+}|\hat{q}| $, respectively, denote the left- and right-derivatives of $ |q| $ with respect to $q$ at $ q=\hat{q} $. 

\begin{assumption}[Asymptotic Linearity of $ \widehat{\theta}_n $]\label{Ass_dC:AsymN}
	\quad
	\begin{enumerate}[label=(\alph*)]
		
  \item\label{Ass_dC:AsymN_IntTheta} $ \theta_o $ is an interior point of $ \Theta $;
	
 \item\label{Ass_dC:AsymN_DiffFU} $ F_{\widetilde{U} \mid \widetilde{\sigma}( [X,Z] )}(\cdot) $ is continuously differentiable with density $ f_{\widetilde{U} \mid \widetilde{\sigma}( [X,Z] )}(\cdot) $, and there exists a constant $ f_o \in (0,\infty) $ such that, for all $ \epsilon $ in a neighbourhood of zero, $ \displaystyle f_o^{-1} < f_{\widetilde{U} \mid \widetilde{\sigma}( [X,Z] )}(\epsilon) \leq \sup_{e \in \mathbb{R}}f_{\widetilde{U} \mid \widetilde{\sigma}( [X,Z] )}(e) \leq f_o^{1/4} \ a.s. $
		
  \item \label{Ass_dC:AsymN_HNonSing} $ \mathcal{H} $ is non-singular.
	\end{enumerate}
\end{assumption}
\Cref{Ass_dC:AsymN}\Cref{Ass_dC:AsymN_IntTheta} is standard. Conditions similar to \Cref{Ass_dC:AsymN}\Cref{Ass_dC:AsymN_DiffFU} are standard in the quantile regression literature---cf. \textcite[Assumption 3.6]{lee2007endogeneity}, \textcite[Assumption 2 R.4]{chernozhukov2006instrumental}, \textcite[Assumption R.4]{chernozhukov2008instrumental}, \textcite[Assumption C4. (i) and (ii)]{powell1991estimation}, \textcite[Assumption A.14]{oberhofer2016asymptotic}), and \citet[Condition D.1]{xu2021omnibus}. It ensures the Hessian is well-defined. As $ \mathcal{Z} $ has both negative and positive values in its support, the Hessian $ \mathcal{H} = 2\E\big[f_{\widetilde{U} \mid \widetilde{\sigma}( [X,Z] )}(0)\mathcal{Z}\widetilde{X^g}'\widetilde{X^g} \big] + \E\big[\operatorname{sgn}(\widetilde{U}) \mathcal{Z}\widetilde{X^{gg}}\big] $ cannot be positive definite by construction; non-singularity (\Cref{Ass_dC:AsymN}\Cref{Ass_dC:AsymN_HNonSing}) is thus required---cf. \citet[Assumption N2]{honore1994pairwise}. The second term in the Hessian disappears if \Cref{Ass_dC:Ident_Exog_II} holds (which ensures $ \E\big[\sgn(\widetilde{U}) \mid \widetilde{\sigma}([X,Z]) \big] = 0 \, a.s. $ ) or the model is linear (which implies $ \widetilde{X^{gg}}(\theta) = 0 $ for all $ \theta \in \Theta $ ).

Define $ \Omega := 4\E[\psi^{(1)}(W)\psi^{(1)}(W)'] $. The following theorem states the asymptotic linearity and normality of the MDep estimator.      
\begin{theorem}\label{Theorem:Normality}
Suppose that \Cref{Ass_dC:AsymN} holds in addition to the conditions of \Cref{Theorem:Consistency}. Then the MDep estimator \( \widehat{\theta}_n \) satisfies:
\begin{enumerate}[(a)]
    \item asymptotic linearity:
    \begin{equation*}
        \sqrt{n}(\widehat{\theta}_n - \theta_o)
        = -\,\mathcal{H}^{-1} \cdot \frac{2}{\sqrt{n}} \sum_{i=1}^{n} \psi^{(1)}(W_i)
        + o_p(1) \quad \text{ and}\, ;
    \end{equation*}
    
    \item asymptotic normality:
    \[
        \sqrt{n}(\widehat{\theta}_n - \theta_o)
        \xrightarrow{d} \mathcal{N}\big(0,\, \mathcal{H}^{-1} \Omega \mathcal{H}^{-1}\big).
    \]
\end{enumerate}
\end{theorem}

\noindent \Cref{Theorem:Normality} establishes the asymptotic normality of the MDep estimator. However, like other MDI estimators, the MDep is not efficient \citep{poirier-2017-efficient}. Although a two-step procedure (not implemented in this paper) for achieving efficiency---along the lines of \citet[Section 4]{dominguez2004consistent}---can be applied, it introduces several complications. Specifically, such an approach would require:  
(1) smoothing the inherently non-smooth moment equations of the MDep,  
(2) non-parametrically estimating components of the efficient GMM objective function, (3) selecting tuning parameters for both smoothing and estimation steps, and (4) accepting the risk of identification failure or inconsistency if the error term \( U \) lacks finite moments.

\subsection{Consistent covariance matrix estimation}\label{SubSect:Covariance_Matrix}

The preceding subsection established the asymptotic normality of the MDep estimator. Building on that result, this subsection introduces a consistent estimator of the asymptotic covariance matrix and proves its consistency. This consistency is crucial for conducting valid statistical inference, including $t$-tests, Wald tests, and the construction of confidence intervals. Define $\displaystyle \widehat{\psi}^{(1)}(W_i) := \frac{1}{n-1} \sum_{j\neq i}^{n} \widehat{\psi}(W_i,W_j) $ where $ \widehat{\psi}(W_i,W_j) := \mathcal{Z}_{ij,n}\sgn\big( \widetilde{U}_{ij}(\widehat{\theta}_n) \big) \widetilde{X^g}_{ij}(\widehat{\theta}_n)'$. The estimators of $ \Omega $ and $ \mathcal{H} $ are given by \( \displaystyle \widehat{\Omega}_n = 4\E_n[\widehat{\psi}^{(1)}(W_i)\widehat{\psi}^{(1)}(W_i)'] \) \quad and
	\begin{equation*}
	\widehat{\mathcal{H}}_n = \frac{1}{n(n-1)}\sum_{i=1}^{n}\sum_{j\neq i}^{n} \Big\{ \frac{\indicator{ |\widetilde{U}_{ij}(\widehat{\theta}_n)| \leq \hat{c}_n }}{\hat{c}_n} \mathcal{Z}_{ij,n} \widetilde{X^g}_{ij}(\widehat{\theta}_n)'\widetilde{X^g}_{ij}\big(\widehat{\theta}_n\big) + \sgn\big(\widetilde{U}_{ij}(\widehat{\theta}_n)\big) \mathcal{Z}_{ij,n}\widetilde{X^{gg}_{ij}}\big( \widehat{\theta}_n \big) \Big\}
	\end{equation*} respectively, where $ \hat{c}_n $, a possibly random bandwidth sequence, and the uniform kernel, as proposed by \citet{powell1991estimation}, is used to estimate the conditional density in $ \mathcal{H} $.\footnote{The second term in $ \widehat{\mathcal{H}}_n $ is identically zero for the linear model.} The estimator of the covariance matrix is $ \widehat{\mathcal{H}}_n^{-1} \widehat{\Omega}_n \widehat{\mathcal{H}}_n^{-1} $. An additional condition is imposed on the bandwidth sequence \( \hat{c}_n \) to ensure the consistency of \( \widehat{\mathcal{H}}_n \).
	\begin{assumption}\label{Ass_dC:Bandwith_Consistent}
		For some non-stochastic sequence $ c_n $ with $ c_n \rightarrow 0 $ and $ \sqrt{n}c_n \rightarrow \infty $, $\displaystyle \plim_{n \rightarrow \infty} (\hat{c}_n/c_n) = 1 $.
	\end{assumption}

	\noindent \Cref{Ass_dC:Bandwith_Consistent} corresponds to \citet[Assumption D1]{powell1991estimation} and is used to establish the consistency of $ \widehat{\mathcal{H}}_n $. It requires that the bandwidth sequence $\hat{c}_n$ satisfy the rate conditions $\hat{c}_n = o_p(1)$ and $\hat{c}_n^{-1} = o_p(\sqrt{n})$. The following theorem states the consistency of the covariance matrix estimator. 
	
	\begin{theorem}\label{Theorem:Consistency_Cov_Matrix}
		Suppose the conditions of \Cref{Theorem:Normality} hold. If, in addition, \Cref{Ass_dC:Bandwith_Consistent} holds, then $\widehat{\mathcal{H}}_n^{-1} \widehat{\Omega}_n \widehat{\mathcal{H}}_n^{-1} \xrightarrow{p} \mathcal{H}^{-1}\Omega\mathcal{H}^{-1} $ as $n \rightarrow \infty$.
	\end{theorem}
	\noindent Estimating the asymptotic covariance matrix involves specifying the bandwidth sequence $ \hat{c}_n $. The bandwidth sequence used throughout this paper follows the approach in \citet[Sect. 3.4.2]{koenker2005quantile} and is given by \( \displaystyle \hat{c}_n = \sqrt{2} k_n \min\Big\{\widehat{\sigma}_{\widehat{U}},\ \frac{\mathrm{IQR}(\widehat{U})}{1.34}\Big\} \) where $k_n: = n^{-1/3}\Big(\frac{3}{4\pi}\big(\Phi^{-1}(0.975)\big)^2\Big)^{1/3} $ is the \citet{hall1988distribution} bandwidth sequence. The terms $\widehat{\sigma}_{\widehat{U}}$ and $\mathrm{IQR}(\widehat{U})$ denote the sample standard deviation and inter-quartile range, respectively, of the residuals $ \big\{\widehat{U}_i, \ i \in [n] \big\} $.

\subsection{Testing the MDep relevance condition}
The weak relevance condition (\Cref{Ass_dC:Ident_Relv}) makes the MDep a powerful tool in a practitioner's toolkit, especially in dealing with unavailable or weak instruments. The practical usefulness of the MDep thus lies crucially in its testability. This subsection demonstrates the testability of the MDep relevance condition (\Cref{Ass_dC:Ident_Relv}) within the class of single-index models.\footnote{MDep relevance in the more general class in \Cref{Rem:mod_class} is left for future work.} Partition $ X $ as $ X = [D, \, Z_{-D}]$ where $ D \in \mathbb{R}^{p_D} $ and $Z_{-D} \in \mathbb{R}^{p_X - p_D} $, respectively, collect endogenous and exogenous covariates. Define $ \mathcal{D}_l(\gamma) := D_l - [D_{-l},\ Z_{-D}]\gamma, \ l\in [p_D] $ where $D_l$ denotes the $l$'th element of $D$ and $D_{-l} \in \mathbb{R}^{p_D - 1} $ excludes $D_l$ from $D$. Let $\mathbb{S}^p$ denote a compact subset of $\mathbb{R}^p,\, p\geq 1$. The following theorem shows the testability of the MDep relevance condition.

	\begin{theorem}\label{Theorem:Test_LC}
	Suppose \Cref{Ass_dC:Reg}\Cref{Ass_dC:Reg_Dominance} holds, then a test of MDep relevance (\Cref{Ass_dC:Ident_Relv}) can be formulated via the following hypotheses:
		\begin{align*}
			\mathbb{H}_o &: \mathcal{D}_{l^*}(\gamma^*) \independent Z \text{ for some } \{\gamma^*, \ l^* \} \in \mathbb{S}^{p_X - 1} \times [p_D]; \text{ and} \\
			\mathbb{H}_a &: \mathcal{D}_l(\gamma) \not \independent Z \text{ for all } \{\gamma, \ l \} \in \mathbb{S}^{p_X - 1} \times [p_D].
		\end{align*}
	\end{theorem}
	\noindent Thanks to Properties \Cref{Property:dCov_geq0} and \Cref{Property:dCov_indep0} of the dCov measure, $\mathbb{H}_o$ and $\mathbb{H}_a$ can be equivalently cast as
    \begin{align*}
        \widetilde{\mathbb{H}}_o: \ \min_{\{\gamma, l\} \in \mathbb{S}^{p_X - 1} \times [p_D]} \mathcal{V}^2\big(\mathcal{D}_l(\gamma), \ Z\big) = 0 \quad \text{ v.s. } \quad \widetilde{\mathbb{H}}_a: \ \min_{\{\gamma,l\} \in \mathbb{S}^{p_X - 1} \times [p_D] } \mathcal{V}^2\big(\mathcal{D}_l(\gamma),\ Z\big) > 0.
    \end{align*}
    
\noindent It follows from the above reformulation that \Cref{Ass_dC:Ident_Relv} is testable using tests of independence between MDep regression disturbance terms $ \mathcal{D}_l(\gamma), \, l\in [p_D] $ and $Z$, e.g., \textcite{sen2014testing,davis2018applications,xu2021omnibus}.

\section{Simulation Experiments}\label{Sect:Sim}
This section examines the finite sample performance of the MDep using simulations. $Y = [X_1, X_2]\theta_o + U$ is the data-generating process, where $\theta_o = [0.5, \ -0.5]'$. Auxiliary variables include $\dot{X} \sim \mathcal{N}(0,\mathrm{I}_2) $, $V = Ua + \dot{U}\sqrt{1-a^2}, \, a=-0.2$, $\dot{U} \sim \mathcal{U}[-\sqrt{3},\sqrt{3}] $, and $ U\independent \dot{U} $. $U\sim (\chi_1^2-1)/\sqrt{2} $ unless otherwise specified. The data-generating processes (DGPs) considered are the following. 
\begin{description}
    \item[LM--0A:] $ U \sim \mathcal{N}(0,1), \ Z=X=\dot{X}$;
    \item[LM--0B:] $U\mid X \sim \mathcal{C}\big(0,\ 0.1+|X_1|\big) $, $Z=X=\dot{X}$;
    \item[LM--1A:] $X_1 = \dot{X}_1 + V$, $X_2=\dot{X}_2$, $Z = \big[\indicator{ |\dot{X}_1| < -\Phi^{-1}(0.25) }, \, X_2 \big]$;
    \item[LM--1B:] $X_1 = \indicator{V < - |\dot{X}_1| - \Phi^{-1}(0.25) } $, $X_2=\dot{X}_2$, $Z = \dot{X}$;
    \item[LM--1C:] $U\mid X \sim \mathcal{N}\big(0,\, (0.1+|X_1|)^{-2}\, \big) $, $Z=\dot{X}$, $ X_1 = \dot{X}_1 + \dot{U} $, $X_2 = \dot{X}_2$;
    \item[LM--2A:] $\dot{Z} \sim \mathcal{N}(0,1) $, $X_1 = \dot{Z} + V$, $ Z = \dot{Z}^2 - a\dot{Z} $, $X_2=Z$;
    \item[LM--2B:] $\Ddot{X} = \dot{X}/||\dot{X}|| $, $X_1 = \Ddot{X}_1 - aU$, $ Z = X_2 = \Ddot{X}_2 $;
    \item[LM--3:] $Z\sim \mathcal{N}(0,1)$, $X_1 = \dot{U}Z^2 - aU$, $X_2=Z$.
\end{description}

$ X:=[X_1, X_2] $ is exogenous in DGPs LM--0A and LM--0B, while $X_1$ is endogenous in the remaining DGPs.\footnote{Specifically, it is scale-endogenous in DGP LM--1C.} DGPs LM--1A, LM--1B, LM--2A, LM--2B, and LM--3 have non-monotone forms of relevance (see \Cref{ex:illus_no_excl_nmono,ex:illus_transform,ex:illus_no_excl_skedastic}). A transformation of $Z$ in DGPs LM--1A and LM--2B that induces correlation (or mean-dependence) between $X_1$ and $Z$ is impossible. The identifying variation in LM--2B is implicit; $Z$ and the exogenous variation in $X_1$, namely $\Ddot{X}_1$ are defined on the unit circle, and one can only determine the other up to sign. Instrument relevance in LM--3 is in the ``first-stage" skedastic function (see \Cref{ex:illus_no_excl_skedastic}). There is MDep identification without excludability in LM--2A through LM--3 (see \Cref{ex:illus_no_excl_nmono}). In LM--1A, the excluded instrument is discrete; in LM--1B, the endogenous covariate is discrete; and in both cases, the first-stage relationships are non-monotone (see \Cref{ex:illus_transform}). Conditional scale heterogeneity in LM--0B and LM--1C does not violate \Cref{Ass_dC:Ident_Exog_II}, and the first moment of $U$ in LM--0B does not exist (see \Cref{ex:illus_moment}). 

\begin{table}[!htbp]
	\centering
	\setlength{\tabcolsep}{4pt}
	\caption{Simulation Results - Linear Models I}
	\footnotesize
	\begin{tabular}{lcccccccccccc}
\toprule
& \multicolumn{4}{c}{\underline{$ n = 50 $}} & \multicolumn{4}{c}{\underline{$ n = 100 $}} & \multicolumn{4}{c}{\underline{$ n = 200 $}} \\ 
& M-$t$ & MAD & RMSE & Rej. & M-$t$ & MAD & RMSE & Rej. & M-$t$ & MAD & RMSE & Rej. \\ 
\cmidrule{2-13}
LM--0A & \multicolumn{12}{c}{\underline{$ U \sim \mathcal{N}(0,1), \ Z=X=\dot{X}$}} \\ 
\cmidrule{1-1}
MDep & -0.017 & 0.108 & 0.162 & 0.086 & -0.035 & 0.067 & 0.108 & 0.059 & -0.065 & 0.052 & 0.078 & 0.058 \\ 
MMD  & -0.007 & 0.105 & 0.157 & 0.079 & -0.023 & 0.068 & 0.105 & 0.059 & -0.069 & 0.048 & 0.075 & 0.060 \\ 
ESC6 & -0.019 & 0.110 & 0.159 & 0.072 & -0.006 & 0.069 & 0.106 & 0.058 & -0.055 & 0.049 & 0.076 & 0.062 \\ 
OLS  & -0.008 & 0.101 & 0.154 & 0.060 & -0.021 & 0.068 & 0.103 & 0.052 & -0.082 & 0.047 & 0.075 & 0.061 \\ 
\midrule
LM--0B & \multicolumn{12}{c}{\underline{$U\mid X \sim \mathcal{C}\big(0,\ 0.1+|X_1|\big)$, $Z=X=\dot{X}$}} \\ 
\cmidrule{1-1}
MDep & 0.004 & 0.287 & 0.454 & 0.070 & 0.029 & 0.192 & 0.287 & 0.054 & 0.051 & 0.129 & 0.192 & 0.049 \\ 
MMD  & -0.082 & 1.155 & 51.412 & 0.035 & -0.072 & 1.138 & 43.186 & 0.041 & 0.071 & 1.106 & 41.091 & 0.019 \\ 
ESC6 & -0.100 & 1.173 & 39.650 & 0.029 & -0.057 & 1.113 & 42.501 & 0.036 & 0.087 & 1.086 & 47.450 & 0.019 \\ 
OLS  & -0.086 & 1.186 & 34.230 & 0.023 & -0.047 & 1.137 & 42.029 & 0.025 & 0.012 & 1.137 & 33.921 & 0.017 \\ 
\midrule
LM--1A & \multicolumn{12}{c}{\underline{$X_1 = \dot{X}_1 + V$, $X_2=\dot{X}_2$, $Z = [\indicator{|\dot{X}_1| < -\Phi^{-1}(0.25)}, X_2]$}} \\ 
\cmidrule{1-1}
MDep & 0.057 & 0.109 & 0.205 & 0.029 & 0.131 & 0.092 & 0.165 & 0.036 & 0.135 & 0.088 & 0.142 & 0.045 \\ 
MMD  & -0.314 & 0.241 & 2.349 & 0.025 & -0.281 & 0.227 & 0.991 & 0.017 & -0.279 & 0.231 & 1.856 & 0.023 \\ 
ESC6 & -0.354 & 0.215 & 3.680 & 0.031 & -0.265 & 0.185 & 0.702 & 0.018 & -0.262 & 0.201 & 0.602 & 0.019 \\ 
TSLS & -0.067 & 0.704 & 8.180 & 0.001 & -0.083 & 0.667 & 7.555 & 0.000 & -0.062 & 0.718 & 47.917 & 0.000 \\ 
\midrule
LM--1B & \multicolumn{12}{c}{\underline{$X_1 = \indicator{V < -|\dot{X}_1| - \Phi^{-1}(0.25)}$, $X_2=\dot{X}_2$, $Z = \dot{X}$}} \\ 
\cmidrule{1-1}
MDep & 0.081 & 0.205 & 0.437 & 0.034 & 0.108 & 0.170 & 0.337 & 0.045 & 0.128 & 0.147 & 0.247 & 0.042 \\ 
MMD  & 0.129 & 0.663 & 17.784 & 0.024 & 0.101 & 0.523 & 1.446 & 0.032 & 0.041 & 0.380 & 0.643 & 0.048 \\ 
ESC6 & 0.118 & 0.606 & 4.280 & 0.022 & 0.065 & 0.496 & 1.465 & 0.016 & 0.049 & 0.368 & 0.612 & 0.035 \\ 
TSLS & 0.067 & 2.044 & 36.640 & 0.001 & 0.025 & 2.267 & 85.231 & 0.000 & -0.002 & 1.940 & 57.465 & 0.000 \\ 
\midrule
LM--1C & \multicolumn{12}{c}{\underline{$U\mid X \sim \mathcal{N}\big(0,(0.1+|X_1|)^{-2}\big)$, $Z=\dot{X}$, $X_1 = \dot{X}_1 + \dot{U}$, $X_2 = \dot{X}_2$}} \\ 
\cmidrule{1-1}
MDep & -0.014 & 0.048 & 0.074 & 0.036 & 0.033 & 0.032 & 0.046 & 0.046 & 0.046 & 0.023 & 0.032 & 0.037 \\ 
MMD  & -0.011 & 0.061 & 0.095 & 0.032 & 0.025 & 0.040 & 0.059 & 0.032 & 0.073 & 0.027 & 0.040 & 0.034 \\ 
ESC6 & -0.006 & 0.063 & 0.102 & 0.029 & 0.032 & 0.043 & 0.063 & 0.034 & 0.050 & 0.029 & 0.043 & 0.032 \\ 
TSLS & 0.026 & 0.053 & 0.086 & 0.012 & 0.035 & 0.036 & 0.053 & 0.019 & 0.060 & 0.024 & 0.036 & 0.030 \\ 
\midrule
LM--2A & \multicolumn{12}{c}{\underline{$\dot{Z} \sim \mathcal{N}(0,1)$, $X_1 = \dot{Z} + V$, $Z = \dot{Z}^2 - a\dot{Z}$, $X_2=Z$}} \\ 
\cmidrule{1-1}
MDep & 0.168 & 0.102 & 0.199 & 0.046 & 0.222 & 0.090 & 0.169 & 0.059 & 0.235 & 0.074 & 0.127 & 0.045 \\ 
MMD  & -0.032 & 0.376 & 15.543 & 0.015 & -0.031 & 0.383 & 2596.770 & 0.006 & -0.047 & 0.390 & 4.994 & 0.007 \\ 
ESC6 & -0.171 & 0.378 & 4.595 & 0.018 & -0.175 & 0.416 & 15.567 & 0.006 & -0.169 & 0.507 & 7.385 & 0.004 \\ 
\midrule
LM--2B & \multicolumn{12}{c}{\underline{$\Ddot{X} = \dot{X}/||\dot{X}||$, $X_1 = \Ddot{X}_1 - aU$, $Z = X_2 = \Ddot{X}_2$}} \\ 
\cmidrule{1-1}
MDep & 0.251 & 0.199 & 0.389 & 0.063 & 0.318 & 0.183 & 0.292 & 0.061 & 0.422 & 0.136 & 0.214 & 0.066 \\ 
MMD  & 0.382 & 0.477 & 1.012 & 0.036 & 0.414 & 0.480 & 1.083 & 0.023 & 0.419 & 0.506 & 1.130 & 0.017 \\ 
ESC6 & 0.450 & 0.537 & 1.152 & 0.040 & 0.436 & 0.566 & 1.280 & 0.033 & 0.424 & 0.561 & 8.208 & 0.024 \\ 
\midrule
LM--3 & \multicolumn{12}{c}{\underline{$Z\sim \mathcal{N}(0,1)$, $X_1 = \dot{U}Z^2 - aU$, $X_2=Z$}} \\ 
\cmidrule{1-1}
MDep & 0.263 & 0.076 & 0.139 & 0.083 & 0.360 & 0.058 & 0.099 & 0.070 & 0.319 & 0.038 & 0.068 & 0.078 \\ 
MMD  & 0.247 & 0.171 & 1.537 & 0.057 & 0.231 & 0.147 & 1.572 & 0.029 & 0.191 & 0.128 & 23.719 & 0.010 \\ 
ESC6 & 0.363 & 0.247 & 6.662 & 0.030 & 0.329 & 0.237 & 3.891 & 0.012 & 0.350 & 0.232 & 27.974 & 0.007 \\ 
\bottomrule
\end{tabular}
	\footnotesize
	\label{Tab:Sim_LM1}
\end{table}

For each of the DGPs, \Cref{Tab:Sim_LM1} reports the median $t$-statistic (M-$t$), the median absolute deviation (MAD), the root mean squared error (RMSE), and the 5\% rejection rate of the $t$-test of the null hypothesis $\theta_1 = 0.5$ across 1000 random samples with sample sizes \( n \in \{ 50, 100,200 \} \). Simulation results with larger samples and non-linear models are available in \Cref{Appendix_Sect:Simulations} of the Supplemental Appendix. Competing estimators include the proposed MDep, conventional instrumental variables (IV) estimators---namely, two-stage least squares (TSLS) and Ordinary Least Squares (OLS)---as well as ICM estimators MMD and ESC6 of \citet{tsyawo2023feasible} and \citet{escanciano2006consistent}, respectively. Across all DGPs, the MDep exhibits stable fine-sample performance and clear robustness to weak or non-monotone instrument relevance, heavy-tailed distributions, heteroskedastic disturbances, and scale endogeneity in $U$ subject to \Cref{Ass_dC:Ident_Exog_I} or~\ref{Ass_dC:Ident_Exog_II}. All estimators perform well in the baseline scenario LM--0A without endogeneity. However, in LM--0B,  where the first moment of $U$ does not exist and its scale is heterogeneous in $X_1$, only the MDep estimator remains reliable---its bias and RMSE shrink steadily with the sample size---while all competing estimators exhibit explosive RMSEs and unreliable inference, underscoring MDep's robustness to infinite-variance disturbances. 

In the \emph{weak, non-monotone, and discontinuous-covariate or instrument designs} (LM--1A and LM--1B), MDep continues to dominate: its median bias and RMSE are modest and improve with $n$, whereas \textsc{MMD}, \textsc{ESC6}, and especially \textsc{TSLS} display severe finite-sample distortions and oversized rejection rates. Under \emph{conditional heteroskedasticity} with scale endogeneity (LM--1C), all estimators improve markedly, but MDep achieves the lowest RMSE overall and the most stable rejection rates across sample sizes. Under endogeneity without excludability, where there is only one instrument for the two covariates (LM--2A, LM--2B, and LM--3), MDep again outperforms: its RMSEs remain small and converge rapidly, while competing estimators become erratic---showing extremely large RMSEs and severe over- or under-rejection. Overall, the simulations confirm that MDep provides accurate, numerically stable, and size-correct inference even in models featuring weak, non-monotone, or endogeneity without excludability, whereas the alternative estimators display unreliable behaviour under those conditions.

\section{Conclusion}\label{Sect:Conclusion}
This paper introduces the MDep estimator, which weakens the relevance condition of conventional IV, ICM, CM, and non-parametric IV methods to stochastic dependence between non-trivial linear combinations of $X$ and $Z$. Thus, under the MDep framework, one can exploit the maximum number of relevant instruments possible in any given empirical setting, subject to either of two non-nested exogeneity conditions.

The MDep framework offers a fundamentally distinct and practically valuable approach to addressing several challenges: (1) the absence of excluded instruments, (2) the weak instrument problem, and (3) the non-existence or contamination of the disturbance term due to outliers or random noise with potentially undefined moments. Moreover, the use of bounded one-to-one transformations of \( Z \) obviates moment bounds on \( Z \), further enhancing robustness. Consistent estimation and reliable inference are feasible without excludability, provided endogenous covariates are non-linearly dependent (in the distributional sense) on exogenous covariates. The MDep handles the weak IV problem by admitting instruments of which endogenous covariates may be uncorrelated or even mean-independent but \emph{not} independent.

Identification, consistency, and asymptotic normality hold in the MDep framework under mild regularity conditions. Moreover, the MDep covariance matrix estimator is shown to be consistent. To ensure the practical usefulness of the MDep, this paper shows the testability of the weak relevance condition. Illustrative examples backed by simulations showcase the remarkable properties of the MDep estimator vis-\`{a}-vis existing conventional IV and ICM methods.

\newpage 

\appendix
\begin{center}
    \LARGE{\bf Appendix}
\end{center}

The proofs of the results in the main text are organised in building blocks of lemmata.

\section{Proof of \Cref{Theorem:Identification}}

Maintaining \Cref{Ass_dC:Reg,Ass_dC:Ident_Relv}, the proof proceeds by first establishing identification under \Cref{Ass_dC:Ident_Exog_I}, followed by identification under \Cref{Ass_dC:Ident_Exog_II}.

\subsection*{Identification under Assumptions \ref{Ass_dC:Reg}, \ref{Ass_dC:Ident_Relv}, and \ref{Ass_dC:Ident_Exog_I}} The result is provided in the following lemma.
\begin{lemma}\label{lem:ident_I}
    Suppose \Cref{Ass_dC:Reg,Ass_dC:Ident_Relv,Ass_dC:Ident_Exog_I} hold, then for every $\varepsilon > 0$, there exists a constant $\delta_\varepsilon > 0$ such that
    \[
        \inf_{\{\theta \in \Theta : \|\theta - \theta_o\| \ge \varepsilon\}} Q(\theta) > \delta_\varepsilon.
    \]
\end{lemma}
\begin{proof}
If $\theta=\theta_o$, then $Q(\theta) = \E\big[\mathcal{Z}\big(|\widetilde{U}(\theta_o)|-|\widetilde{U}|\big)\big] = 0$ trivially. Conversely, suppose $\theta \in \Theta_{\varepsilon}:= \{\dot{\theta} \in \Theta : \|\dot{\theta} - \theta_o\| \ge \varepsilon\}$ for some $ \varepsilon>0$. The rest of the proof proceeds by contraposition. Under \Cref{Ass_dC:Reg} and that no non-trivial linear combination of $X$ is independent of $Z$ (\Cref{Ass_dC:Ident_Relv}), 
\[
    U(\theta) = U - X^g(\bar{\theta})(\theta-\theta_o) = U - g'(X\bar{\theta})X(\theta-\theta_o) \not\independent Z
\]
for some $ \bar{\theta} $ that satisfies $ || \bar{\theta} - \theta_o || \leq || \theta - \theta_o || $. Indeed, for any \( \theta \in \Theta_{\varepsilon} \), it follows from \Cref{Ass_dC:Reg}\ref{Ass_dC:Reg_UdiffXMeas} that \( (\theta-\theta_o)g'(X\bar{\theta}) \neq 0 \) with positive probability. This implies by Properties \Cref{Property:dCov_geq0} and \Cref{Property:dCov_indep0} that $ \displaystyle \delta_{\varepsilon}:= (1/2)\inf_{\theta \in \Theta_{\varepsilon}}Q(\theta)$ is positive, and hence $ \displaystyle \inf_{\theta \in \Theta_{\varepsilon}}Q(\theta) - Q(\theta_o) > \delta_{\varepsilon}$. This completes the proof under the stated conditions.
\end{proof}

\subsection*{Identification under Assumptions \ref{Ass_dC:Reg}, \ref{Ass_dC:Ident_Relv}, and \ref{Ass_dC:Ident_Exog_II}}
The proof of identification under \Cref{Ass_dC:Reg,Ass_dC:Ident_Relv,Ass_dC:Ident_Exog_II} requires the following preliminary decomposition result. For any \( \theta \in \Theta \), define \( q(W,W^\dagger;\theta):= \mathcal{Z} \big(|\widetilde{U}(\theta)| - |\widetilde{U}|\big) \) and
\[
\mathcal{T}_\theta := F_{\widetilde{U} \mid \widetilde{\sigma}( [X,Z] )}\big( \lambda \widetilde{X^g}(\bar{\theta})(\theta - \theta_o) \big)
\]where $ F_{\widetilde{U} \mid \widetilde{\sigma}( [X,Z] )}(\cdot) $ is the conditional distribution function, $ \bar{\theta} $ satisfies $ \widetilde{U}(\theta) = \widetilde{U} - \widetilde{X^g}(\bar{\theta})(\theta-\theta_o) $, and $ \lambda \in (0,1) $ under \Cref{Ass_dC:Reg} satisfies $ \int_{0}^{x}F_{\widetilde{U} \mid \widetilde{\sigma}( [X,Z] )}( \eta)d\eta=F_{\widetilde{U} \mid \widetilde{\sigma}( [X,Z] )}(\lambda x)x $ by the (integral) Mean Value Theorem (MVT). Under \Cref{Ass_dC:Ident_Exog_II}, $\mathcal{T}_\theta=1/2 \ a.s.$ at $\theta=\theta_o$.
		
\begin{lemma}\label{Lemma:dCov_Decomp}
Suppose Assumptions \ref{Ass_dC:Reg}\Cref{Ass_dC:Reg_UdiffXMeas} and \ref{Ass_dC:Ident_Exog_II} hold, then for any $ W,W^\dagger $ defined on the support of $ W_i $, 

\[
\E[q(W,W^\dagger;\theta)] = \int_{0}^{1}|2\tau-1| \E\big[(\mathcal{Z}| \widetilde{X^g}(\bar{\theta})(\theta-\theta_o)|) \mid \mathcal{T}_\theta = \tau \big] dF_{\mathcal{T}_\theta}(\tau).
\]
\end{lemma}
		
The inner integrand $ \mathcal{Z}\big| \widetilde{X^g}(\bar{\theta})(\theta-\theta_o)\big| $ has the same form as that of the distance covariance measure $ \mathcal{V}^2(U,Z) = \E[\mathcal{Z}|\widetilde{U}|] $. This is important because the expectation of the normalised minimand can be expressed in terms of the distance covariance between $ X^g(\bar{\theta})(\theta-\theta_o) $ and $ Z $. The proof is provided next.

\begin{proof}[\textbf{Proof of \Cref{Lemma:dCov_Decomp}}]
By \Cref{Ass_dC:Reg}\Cref{Ass_dC:Reg_UdiffXMeas}, the equality $ \widetilde{U}(\theta) = \widetilde{U} - \widetilde{X^g}(\bar{\theta})(\theta-\theta_o) $ holds by the MVT for any pair of random vectors $ W, W^\dagger $ where $ \bar{\theta} $ satisfies $ ||\theta-\bar{\theta}|| \leq ||\theta-\theta_o|| $. Knight's identity \citep{knight1998limiting} is given by
\[
  |\xi-b|-|\xi| = -b\big(\indicator{ \xi > 0 } - \indicator{ \xi < 0 } \big)  + 2\int_{0}^{b}\big(\indicator{ \xi \leq \eta } - \indicator{ \xi \leq 0 } \big)d\eta.
 \]Applying expectations to a continuously distributed $\xi$, one has 
 \begin{align*}
     \E[|\xi-b|-|\xi|] &= (2F_{\xi}(0) - 1)b + 2\int_0^b(F_{\xi}(\eta) - F_{\xi}(0))d\eta\\
     &= (2F_{\xi}(0) - 1)b + 2(F_{\xi}(\lambda b) - F_{\xi}(0))b\\
     &= (2F_{\xi}(\lambda b) - 1)b
 \end{align*}by the MVT and Knight's identity for some $ \lambda\in (0,1) $. It follows from the foregoing and the Law of Iterated Expectations (LIE) that
		\begin{equation}\label{eqn:Mean_Val_Vntheta}
		\begin{split}
		\E[q(W,W^\dagger;\theta)] &= \E[\mathcal{Z}(|\widetilde{U}(\theta)|-|\widetilde{U}|)] \\ 
		&= \E\big[\mathcal{Z}\big(|\widetilde{U} - \widetilde{X^g}(\bar{\theta})(\theta-\theta_o)|-|\widetilde{U}|\big)\big] \\
		& = \E\Big[\mathcal{Z}\Big(2F_{\widetilde{U} \mid \widetilde{\sigma}( [X,Z] )}\big( \lambda\widetilde{X^g}(\bar{\theta})(\theta-\theta_o)\big) - 1\Big) \widetilde{X^g}(\bar{\theta})(\theta-\theta_o)\Big]
	\end{split}
        \end{equation}
\noindent for some $ \lambda\in (0,1) $ thanks to the MVT and the LIE.	
		
It is claimed that $ \Big(2F_{\widetilde{U} \mid \widetilde{\sigma}( [X,Z] )}(\lambda b)-1\Big) b \geq 0 $ for all $ (b,\lambda) \in \mathbb{R} \times (0,1)$ under \Cref{Ass_dC:Ident_Exog_II}. If $b>0$, $\lambda b > 0$, $F_{\widetilde{U} \mid \widetilde{\sigma}( [X,Z] )}(\lambda b) \geq 1/2$ by \Cref{Ass_dC:Ident_Exog_II} and the monotonicity property of (conditional) cumulative distribution functions, thus $ \Big(2F_{\widetilde{U} \mid \widetilde{\sigma}( [X,Z] )}(\lambda b)-1\Big) b \geq 0 $ if $b\geq 0$. The same sequence of arguments shows that $ \Big(2F_{\widetilde{U} \mid \widetilde{\sigma}( [X,Z] )}(\lambda b)-1\Big) b \geq 0 $ if $b < 0$. Hence, $ \Big(2F_{\widetilde{U} \mid \widetilde{\sigma}( [X,Z] )}( \lambda\widetilde{X^g}(\bar{\theta})(\theta-\theta_o))-1\Big) \widetilde{X^g}(\bar{\theta})(\theta-\theta_o) = \Big| \Big(2F_{\widetilde{U} \mid \widetilde{\sigma}( [X,Z] )}( \lambda\widetilde{X^g}(\bar{\theta})(\theta-\theta_o))-1\Big) \widetilde{X^g}(\bar{\theta})(\theta-\theta_o) \Big|$ under \Cref{Ass_dC:Ident_Exog_II}. It therefore follows from \eqref{eqn:Mean_Val_Vntheta} that
		\begin{equation*}\label{eqn:Mean_Val_Vntheta2}
		\begin{split}
		\E[q(W,W^\dagger;\theta)] & =  \E\Big[\mathcal{Z}\Big|\Big(2F_{\widetilde{U} \mid \widetilde{\sigma}( [X,Z] )}( \lambda\widetilde{X^g}(\bar{\theta})(\theta-\theta_o))-1\Big) \widetilde{X^g}(\bar{\theta})(\theta-\theta_o)\Big|\Big]\\
		& = \E\big[\mathcal{Z}\big|2F_{\widetilde{U} \mid \widetilde{\sigma}( [X,Z] )}( \lambda\widetilde{X^g}(\bar{\theta})(\theta-\theta_o))-1\big|\times \big| \widetilde{X^g}(\bar{\theta})(\theta-\theta_o)\big|\big]\\
		& = \E\big[|2\mathcal{T}_\theta-1| \times \mathcal{Z} | \widetilde{X^g}(\bar{\theta})(\theta-\theta_o) |\big]\\
		& = \int_{0}^{1}|2\tau-1| \E\big[(\mathcal{Z}| \widetilde{X^g}(\bar{\theta})(\theta-\theta_o)|) \mid \mathcal{T}_\theta = \tau \big] dF_{\mathcal{T}_\theta}(\tau).
		\end{split}
		\end{equation*} 
	\noindent The fourth equality follows from the LIE.  
	\end{proof}
 
The proof of this part is completed in the following lemma.
\begin{lemma}\label{lem:ident_II}
    Suppose \Cref{Ass_dC:Reg,Ass_dC:Ident_Relv,Ass_dC:Ident_Exog_II} hold, then for every $\varepsilon > 0$, there exists a constant $\delta_\varepsilon > 0$ such that
    \[
        \inf_{\{\theta \in \Theta : \|\theta - \theta_o\| \ge \varepsilon\}} Q(\theta) > \delta_\varepsilon.
    \]
\end{lemma}
\begin{proof}
		Under the assumptions of \Cref{Lemma:dCov_Decomp}, namely Assumptions \ref{Ass_dC:Reg}\Cref{Ass_dC:Reg_UdiffXMeas} and \ref{Ass_dC:Ident_Exog_II}, 
		\begin{align*}
		Q(\theta) &= \E[q(W,W^\dagger;\theta)]\\
		& =  \int_{0}^{1}|2\tau-1|\mathcal{V}_{\cdot,\tau}^2\big(X^g(\bar{\theta})(\theta-\theta_o),Z\big) dF_{\mathcal{T}_\theta}(\tau)
		\end{align*} where $\mathcal{V}_{\cdot,\tau}^2\big(X^g(\bar{\theta})(\theta-\theta_o),Z\big) := \E\big[(\mathcal{Z}| \widetilde{X^g}(\bar{\theta})(\theta-\theta_o)|) \mid \mathcal{T}_\theta = \tau \big]$ is the distance covariance between $ X^g(\bar{\theta})(\theta-\theta_o)$ and  $ Z$ given the event \( \{ \mathcal{T}_\theta = \tau \} \). 
		
	By the LIE,
	\begin{equation}\label{eqn:Vtau_Int_1}
			\begin{split}
		 \int_{0}^{1}\mathcal{V}_{\cdot,\tau}^2(X^g(\bar{\theta})(\theta-\theta_o),Z) dF_{\mathcal{T}_\theta}(\tau) & = \int_{0}^{1}\E\big[(\mathcal{Z}| \widetilde{X^g}(\bar{\theta})(\theta-\theta_o)|)|\mathcal{T}_\theta = \tau \big] dF_{\mathcal{T}_\theta}(\tau)\\
		&= \E\big[\mathcal{Z}\big| \widetilde{X^g}(\bar{\theta})(\theta-\theta_o)\big|\big]\\
		& =: \mathcal{V}^2\big(X^g(\bar{\theta})(\theta-\theta_o),Z\big).
		\end{split}
		\end{equation} 
    
    \noindent \Cref{Ass_dC:Ident_Relv} (by Properties \Cref{Property:dCov_geq0,Property:dCov_indep0}) implies that for any $\varepsilon>0$ there exists a $ \tilde{\delta}_\varepsilon > 0 $ such that 
        \begin{equation}\label{eqn:dCov_pos}
            \inf_{\{\theta\in \Theta:||\theta-\theta_o|| \geq \varepsilon\}} \mathcal{V}^2\big(X^g(\bar{\theta})(\theta-\theta_o),Z\big) = \inf_{\{\theta\in \Theta:||\theta-\theta_o|| \geq \varepsilon\}} \mathcal{V}^2\big(g'(X\bar{\theta})X(\theta-\theta_o),Z\big) > \tilde{\delta}_\varepsilon
        \end{equation} observing that \( (\theta-\theta_o)g'(X\bar{\theta}) \neq 0 \) with positive probability under \Cref{Ass_dC:Reg}\ref{Ass_dC:Reg_UdiffXMeas}.
         
        \noindent Also,  
		\begin{equation}\label{eqn:dCov_cond_nneg}
		    \inf_{\{\theta\in \Theta:||\theta-\theta_o|| \geq \varepsilon\}} \mathcal{V}_{\cdot,\tau}^2\big(X^g(\bar{\theta})(\theta-\theta_o),Z\big) \geq 0
		\end{equation}
	 by Property \cref{Property:dCov_geq0} for any $ \tau \in [0,1] $.

    The remainder of the proof is to show that
\[
Q(\theta) = \int_{0}^{1} |2\tau - 1| \, \mathcal{V}_{\cdot,\tau}^2\big(X^g(\bar{\theta})(\theta - \theta_o), Z\big) \, dF_{\mathcal{T}_\theta}(\tau) > 0
\]
for all $\theta \in \Theta\setminus \theta_o $, under the conditions of the theorem. For contradiction, suppose
\[
Q(\theta^*) := \int_{0}^{1} |2\tau - 1| \, \mathcal{V}_{\cdot,\tau}^2\big(X^g(\bar{\theta}^*)(\theta^* - \theta_o), Z\big) \, dF_{\mathcal{T}_{\theta^*}}(\tau) = 0,
\] for some $\theta^* \in \Theta_{\varepsilon}:= \{\dot{\theta} \in \Theta : \|\dot{\theta} - \theta_o\| \ge \varepsilon\} $ where $\bar{\theta}^*$ satisfies $ \lVert \bar{\theta}^* - \theta_o \rVert \leq \lVert \theta^* - \theta_o \rVert $, $\mathcal{T}_{\theta^*} := F_{\widetilde{U} \mid \widetilde{\sigma}( [X,Z] )}\big(\lambda \widetilde{X^g}(\bar{\theta}^*)(\theta^* - \theta_o)\big)$ and $\mathcal{V}_{\cdot,\tau}^2(\cdot,\cdot)$ is non-negative for all $\tau \in [0,1]$---\eqref{eqn:dCov_cond_nneg}.  $Q(\theta^*) = 0$ if and only if at least one of the following conditions holds:
\begin{enumerate}
    \item $\mathcal{V}_{\cdot,\tau}\big(X^g(\bar{\theta}^*)(\theta^* - \theta_o), Z\big) = 0$ for $F_{\mathcal{T}_{\theta^*}}$-almost every $\tau \in [0,1]$;
    \item $\mathcal{T}_{\theta^*} = 1/2 \ a.s. $, i.e., $F_{\mathcal{T}_{\theta^*}}(\{1/2\}) = 1$.
\end{enumerate} Point (1) contradicts \eqref{eqn:dCov_pos}; hence, it contradicts \Cref{Ass_dC:Ident_Relv} by Properties \Cref{Property:dCov_geq0} and \Cref{Property:dCov_indep0}. Point (2) implies $\mathcal{T}_{\theta^*} := F_{\widetilde{U} \mid \widetilde{\sigma}( [X,Z] )}\big(\lambda \widetilde{X^g}(\bar{\theta}^*)({\theta^*} - \theta_o)\big)=1/2 \ a.s.$, which under \Cref{Ass_dC:Ident_Exog_II} further implies $\widetilde{X^g}(\bar{\theta}^*)({\theta^*} - \theta_o)=0 \ a.s.$ for ${\theta^*} \in \Theta_{\varepsilon} \not\ni \theta_o$---a contradiction of \Cref{Ass_dC:Ident_Relv} since \Cref{Ass_dC:Ident_Relv} together with \Cref{Ass_dC:Reg}\ref{Ass_dC:Reg_UdiffXMeas} rule out the degeneracy of $ X^g(\theta):=g'(X\theta)X$ for all $ \theta\in\Theta$.

Thus, by contradiction, it must be that for any $\varepsilon > 0$, there exists a constant $\delta_\varepsilon > 0$ such that
\[
\inf_{\{\theta \in \Theta : \|\theta - \theta_o\| \geq \varepsilon\}} Q(\theta) > \delta_\varepsilon,
\]
under the conditions of the lemma.
\end{proof}

\subsection*{Conclusion:} Combining \Cref{lem:ident_I,lem:ident_II} above completes the proof of the theorem.

\qed

\section{Proof of \Cref{Theorem:Consistency}}

The following lemma essentially verifies the conditions of \textcite[Theorem 1]{honore1994pairwise}. Recall $q(W,W^\dagger;\theta):= \mathcal{Z} \big(|\widetilde{U}(\theta)| - |\widetilde{U}|\big)$.
	\begin{lemma}\label{Lem:Unif_Conv_Obj_Fun}
		Suppose \Cref{Ass_dC:Reg,Ass_dC:Sampling} hold, then 
		\begin{enumerate}[label=(\alph*)]
			\item there exists a function $ \mathcal{B} : \mathcal{W} \times \mathcal{W} \to \mathbb{R}_+ $ with $ \E[\mathcal{B}(X,X^\dagger,Z,Z^\dagger)] \leq C^{1/4} $ such that for any $ \theta_1, \theta_2 \in \Theta $, $ \displaystyle |q(W,W^\dagger;\theta_1) - q(W,W^\dagger;\theta_2)| \leq \mathcal{B}(X,X^\dagger,Z,Z^\dagger)||\theta_1 - \theta_2|| $;
			\item $ Q(\theta) $ is continuous in $ \theta $ uniformly, and $ \displaystyle \sup_{\theta \in \Theta } \big|Q_n(\theta) - Q(\theta)\big| \xrightarrow{a.s.} 0 $.
		\end{enumerate}
		\end{lemma}
	
	\begin{proof}[\textbf{Proof of \Cref{Lem:Unif_Conv_Obj_Fun}}]
		\quad
		
		\textbf{Part (a)}: First, let $ \displaystyle \mathcal{B}(X,X^\dagger,Z,Z^\dagger) := \sup_{\theta \in \Theta }||\mathcal{Z}\widetilde{X^g}(\theta)|| $. By Lyapunov's inequality and \Cref{Ass_dC:Reg}\Cref{Ass_dC:Reg_Dominance}, 
		\[\E[\mathcal{B}(X,X^\dagger,Z,Z^\dagger)] = \E[\sup_{\theta \in \Theta }||\mathcal{Z}\widetilde{X^g}(\theta)||] \leq \big(\E[\sup_{\theta \in \Theta }||\mathcal{Z}\widetilde{X^g}(\theta)||^4]\big)^{1/4} \leq C^{1/4}.\]
		
		Second, for any $ W,W^\dagger $ defined on the support of $ W_i $ and $ \theta_1, \theta_2, \bar{\theta}_{1,2} \in \Theta $ where $ \bar{\theta}_{1,2} $, by \Cref{Ass_dC:Reg}\Cref{Ass_dC:Reg_UdiffXMeas} and the Mean-Value Theorem (MVT), satisfies $ \widetilde{U}(\theta_1) - \widetilde{U}(\theta_2) = -\widetilde{X^g}(\bar{\theta}_{1,2})(\theta_1-\theta_2) $,
		\begin{equation*}
		\begin{split}
		\mathcal{B}(X,X^\dagger,Z,Z^\dagger) \cdot ||\theta_1 - \theta_2|| &= \sup_{\theta \in \Theta }||\mathcal{Z}\widetilde{X^g}(\theta)|| \cdot ||\theta_1 - \theta_2|| \\
        &\geq |\mathcal{Z}| \cdot  \big| \widetilde{X^g}(\bar{\theta}_{1,2})(\theta_2-\theta_1) \big|\\
		& = |\mathcal{Z}| \cdot \big| \widetilde{U}(\theta_1) - \widetilde{U}(\theta_2) \big|\\
		& \geq \Big|\mathcal{Z} \Big((|\widetilde{U}(\theta_1)| - |\widetilde{U}(\theta_o)|) - (|\widetilde{U}(\theta_2)| - |\widetilde{U}(\theta_o)|)\Big)\Big|\\
		& = \big| q(W,W^\dagger;\theta_1) - q(W,W^\dagger;\theta_2) \big|.
		\end{split}
		\end{equation*} The first and second inequalities follow from the Schwarz and the reverse triangle inequalities, respectively.
		
		\textbf{Part (b)}: From \Cref{Ass_dC:Reg}\Cref{Ass_dC:Reg_CompactTheta}, there exists a constant $ C_\theta < \infty $ such that $ ||\theta_1-\theta_2|| < C_\theta $ for all $ \theta_1, \theta_2 \in \Theta $. It thus follows from part (a) above that  $ |q(W,W^\dagger;\theta)| = |q(W,W^\dagger;\theta)-q(W,W^\dagger;\theta_0)| < C_\theta\mathcal{B}(X,X^\dagger,Z,Z^\dagger) $, and this verifies \textcite[Assumption C3]{honore1994pairwise}. 
		
		\Cref{Ass_dC:Reg}\Cref{Ass_dC:Reg_UdiffXMeas} implies the measurability of $ q(W,W^\dagger;\theta) := \mathcal{Z} (|\widetilde{U}(\theta)| - |\widetilde{U}|) $ in $ [W,W^\dagger] $ for all $ \theta \in \Theta $. \Cref{Ass_dC:Reg}\Cref{Ass_dC:Reg_UdiffXMeas} and the continuity of the absolute value function imply $ q(W,W^\dagger;\theta) $ is continuous in $ \theta \in \Theta $ on the support of $ [W,W^\dagger] $. This further implies $ Q(\theta) $ is continuous since the expectation operator preserves continuity and $ \E[Q_n(\theta)] = Q(\theta) := \E[q(W,W^\dagger;\theta)] $ by \Cref{Ass_dC:Sampling} and Property \Cref{Property:dCov_unbiased}. This verifies \textcite[Assumption C2]{honore1994pairwise}.  In addition to \Cref{Ass_dC:Reg}\Cref{Ass_dC:Reg_CompactTheta},  the conclusion follows from Theorem 1 of \textcite{honore1994pairwise}.
	\end{proof}

 \paragraph{Conclusion:}
    Under the assumptions of Lemma \Cref{Lem:Unif_Conv_Obj_Fun} and \Cref{Theorem:Identification}, the conclusion follows from Corollary 1 of \textcite{honore1994pairwise}.

\qed

\section{Proof of \Cref{Theorem:Normality}}
	
	Define the score functions
	\begin{equation}\label{eqn:Score_Funs}
		\begin{split}
			\widehat{\mathcal{S}}_n(\theta) :&= \frac{\partial Q_n(\theta) }{\partial \theta} = \frac{1}{n(n-3)} \sum_{i=1}^n\sum_{j\neq i}^n \big[(1-2\indicator{ \widetilde{U}_{ij}(\theta)\leq 0 })\mathcal{Z}_{ij,n}\widetilde{X^g}_{ij}(\theta) \big] \text{ and }\\
		\mathcal{S}_n(\theta) :&= \E_n\big[(1-2\indicator{ \widetilde{U}_{ij}(\theta)\leq 0 })\mathcal{Z}_{ij}\widetilde{X^g}_{ij}(\theta) \big],
		\end{split}
	\end{equation} noting that $ \mathcal{S}_n(\theta) $ uses $ \mathcal{Z}_{ij} = h(Z_i,Z_i) $ instead of $ \mathcal{Z}_{ij,n} = h_n(Z_i,Z_j) $. The following result provides convergence rates on the score functions evaluated at the estimator $\widehat{\theta}_n$.
	\begin{lemma}\label{Lem:FOC_op1}
		Under \Cref{Ass_dC:Reg}(\Cref{Ass_dC:Reg_UdiffXMeas,Ass_dC:Reg_Dominance}), \Cref{Ass_dC:Sampling}, and \Cref{Ass_dC:AsymN}(\Cref{Ass_dC:AsymN_IntTheta,Ass_dC:AsymN_DiffFU}), 
		(a) $ \sqrt{n}||\mathcal{S}_n(\widehat{\theta}_n)|| = \mathcal{O}_p(n^{-3/2}) $, (b) $\sqrt{n}|| \widehat{\mathcal{S}}_n(\widehat{\theta}_n) - \mathcal{S}_n(\widehat{\theta}_n)|| = \mathcal{O}_p(n^{-1})$, and (c) $ \sqrt{n}||\widehat{\mathcal{S}}_n(\widehat{\theta}_n)|| = \mathcal{O}_p(n^{-1}) $.
	\end{lemma}
	\begin{proof}
    \quad
		
	\textbf{Part (a)}: Applying the chain rule, 
		\[
        \frac{\partial^{-}(q(W_i,W_j;\theta))}{\partial \theta } = \mathcal{Z}_{ij}\partial^{-}|\widetilde{U}_{ij}(\theta)|\times \frac{\partial^{-}\widetilde{U}_{ij}(\theta)}{\partial \theta} = \partial^{-}|\widetilde{U}_{ij}(\theta)|\mathcal{Z}_{ij}\widetilde{X^g}_{ij}(\theta)
        \]
	where the last equality follows by the continuous differentiability of $ \widetilde{U}(\theta) $ (\Cref{Ass_dC:Reg}\Cref{Ass_dC:Reg_UdiffXMeas}). By the consistency of the MDep (\Cref{Theorem:Consistency}) and the left- and right-differentiability of the absolute value function, the left and right derivatives of $ Q_n(\theta) $ at $ \widehat{\theta}_n $ are of opposite signs. It follows from the Markov inequality that 
	\begin{align*}
	||\sqrt{n}\mathcal{S}_n(\widehat{\theta}_n)|| &\leq \frac{1}{2n^{3/2}} \sum_{i=1}^{n}\sum_{j\neq i}^{n} \big| \big(\partial^{-}|\widetilde{U}_{ij}(\widehat{\theta}_n)| -  \partial^{+}|\widetilde{U}_{ij}(\widehat{\theta}_n)|\big) \big|\cdot||\mathcal{Z}_{ij}\widetilde{X^g}_{ij}(\widehat{\theta}_n)||\\
    &= \frac{1}{n^{3/2}} \sum_{i=1}^{n}\sum_{j\neq i}^{n} \indicator{ \partial^{-}|\widetilde{U}_{ij}(\widehat{\theta}_n)| \neq  \partial^{+}|\widetilde{U}_{ij}(\widehat{\theta}_n)| }\cdot||\mathcal{Z}_{ij}\widetilde{X^g}_{ij}(\widehat{\theta}_n)||\\
    &= \frac{1}{n^{3/2}} \sum_{i=1}^{n}\sum_{j\neq i}^{n} \indicator{ \widetilde{U}_{ij}(\widehat{\theta}_n) = 0 }\cdot||\mathcal{Z}_{ij}\widetilde{X^g}_{ij}(\widehat{\theta}_n)||\\
	&= \mathcal{O}_p(n^{-3/2}).
	\end{align*} The first inequality follows from the inequality \( \Big| \frac{d}{dq}|q| \Big| \leq \frac{1}{2} \big| \partial^{-}|q| -  \partial^{+}|q|\big| \) and the triangle inequality, the first equality holds because $\big| \partial^{-}|q| -  \partial^{+}|q|\big| \in \{0,2\} $, and the second equality follows because $\partial^{-}|q| \neq  \partial^{+}|q| \iff q=0 $. For the last equality, observe that thanks to \Cref{Theorem:Consistency} for $n$ sufficiently large, 
    \begin{align*}
        \E\Big[ \sum_{i=1}^{n}\sum_{j\neq i}^{n} \indicator{ \widetilde{U}_{ij}(\widehat{\theta}_n) = 0 }\cdot & ||\mathcal{Z}_{ij}\widetilde{X^g}_{ij}(\widehat{\theta}_n)|| \Big]\\ 
        &\leq \sup_{\theta\in\Theta_o} \E\Big[ \sum_{i=1}^{n}\sum_{j\neq i}^{n} \indicator{ \widetilde{U}_{ij}(\theta) = 0 }\cdot||\mathcal{Z}_{ij}\widetilde{X^g}_{ij}(\theta)|| \Big]\\ 
        &= n(n-1)\sup_{\theta\in\Theta_o} \E\Big[\indicator{ \widetilde{U}(\theta) = 0 }\cdot||\mathcal{Z}\widetilde{X^g}(\theta)||\Big]\\
        &= n(n-1)\sup_{\theta\in\Theta_o} \E\Big[ \Prob\Big( \widetilde{U} = \widetilde{X^g}(\bar{\theta})(\theta-\theta_o) \big| \widetilde{\sigma}( [X,Z] )\Big) \cdot||\mathcal{Z}\widetilde{X^g}(\theta)||\Big]\\
        &=0
    \end{align*}for some open neighbourhood $\Theta_o \ni \theta_o $ in $\Theta$, and $ \bar{\theta} $ satisfying $ ||\bar{\theta}-\theta_o|| \leq ||\theta-\theta_o||$. The first inequality follows from the strong consistency of the MDep (\Cref{Theorem:Consistency}). Next, the first equality follows from the $i.i.d.$ sampling of data \Cref{Ass_dC:Sampling}, and the second equality follows from the LIE. The last equality results from the continuous distribution of $\widetilde{U}$ conditional on $\widetilde{\sigma}( [X,Z] )$ and the MVT---\Cref{Ass_dC:AsymN}\Cref{Ass_dC:AsymN_DiffFU}) and \Cref{Ass_dC:Reg}\Cref{Ass_dC:Reg_Dominance}, respectively.

    \textbf{Part (b)}: Uniformly in $\Theta$, $ \displaystyle \widehat{\mathcal{S}}_n(\theta)  - \E_n\big[(1-2\indicator{ \widetilde{U}_{ij}(\theta)\leq 0 })\mathcal{Z}_{ij,n}\widetilde{X^g}_{ij}(\theta) \big] = \frac{2}{n(n-1)(n-3)}\E_n\big[(1-2\indicator{ \widetilde{U}_{ij}(\theta)\leq 0 })\mathcal{Z}_{ij,n}\widetilde{X^g}_{ij}(\theta) \big] = \mathcal{O}_p(n^{-3}) $ under \Cref{Ass_dC:Reg}\ref{Ass_dC:Reg_Dominance} and \Cref{Ass_dC:Reg}\ref{Ass_dC:Reg_CompactTheta}. From the foregoing and the Cauchy-Schwarz (CS) inequality,
	\begin{align*}
	\sqrt{n}&|| \widehat{\mathcal{S}}_n(\widehat{\theta}_n) - \mathcal{S}_n(\widehat{\theta}_n)||\\ 
    &\leq \frac{1}{2n^{3/2}} \sum_{i=1}^{n}\sum_{j\neq i}^{n} \big| \big(\partial^{-}|\widetilde{U}_{ij}(\widehat{\theta}_n)| -  \partial^{+}|\widetilde{U}_{ij}(\widehat{\theta}_n)|\big)\big|\cdot|\mathcal{Z}_{ij,n} - \mathcal{Z}_{ij} |\cdot||\widetilde{X^g}_{ij}(\widehat{\theta}_n)|| + \mathcal{O}_p(n^{-3})\\
		 &= \frac{1}{n^{3/2}} \sum_{i=1}^{n}\sum_{j\neq i}^{n} \indicator{ \widetilde{U}_{ij}(\widehat{\theta}_n) = 0 } \cdot||\widetilde{X^g}_{ij}(\widehat{\theta}_n)|| \cdot |\mathcal{Z}_{ij,n} - \mathcal{Z}_{ij} | + \mathcal{O}_p(n^{-3})\\
         &\leq \frac{\sqrt{n(n-1)}}{n^{3/2}} \bigg( \sum_{i=1}^{n}\sum_{j\neq i}^{n} \indicator{ \widetilde{U}_{ij}(\widehat{\theta}_n) = 0 } \cdot||\widetilde{X^g}_{ij}(\widehat{\theta}_n)||^2 \bigg)^{1/2} \cdot \bigg( \frac{1}{n(n-1)} \sum_{i=1}^{n}\sum_{j\neq i}^{n} \big(\mathcal{Z}_{ij,n} - \mathcal{Z}_{ij} \big)^2 \bigg)^{1/2}\\
         &+ \mathcal{O}_p(n^{-3})\\
		&= \mathcal{O}_p(n^{-1})
	\end{align*} where $\displaystyle \frac{1}{n(n-1)} \sum_{i=1}^{n}\sum_{j\neq i}^{n} (\mathcal{Z}_{ij,n} - \mathcal{Z}_{ij} )^2 = \mathcal{O}_p(n^{-1}) $ under the conditions of \Cref{Lem:Zijn_Zij_op1}.

    \textbf{Part (c)}: By the triangle inequality, 
	\begin{align*}
		||\sqrt{n}\widehat{\mathcal{S}}_n(\widehat{\theta}_n)|| \leq \sqrt{n}|| \widehat{\mathcal{S}}_n(\widehat{\theta}_n) - \mathcal{S}_n(\widehat{\theta}_n)|| + ||\sqrt{n}\mathcal{S}_n(\widehat{\theta}_n)|| = \mathcal{O}_p(n^{-1}).
	\end{align*}
	\end{proof}

The next result obtains an asymptotically linear expression for the MDep $\widehat{\theta}_n$.
	\begin{lemma}\label{Lem:Asymp_Linearity}
		Under the conditions of Lemmata \ref{Lem:FOC_op1}, \ref{Lem:Hessian_Derivation}, and \ref{Lem:Verif_AssStochEqui}, the MDep $ \widehat{\theta}_n $ has the asymptotically linear representation
	\[
   \sqrt{n}(\widehat{\theta}_n-\theta_o) = -\mathcal{H}^{-1}\frac{2}{\sqrt{n}}\sum_{i=1}^{n} \psi^{(1)}(W_i) + o_p(1).
   \]
	\end{lemma}
	\begin{proof}

Under interior point and differentiability conditions (\Cref{Ass_dC:Reg}\ref{Ass_dC:Reg_UdiffXMeas} and \Cref{Ass_dC:AsymN}\ref{Ass_dC:AsymN_IntTheta}, respectively), \( \displaystyle \theta_o:= \argmin_{\theta\in\Theta}Q(\theta) \) satisfies the first-order condition, namely
		\begin{align}\label{eqn:FOC_zero}
			\mathcal{S}(\theta_o) := \frac{\partial Q(\theta)}{\partial \theta}\Big|_{\theta=\theta_o} = \E\big[(1-2\indicator{ \widetilde{U}\leq 0 })\mathcal{Z}\widetilde{X^g}\big] = \E\big[(1-2F_{\widetilde{U} \mid \widetilde{\sigma}( [X,Z] )}(0))\mathcal{Z}\widetilde{X^g}\big] = 0,
		\end{align}where the second equality follows by the LIE. Under the conditions of \Cref{Lem:Hessian_Derivation}, $ \mathcal{S}(\theta) $ is differentiable. Expanding around $ \theta_o $, one has $ \mathcal{S}(\widehat{\theta}_n) = \mathcal{S}(\theta_o) + \mathcal{H}(\bar{\theta}_n)(\widehat{\theta}_n-\theta_o) = \mathcal{H}(\bar{\theta}_n)(\widehat{\theta}_n-\theta_o) $ where $\bar{\theta}_n$ satisfies $||\bar{\theta}_n - \theta_o|| \leq ||\widehat{\theta}_n - \theta_o|| $ and 
        \begin{align*}
	\mathcal{H}(\theta):& = 2\E\Big[f_{\widetilde{U} \mid \widetilde{\sigma}( [X,Z] )}\big( \widetilde{X^g}(\bar{\theta})(\theta-\theta_o) \big) \mathcal{Z}\widetilde{X^g}(\theta)'\widetilde{X^g}(\bar{\theta})\Big] + \E\Big[\big(1-2F_{\widetilde{U} \mid \widetilde{\sigma}( [X,Z] )}\big( \widetilde{X^g}(\bar{\theta})(\theta-\theta_o) \big)\big) \mathcal{Z}\widetilde{X^{gg}}(\theta)\Big]
	\end{align*} from the proof of \Cref{Lem:Hessian_Derivation}.
    
\begin{align*}
   \underbrace{\sqrt{n}\widehat{\mathcal{S}}_n(\widehat{\theta}_n)}_{-R_{0n}} &= \sqrt{n}\mathcal{S}_n(\widehat{\theta}_n) + \underbrace{\sqrt{n}\big(\widehat{\mathcal{S}}_n(\widehat{\theta}_n) - \mathcal{S}_n(\widehat{\theta}_n)\big)}_{R_{1n}}\\
     &= \sqrt{n}\mathcal{S}(\widehat{\theta}_n) + \underbrace{\sqrt{n}\big(\mathcal{S}_n(\widehat{\theta}_n) - \mathcal{S}(\widehat{\theta}_n)\big)}_{v_n(\widehat{\theta}_n)} + R_{1n}\\
     &= \mathcal{H}(\bar{\theta}_n)\sqrt{n}(\widehat{\theta}_n-\theta_o) + v_n(\widehat{\theta}_n) + R_{1n}\\
     &= \mathcal{H}(\bar{\theta}_n)\sqrt{n}(\widehat{\theta}_n-\theta_o) + v_n(\theta_o) + \underbrace{\big(v_n(\widehat{\theta}_n) - v_n(\theta_o)\big)}_{R_{2n}} + R_{1n}\\
     &=\mathcal{H}(\bar{\theta}_n)\sqrt{n}(\widehat{\theta}_n-\theta_o) + \sqrt{n}\binom{n}{2}^{-1} \sum_{i<j}^{n}\psi(W_i,W_j) + R_{1n} + R_{2n}\\
     &= \mathcal{H}(\bar{\theta}_n)\sqrt{n}(\widehat{\theta}_n-\theta_o) + \frac{2}{\sqrt{n}}\sum_{i=1}^{n} \psi^{(1)}(W_i) +  R_{1n} + R_{2n} + R_{3n}.
\end{align*}The last equality uses Hoeffding's decomposition with \( \displaystyle R_{3n} := \frac{2}{n^{1/2}(n-1)}\sum_{i < j}^{n} \big[ \psi(W_i,W_j) - \psi^{(1)}(W_i) - \psi^{(1)}(W_j) \big] \) where $ R_{3n} = \mathcal{O}_p(n^{-1/2}) $ by, e.g., \citet[Theorem 3, Sect. 1.3]{lee-1990-Ustats}. By \Cref{Lem:FOC_op1}, $R_{0n}=\mathcal{O}_p(n^{-1})$ and $R_{1n}=\mathcal{O}_p(n^{-1})$. 

It remains to study the term $R_{2n}$. From \Cref{Lem:FOC_op1}(a), $ \sqrt{n}||\mathcal{S}(\widehat{\theta}_n)|| = \mathcal{O}_p(n^{-3/2}) $ since $ \mathcal{S}(\theta) = \E[\mathcal{S}_n(\theta)] $. The consistency of the MDep $ \widehat{\theta}_n $ (\Cref{Theorem:Consistency}) and the stochastic equi-continuity condition (\Cref{Lem:Verif_AssStochEqui}) imply that $||v_n(\widehat{\theta}_n) - v_n(\theta_o)||=o_p(1)\times (1+\sqrt{n}||\mathcal{S}(\widehat{\theta}_n)||)~= o_p(1)$. It then follows that $ R_{2n} = o_p(1) $.

Recall $\mathcal{H}:= \mathcal{H}(\theta_o) $, and let $R_{4n}:= \mathcal{H}^{-1}\big(\mathcal{H}(\bar{\theta}_n) - \mathcal{H} \big)$. Using $A^{-1} - B^{-1} = B^{-1}[B-A]A^{-1} $ in addition to \Cref{Ass_dC:AsymN}\Cref{Ass_dC:AsymN_HNonSing}, it follows from the above that
		\begin{align}\label{eqn:Taylor_Expand_1}
			\sqrt{n}(\widehat{\theta}_n-\theta_o) &= -\big( \mathrm{I}_{p_\theta} + R_{4n} \big)^{-1} \mathcal{H}^{-1} \Big(\frac{2}{\sqrt{n}}\sum_{i=1}^{n} \psi^{(1)}(W_i) + \sum_{l=0}^3 R_{ln} \Big) \nonumber\\
            &= -\mathcal{H}^{-1}\frac{2}{\sqrt{n}}\sum_{i=1}^{n} \psi^{(1)}(W_i) - R_{4n}\big( \mathrm{I}_{p_\theta} + R_{4n} \big)^{-1} \mathcal{H}^{-1}\frac{2}{\sqrt{n}}\sum_{i=1}^{n} \psi^{(1)}(W_i) + o_p(1). 
		\end{align} 
        Under \Cref{Ass_dC:Reg}\ref{Ass_dC:Reg_UdiffXMeas} and \Cref{Ass_dC:AsymN}\ref{Ass_dC:AsymN_DiffFU}, the Hessian function $\mathcal{H}(\theta)$ is continuous in $\theta$. In addition to \Cref{Theorem:Consistency}, that $ ||\bar{\theta}_n - \theta_o|| \leq ||\widehat{\theta}_n - \theta_o|| $, the non-singularity of $\mathcal{H}$ (\Cref{Ass_dC:AsymN}\ref{Ass_dC:AsymN_HNonSing}) and the continuous mapping theorem, $R_{4n} = o_p(1) $.

    Next, $\frac{1}{\sqrt{n}}\sum_{i=1}^{n} \psi^{(1)}(W_i) = \mathcal{O}_p(1) $ by Chebyshev's inequality:
    \begin{equation}\label{eqn:bnd_psi1}
    \begin{split}
        \Big\lVert \mathrm{var}\Big[ \frac{1}{\sqrt{n}}\sum_{i=1}^{n} \psi^{(1)}(W_i) \Big] \Big\lVert &= \big\lVert \E\big[ \psi^{(1)}(W)\psi^{(1)}(W)' \big] \big\rVert \\
        &\leq \E[||\psi^{(1)}(W)||^2] \\ 
        &= \E\big[\big|\big|\E[\psi(W,W^\dagger)|W]\big|\big|^2\big] \\ 
        &\leq \E[||\psi(W,W^\dagger)||^2] \\
        &\leq \big( \E[||\psi(W,W^\dagger)||^4] \big)^{1/2} \\
        &= \big(\E\big[||\mathcal{Z}\big(1-2\indicator{\widetilde{U} \leq 0}\big) \widetilde{X}^{g\prime}||^4\big] \big)^{1/2} \\
    &=\big(\E\big[||\mathcal{Z}\widetilde{X}^{g\prime}||^4\big] \big)^{1/2}\\
    &\leq C^{1/2}
    \end{split}
    \end{equation}
    where the first inequality follows from the Cauchy-Schwarz inequality, the second inequality follows from the conditional Jensen's inequality, the third inequality follows from Lyapunov's inequality, and the last inequality holds by \Cref{Ass_dC:Reg}\Cref{Ass_dC:Reg_Dominance}.
    
    Thus, 
		\begin{equation*}
			\sqrt{n}(\widehat{\theta}_n-\theta_o) = -\mathcal{H}^{-1}\frac{2}{\sqrt{n}}\sum_{i=1}^{n} \psi^{(1)}(W_i) + o_p(1)
		\end{equation*} by combining \Cref{eqn:Taylor_Expand_1} with the foregoing.
	\end{proof}

\Cref{Lem:Asymp_Linearity} proves the Part (a) of \Cref{Theorem:Normality}.
In addition to \Cref{Ass_dC:Sampling} and the second moment bound in \eqref{eqn:bnd_psi1}, the Lindeberg-L\'evy Central Limit Theorem applies. Part (b) then follows from the continuous mapping theorem.

\section{Proof of \Cref{Theorem:Consistency_Cov_Matrix}}

The proof of consistency of the covariance matrix estimator (\Cref{Theorem:Consistency_Cov_Matrix}) is organised in two parts, establishing in turn the consistency of \( \widehat{\Omega}_n \) and of \( \widehat{\mathcal{H}}_n \).

\subsection{Consistency of $\widehat{\Omega}_n$ }
  The result is stated in the following lemma.
	
		\begin{lemma}\label{lem:Omega_consistency}
			Under the conditions of \Cref{Theorem:Consistency} and \Cref{Ass_dC:AsymN}, $ \displaystyle \plim_{n \rightarrow \infty}\widehat{\Omega}_n = \Omega $.
		\end{lemma}
		
		\begin{proof}[\textbf{Proof}]
		Recall $ \widehat{\Omega}_n = 4\E_n[\widehat{\psi}^{(1)}(W_i)\widehat{\psi}^{(1)}(W_i)'] $ where 
		\[
        \widehat{\psi}^{(1)}(W_i) := \frac{1}{n-1} \sum_{j\neq i}^{n} \mathcal{Z}_{ij,n}\big(1-2\indicator{ \widetilde{U}_{ij}(\widehat{\theta}_n) \leq 0 }\big) \widetilde{X^g}_{ij}(\widehat{\theta}_n)'.
        \] Define $ \widetilde{\Omega}_n = 4\E_n[\widetilde{\psi}^{(1)}(W_i)\widetilde{\psi}^{(1)}(W_i)'] $ where 
		\[\widetilde{\psi}^{(1)}(W_i) := \frac{1}{n-1} \sum_{j\neq i}^{n} \mathcal{Z}_{ij}\big(1-2\indicator{\widetilde{U}_{ij} \leq 0}\big) \widetilde{X}_{ij}^{g\prime}.\]
		\noindent Since $ ||\widehat{\Omega}_n-\Omega|| \leq ||\widehat{\Omega}_n-\widetilde{\Omega}_n|| + ||\widetilde{\Omega}_n-\Omega|| $ by the triangle inequality, the proof proceeds by showing that $ ||\widehat{\Omega}_n-\widetilde{\Omega}_n|| = o_p(1) $ and $ ||\widetilde{\Omega}_n-\Omega|| = o_p(1) $ under the conditions of the theorem. 
        
        First, 		
		\begin{align*}
			\lVert \widehat{\Omega}_n-\widetilde{\Omega}_n \rVert &= 4\big\lVert \E_n[\widehat{\psi}^{(1)}(W_i)\widehat{\psi}^{(1)}(W_i)' - \widetilde{\psi}^{(1)}(W_i)\widetilde{\psi}^{(1)}(W_i)'] \big\rVert\\
			&= 4\big\lVert \E_n[\widehat{\psi}^{(1)}(W_i)(\widehat{\psi}^{(1)}(W_i) - \widetilde{\psi}^{(1)}(W_i))' + (\widehat{\psi}^{(1)}(W_i) - \widetilde{\psi}^{(1)}(W_i))\widetilde{\psi}^{(1)}(W_i)'] \big\rVert\\
			&\leq 4\E_n\big[\big(\lVert \widehat{\psi}^{(1)}(W_i) \rVert + \lVert \widetilde{\psi}^{(1)}(W_i) \rVert \big) \lVert \widehat{\psi}^{(1)}(W_i) - \widetilde{\psi}^{(1)}(W_i) \rVert \big]\\
			&\leq 4 \big(\E_n\big[(\lVert \widehat{\psi}^{(1)}(W_i) \rVert + \lVert \widetilde{\psi}^{(1)}(W_i) \rVert )^2\big]\big)^{1/2} \times \big(\E_n\big[\lVert \widehat{\psi}^{(1)}(W_i) - \widetilde{\psi}^{(1)}(W_i) \rVert^2\big]\big)^{1/2}
		\end{align*}by Jensen's and the CS inequalities.

  Second, obtain the following upper bound:
		\begin{align*}
			||\widehat{\psi}^{(1)}(W_i) - \widetilde{\psi}^{(1)}(W_i)|| \leq & \frac{1}{n-1}\sum_{j\neq i}^{n}||\mathcal{Z}_{ij,n}\big(1-2\indicator{ \widetilde{U}_{ij}(\widehat{\theta}_n) \leq 0 }\big) \widetilde{X^g}_{ij}(\widehat{\theta}_n)' - \mathcal{Z}_{ij}\big(1-2\indicator{\widetilde{U}_{ij} \leq 0}\big) \widetilde{X}_{ij}^{g\prime}||\\
			\leq & \frac{1}{n-1}\sum_{j\neq i}^{n}||(\mathcal{Z}_{ij,n} - \mathcal{Z}_{ij})\big(1-2\indicator{ \widetilde{U}_{ij}(\widehat{\theta}_n) \leq 0 }\big) \widetilde{X^g}_{ij}(\widehat{\theta}_n)||\\
             &+ \frac{2}{n-1}\sum_{j\neq i}^{n}||\big(\indicator{ \widetilde{U}_{ij}(\widehat{\theta}_n) \leq 0 } - \indicator{\widetilde{U}_{ij} \leq 0}\big) \mathcal{Z}_{ij} \widetilde{X^g}_{ij}(\widehat{\theta}_n)||\\
			& + \frac{1}{n-1}\sum_{j\neq i}^{n}||\big(1-2\indicator{\widetilde{U}_{ij} \leq 0}\big) \mathcal{Z}_{ij}(\widetilde{X^g}_{ij}(\widehat{\theta}_n) - \widetilde{X^g}_{ij})|| \\
			\leq & \frac{1}{n-1}\sum_{j\neq i}^{n}|\mathcal{Z}_{ij,n} - \mathcal{Z}_{ij}|\cdot\sup_{\theta \in \Theta }||\widetilde{X^g}_{ij}(\theta)||\\
			&+ \frac{2}{n-1}\sum_{j\neq i}^{n}|\indicator{ \widetilde{U}_{ij}(\widehat{\theta}_n) \leq 0 } - \indicator{\widetilde{U}_{ij} \leq 0}|\cdot\sup_{\theta \in \Theta }||\mathcal{Z}_{ij} \widetilde{X^g}_{ij}(\theta)||\\
			& + \frac{1}{n-1}\sum_{j\neq i}^{n}||\mathcal{Z}_{ij}(\widetilde{X^g}_{ij}(\widehat{\theta}_n) - \widetilde{X^g}_{ij})||\\
			=&: B_{1i,n} + 2B_{2i,n} + B_{3i,n}.
		\end{align*}
		
		By the $ c_r $-inequality, 
        \[
        \E_n[||\widehat{\psi}^{(1)}(W_i) - \widetilde{\psi}^{(1)}(W_i)||^2] \leq 3\E_n[B_{1i,n}^2] + 6\E_n[B_{2i,n}^2] + 3\E_n[B_{3i,n}^2].
        \]
        It can be observed that $ \E_n[B_{1i,n}^2] = \mathcal{O}_p(n^{-1}) $ under the conditions of \Cref{Lem:Zijn_Zij_op1} and \Cref{Ass_dC:Reg}\Cref{Ass_dC:Reg_Dominance}, while $ \E_n[B_{3i,n}^2] = o_p(1) $ by \Cref{Ass_dC:Reg}\ref{Ass_dC:Reg_UdiffXMeas}, the Continuous Mapping Theorem (CMT), and \Cref{Theorem:Consistency}. It remains to show that $ \E_n[B_{2i,n}^2] = o_p(1) $.
	\begin{align}\label{eqn:ItaItb}
			\big|\indicator{ \widetilde{U}(\theta) \leq 0 } - \indicator{ \widetilde{U} \leq 0 }\big| &= \big|\indicator{ \widetilde{U} - \widetilde{X^g}(\bar{\theta})(\theta-\theta_o) \leq 0 } - \indicator{ \widetilde{U} \leq 0 }\big| \nonumber \\
			&= \indicator{ 0 < \widetilde{U} \leq \widetilde{X^g}(\bar{\theta})(\theta-\theta_o) } + \indicator{ \widetilde{X^g}(\bar{\theta})(\theta - \theta_o) \leq \widetilde{U} < 0 } \nonumber \\
			&=: \widetilde{I}^a(\theta) + \widetilde{I}^b(\theta).
		\end{align} Further, 
\begin{align}\label{eqn:Exp_ItaItb}
	\E\big[\widetilde{I}^a(\theta) + \widetilde{I}^b(\theta) \mid \widetilde{\sigma}( [X,Z] )\big]
	&= \Big\{0 \vee \Big(F_{\widetilde{U} \mid \widetilde{\sigma}( [X,Z] )}(\widetilde{X^g}(\bar{\theta})(\theta - \theta_o)) - F_{\widetilde{U} \mid \widetilde{\sigma}( [X,Z] )}(0)\Big) \Big\} \nonumber \\
    &+ \Big\{0 \vee \Big(F_{\widetilde{U} \mid \widetilde{\sigma}( [X,Z] )}(0) -  F_{\widetilde{U} \mid \widetilde{\sigma}( [X,Z] )}(\widetilde{X^g}(\bar{\theta})(\theta - \theta_o))\Big) \Big \} \nonumber \\
	&= \Big|F_{\widetilde{U} \mid \widetilde{\sigma}( [X,Z] )}(\widetilde{X^g}(\bar{\theta})(\theta - \theta_o)) - F_{\widetilde{U} \mid \widetilde{\sigma}( [X,Z] )}(0)\Big| \nonumber \\
	&= f_{\widetilde{U} \mid \widetilde{\sigma}( [X,Z] )}(\lambda\widetilde{X^g}(\bar{\theta})(\theta - \theta_o)) \times \big|\widetilde{X^g}(\bar{\theta})(\theta - \theta_o)\big|
		\end{align}by \Cref{Ass_dC:AsymN}\ref{Ass_dC:AsymN_DiffFU}, the MVT, and the Schwarz inequality for some $ \lambda \in (0,1) $. 
  \begin{equation}\label{eqn:cons_indicator}
      \begin{split}
      \E\big[\E_n[B_{2i,n}^2]\big] &\leq \E_n \Big[ \E \big[|\indicator{ \widetilde{U}_{ij}(\widehat{\theta}_n) \leq 0 } - \indicator{ \widetilde{U}_{ij} \leq 0 }|^2\cdot\sup_{\theta \in \Theta }||\mathcal{Z}_{ij} \widetilde{X^g}_{ij}(\theta)||^2 \big] \Big] \\
      &\leq \E_n \Big[ \big( \E \big[|\indicator{ \widetilde{U}_{ij}(\widehat{\theta}_n) \leq 0 } - \indicator{ \widetilde{U}_{ij} \leq 0 }| \big] \big)^{1/2} \cdot \big(\E\big[\sup_{\theta \in \Theta }||\mathcal{Z}_{ij} \widetilde{X^g}_{ij}(\theta)||^4 \big] \big)^{1/2} \Big] \\
      &\leq 2C^{1/2} \E_n \Big[ \Big( \E \big[f_{\widetilde{U} \mid \widetilde{\sigma}( [X,Z] )}\big(\lambda\widetilde{X^g}_{ij}(\bar{\theta}_n)(\widehat{\theta}_n - \theta_o)\big) \times \big| \widetilde{X^g}_{ij}(\bar{\theta}_n)(\widehat{\theta}_n - \theta_o) \big| \big] \Big)^{1/2} \Big]\\
      &= o(1).
  \end{split}
  \end{equation}
   The first inequality follows from the $c_r$-inequality, and the second follows from the CS using that $|\indicator{ \widetilde{U}(\widehat{\theta}_n) \leq 0 } - \indicator{ \widetilde{U} \leq 0 }|^2=|\indicator{ \widetilde{U}(\widehat{\theta}_n) \leq 0 } - \indicator{ \widetilde{U} \leq 0 }|$. The third inequality uses the LIE, \eqref{eqn:ItaItb}, \eqref{eqn:Exp_ItaItb}, the continuous mapping theorem, the consistency of $\widehat{\theta}_n$ (\Cref{Theorem:Consistency}), and \Cref{Ass_dC:Reg}\Cref{Ass_dC:Reg_Dominance}. Thus, $\E_n[B_{2i,n}^2] = o_p(1)$ by the Markov inequality. Deduce from the above that $||\widehat{\Omega}_n-\widetilde{\Omega}_n|| = o_p(1) $.

By independent and identical sampling (\Cref{Ass_dC:Sampling}) and the LIE, 
	\begin{equation}\label{eqn:Exp_Omgtilde}
		\begin{split}
			\E[\widetilde{\Omega}_n] &= 4\frac{1}{n(n-1)^2}\sum_{i=1}^{n}\sum_{j\neq i}^{n}\sum_{j'\neq i}^{n} \E[\psi(W_i,W_j)\psi(W_i,W_{j'})']\\
             &= 4\frac{1}{n(n-1)^2} \Big( \sum_{i=1}^{n}\sum_{j\neq i}^{n}\sum_{ \substack{j'\neq i \\ j'\neq j} }^{n} \E[\psi(W_i,W_j)\psi(W_i,W_{j'})'] + \sum_{i=1}^{n}\sum_{j\neq i}^{n} \E[\psi(W_i,W_j)\psi(W_i,W_j)'] \Big)\\
            &= 4\E[\psi(W,W^\dagger)\psi(W,W^{\dagger\dagger})'] + 4\frac{1}{(n-1)}\Big(\E[\psi(W,W^\dagger)\psi(W,W^\dagger)'] - \E[\psi(W,W^\dagger)\psi(W,W^{\dagger\dagger})'] \Big)\\
		&= 4\E\big[\E[\psi(W,W^\dagger)\psi(W,W^{\dagger\dagger})'|W]\big] + \mathcal{O}(n^{-1})\\
        &= 4\E\big[\E[\psi(W,W^\dagger)|W]\cdot\E[\psi(W,W^{\dagger\dagger})|W]'\big]+ \mathcal{O}(n^{-1})\\
		&= 4\E[\psi^{(1)}(W)\psi^{(1)}(W)'] + \mathcal{O}(n^{-1})\\
        &=: \Omega + \mathcal{O}(n^{-1}).
		\end{split}
	\end{equation} 

\noindent By the CS and \Cref{Ass_dC:Reg}\Cref{Ass_dC:Reg_Dominance},
	\begin{equation}\label{eqn:Bnd_Kern_Omgtilde}
		\begin{split}
			\E[||\psi(W_i,W_j)\psi(W_i,W_{j'})'||] &\leq \E[||\psi(W_i,W_j)||\cdot||\psi(W_i,W_{j'})||]\\
			& \leq (\E[||\psi(W_i,W_j)||^2]\cdot\E[||\psi(W_i,W_{j'})'||^2])^{1/2}\\
		&=\E[||\psi(W,W^\dagger)||^2] = \E[||\mathcal{Z}\widetilde{X^g}||^2] \leq (\E[||\mathcal{Z}\widetilde{X^g}||^4])^{1/2}\leq C^{1/2}.
		\end{split}
	\end{equation} $ \widetilde{\Omega}_n $ is a U-statistic of order $ 3 $. Combining \eqref{eqn:Exp_Omgtilde} and \eqref{eqn:Bnd_Kern_Omgtilde},  $ ||\widetilde{\Omega}_n-\Omega|| = o_p(1) $ by the strong law of large numbers for U-statistics \citep{hoeffding1961strong}.
\end{proof}

\subsection{Consistency of $\widehat{\mathcal{H}}_n$}
        \begin{lemma}\label{lem:Hessian_consistency}
			Suppose that the conditions of \Cref{Theorem:Consistency} hold. Then, in addition to \Cref{Ass_dC:AsymN,Ass_dC:Bandwith_Consistent} $, \displaystyle \plim_{n \rightarrow \infty}\widehat{\mathcal{H}}_n = \mathcal{H} $.
		\end{lemma}
        \begin{proof}
            Re-express \( \widehat{\mathcal{H}}_n = \widehat{\mathcal{H}}_{1n} + \widehat{\mathcal{H}}_{2n} \) where \( \displaystyle \widehat{\mathcal{H}}_{1n} := \frac{1}{n(n-1)\hat{c}_n}\sum_{i=1}^{n}\sum_{j\neq i}^{n} \Big\{\indicator{ |\widetilde{U}_{ij}(\widehat{\theta}_n)| \leq \hat{c}_n }\mathcal{Z}_{ij,n} \widetilde{X^g}_{ij}(\widehat{\theta}_n)'\widetilde{X^g}_{ij}\big(\widehat{\theta}_n\big) \Big\} \) and \( \displaystyle \widehat{\mathcal{H}}_{2n} := \frac{1}{n(n-1)}\sum_{i=1}^{n}\sum_{j\neq i}^{n} \Big\{\sgn\big(\widetilde{U}_{ij}(\widehat{\theta}_n)\big) \mathcal{Z}_{ij,n}\widetilde{X^{gg}_{ij}}\big( \widehat{\theta}_n \big) \Big\} \). 
        
        \noindent Similarly, define  \(\displaystyle \mathcal{H}_{1n} := \frac{1}{n(n-1)c_n}\sum_{i=1}^{n}\sum_{j\neq i}^{n} \Big\{\indicator{ |\widetilde{U}_{ij}| \leq c_n }\mathcal{Z}_{ij} \widetilde{X}_{ij}^{g\prime}\widetilde{X^g}_{ij}\Big\} \), \\
        \noindent \( \displaystyle \mathcal{H}_{2n} = \frac{1}{n(n-1)}\sum_{i=1}^{n}\sum_{j\neq i}^{n} \Big\{ \sgn(\widetilde{U}_{ij}) \mathcal{Z}_{ij}\widetilde{X^{gg}_{ij}} \Big\} \), \( \mathcal{H}_1 := 2\E\big[f_{\widetilde{U} \mid \widetilde{\sigma}( [X,Z] )}(0)\mathcal{Z}\widetilde{X^g}'\widetilde{X^g}\big] \), and \( \mathcal{H}_2 = \E\big[\sgn(\widetilde{U}) \mathcal{Z}\widetilde{X^{gg}}\big] \),  then notice that $ \mathcal{H} = \mathcal{H}_1 + \mathcal{H}_2 $. 
        
        Consider the following decomposition:
        \begin{align*}
            \widehat{\mathcal{H}}_n - \mathcal{H} = \big(\widehat{\mathcal{H}}_{1n} - \mathcal{H}_{1n} \big) + \big(\mathcal{H}_{1n} - \mathcal{H}_1 \big) + \big(\widehat{\mathcal{H}}_{2n} - \mathcal{H}_{2n} \big) + \big(\mathcal{H}_{2n} - \mathcal{H}_2 \big).
        \end{align*} \( \big\lVert \widehat{\mathcal{H}}_{1n} - \mathcal{H}_{1n} \big\rVert = o_p(1) \), \( \big\rVert \mathcal{H}_{1n} - \mathcal{H}_1 \big\rVert = o_p(1) \), \( \big\lVert \widehat{\mathcal{H}}_{2n} - \mathcal{H}_{2n} \big\rVert = o_p(1) \), and \( \big\rVert \mathcal{H}_{2n} - \mathcal{H}_2 \big\rVert = o_p(1) \) under the conditions of Lemmata \ref{Lem:Conv_op1_H_Terms}, \ref{Lem:Conv_Quadratic_Mean_H}, \ref{lem:Hess_2_consistency}(a), and \ref{lem:Hess_2_consistency}(b), respectively. The conclusion follows from the triangle inequality.
        \end{proof}

\paragraph{Conclusion:} Finally, combining \Cref{lem:Omega_consistency,lem:Hessian_consistency} and noting that the matrix inverse is continuous at the non-singular $\mathcal{H}$ (\Cref{Ass_dC:AsymN}\ref{Ass_dC:AsymN_HNonSing}), the result follows from the CMT.

\qed
		
\section{Proof of \Cref{Theorem:Test_LC}}
	Define $\displaystyle \uptau^* := \arginf_{\{\uptau \in \mathbb{R}^{p_X} : ||\uptau||=1\}}\mathcal{V}^2(X\uptau,Z) $. By Properties \Cref{Property:dCov_geq0,Property:dCov_indep0}, a test of \Cref{Ass_dC:Ident_Relv} can be formulated as
	\begin{align*}
		\mathbb{H}_o' : \mathcal{V}^2(X\uptau^*,Z) = 0 \text{ v.s. } \mathbb{H}_a' : \mathcal{V}^2(X\uptau^*,Z) > 0.
	\end{align*}
	\noindent Partition $ \uptau^* = [\uptau_D^*, \uptau_{-D}^*] $ conformably, then $ X\uptau^* = D \uptau_D^* + Z_{-D}\uptau_{-D}^* $. The first step in the proof rests on the following lemma.
	\begin{lemma}\label{Lem:tau1neq0}
		$ \mathbb{H}_o' $ implies $ ||\uptau_D^*|| > 0 $, while the converse does not hold.
	\end{lemma}
	\begin{proof}
	The first part of the proof proceeds by contradiction. Suppose $ \uptau_D^* = 0 $, then $ \mathcal{V}^2(X\uptau^*,Z) =  \mathcal{V}^2(D \uptau_D^* + Z_{-D}\uptau_{-D}^*,Z) = \mathcal{V}^2(Z_{-D}\uptau_{-D}^*,Z) > 0 $ by Properties ~\ref{Property:dCov_geq0} and ~\ref{Property:dCov_indep0} since $ Z $ contains $ Z_{-D} $, i.e., $ \uptau_D^* = 0 $ implies $ \mathbb{H}_a' $. Thus, $ \mathbb{H}_o' $ implies $ ||\uptau_D^*|| > 0 $.
		
In examining the converse, two cases of $ ||\uptau_D^*|| > 0 $ arise. First, $ ||\uptau_D^*|| \in (0,1) $ implies $ || \uptau_{-D}^*|| > 0 $ hence $ \mathcal{V}^2(X\uptau^*,Z) =  \mathcal{V}^2(D\uptau_D^* + Z_{-D}\uptau_{-D}^*,Z) > 0 $ since $ Z $ contains $ Z_{-D} $, i.e., $ ||\uptau_D^*|| \in (0,1) $ implies $ \mathbb{H}_a' $. Second, $ ||\uptau_D^*|| = 1 $ means $ \uptau_{-D}^* = 0 $, thus $ \mathcal{V}^2(X\uptau^*,Z) =  \mathcal{V}^2(D\uptau_D^* + Z_{-D}\uptau_{-D}^*,Z) = \mathcal{V}^2(D\uptau_D^*,Z) \geq 0 $ by Property \Cref{Property:dCov_geq0}, i.e., $ ||\uptau_D^*|| = 1 $ implies either $ \mathbb{H}_o' $ or $ \mathbb{H}_a' $ depending on whether $ D\uptau_D^* $ is dependent on $ Z $ or not.
	\end{proof}

Next, consider the elements of $\uptau_D^*:= [\uptau_1^*,\ldots,\uptau_{p_D}^*]'$. First, for $l\in [p_D] $ such that $ \uptau_l^* \neq 0 $,
	\begin{align*}
		\mathcal{V}^2(X\uptau^*,Z) = \E\big[\mathcal{Z}|\widetilde{X}\uptau^*|\big] &= \E\big[\mathcal{Z}|(D_l-D_l^\dagger)\uptau_l^* + [D_{-l} - D_{-l}^\dagger,\ Z_{-D}-Z_{-D}^\dagger]\uptau_{-l}^*|\big]\\
		& = |\uptau_l^*|\E\big[\mathcal{Z}\big|(D_l - D_l^\dagger) + [D_{-l} - D_{-l}^\dagger,\ Z_{-D}-Z_{-D}^\dagger]\uptau_{-l}^*/\uptau_l^*\big|\big]\\ 
		& = |\uptau_l^*|\mathcal{V}^2\big(D_l - [D_{-l},Z_{-D}]\gamma_l^*,Z\big)\\
		& = |\uptau_l^*|\mathcal{V}^2\big(\mathcal{D}(\gamma_l^*),Z\big)
	\end{align*}
    where $ \gamma_l^* := - \uptau_{-l}^*/\uptau_l^* $. As $ |\uptau_l^*| > 0 $, $ \mathcal{V}^2(X\uptau^*,Z) = 0 $ if and only if $ \mathcal{V}^2(\mathcal{D}(\gamma_l^*),Z) = 0 $, and $ \mathcal{V}^2(X\uptau^*,Z) > 0 $ if and only if $ \mathcal{V}^2(\mathcal{D}(\gamma_l^*),Z) > 0 $.\\
    
    Second, for $l\in [p_D] $ such that $\uptau_l^*=0$,
    \begin{align*}
		\mathcal{V}^2(X\uptau^*,Z) 
		 \leq \min_{\gamma \in \mathbb{S}^{p_X - 1} } \frac{\mathcal{V}^2\big(D_l - [D_{-l},Z_{-D}]\gamma,\ Z\big)}{||[1,\gamma']'||} =: \mathcal{V}^2\big(\mathcal{D}(\gamma_l^*),Z\big)
	\end{align*} by the definition of $\uptau^*$.

	By \Cref{Lem:tau1neq0} above, there is at least one $l\in [p_D] $ such that $ \mathcal{V}^2(\mathcal{D}(\gamma_l^*),Z) = 0 $ if and only if $ \mathcal{V}^2(X\uptau^*,Z) = 0 $ under $ \mathbb{H}_o' $. Under $ \mathbb{H}_a' $, $0 < \mathcal{V}^2(X\uptau^*,Z) \leq \mathcal{V}^2\big(\mathcal{D}(\gamma_l^*),Z\big)$ for all $l\in [p_D] $. 
    An equivalent expression of the test hypotheses becomes
    \begin{align*}
        \widetilde{\mathbb{H}}_o: \ \min_{\{\gamma, l\} \in \mathbb{S}^{p_X - 1} \times [p_D] } \mathcal{V}^2\big(\mathcal{D}_l(\gamma), \ Z\big) = 0 \text{ v.s. } \widetilde{\mathbb{H}}_a: \ \min_{\{\gamma,l\} \in \mathbb{S}^{p_X - 1} \times [p_D] } \mathcal{V}^2\big(\mathcal{D}_l(\gamma),\ Z\big) > 0
    \end{align*}where $\mathbb{S}^{p_X - 1} \subset \mathbb{R}^{p_X - 1} $. The conclusion follows from Properties \Cref{Property:dCov_geq0,Property:dCov_indep0}.	
\qed

\newpage

\printbibliography

\end{refsection}

\newpage
\begin{refsection}
\setcounter{page}{1}
\setcounter{section}{0}

\renewcommand{\thetable}{S.\arabic{table}}
\renewcommand{\thefigure}{S.\arabic{figure}}
\renewcommand{\thesection}{S.\arabic{section}}
\begin{center}
    \LARGE{\bf Supplemental Appendix:\\ A Distance Covariance-based Estimator}
\end{center}
\begin{center}
    Emmanuel Selorm Tsyawo  ~~~~~~~~  Abdul-Nasah Soale
\end{center}
This supplemental material provides auxiliary lemmata in \Cref{Appendix_Sect:Sufficient_Conditions} used in the proofs of results presented in the main text. \Cref{Appendix_Sect:Useful_Prop} provides the alternative definition of the dCov measure on which the MDep is based, and \Cref{Appendix_Sect:Simulations} supplies supplementary simulation results based on non-linear models and larger samples for both linear and non-linear models.

\section{Supporting Lemmata}\label{Appendix_Sect:Sufficient_Conditions}

\subsection{Convergence in probability of the $ U $-centred $Z$ }
\begin{lemma}\label{Lem:Zijn_Zij_op1}
	Under \Cref{Ass_dC:Reg} and \Cref{Ass_dC:Sampling},
	$ (\mathcal{Z}_{ij,n} -  \mathcal{Z}_{ij})^2 = \mathcal{O}_p(n^{-1}) $ for any $ (i,j) \in [n]\times \{[n]\setminus i\} $.
\end{lemma}

\begin{proof}[\textbf{Proof}]
	\quad 

    For any $ (i,j) \in [n]\times \{[n]\setminus i\} $
\begin{align*}
    \mathcal{Z}_{ij,n} &= ||\widetilde{Z}_{ij}|| - \frac{1}{n-1}\sum_{l=1}^{n}(||\widetilde{Z}_{il}||+||\widetilde{Z}_{lj}||) + \frac{1}{n(n-1)}\sum_{k=1}^n \sum_{l\neq k}^n||\widetilde{Z}_{kl}|| \\
    & - \frac{1}{(n-1)(n-2)}\sum_{l=1}^{n}(||\widetilde{Z}_{il}||+||\widetilde{Z}_{lj}||) + \frac{2}{n(n-1)(n-2)}\sum_{k=1}^n \sum_{l\neq k}^n||\widetilde{Z}_{kl}||\\
    &=: \widetilde{\mathcal{Z}}_{ij,n} - \frac{1}{(n-1)(n-2)}\sum_{l=1}^{n}(||\widetilde{Z}_{il}||+||\widetilde{Z}_{lj}||) + \frac{2}{n(n-1)(n-2)}\sum_{k=1}^n \sum_{l\neq k}^n||\widetilde{Z}_{kl}||.
\end{align*}	
Recall $ \mathcal{Z}_{ij} := h(Z_i,Z_j) $ where $h(z_a,z_b) := ||z_a - z_b|| - \E\big[||z_a - Z||+||Z - z_b||\big] + \E\big[||Z - Z^\dagger||\big]$. $ \E[\widetilde{\mathcal{Z}}_{ij,n}-\mathcal{Z}_{ij}] = 0 $ by the LIE for any $ (i,j) \in [n]^2 $. Under \Cref{Ass_dC:Reg}\Cref{Ass_dC:Reg_Dominance}, $ \mathcal{Z}_{ij,n} - \widetilde{\mathcal{Z}}_{ij,n} = \mathcal{O}_p(n^{-1})$. In addition to the triangle inequality, $|\mathcal{Z}_{ij,n} -  \mathcal{Z}_{ij}| \leq |\mathcal{Z}_{ij,n} - \widetilde{\mathcal{Z}}_{ij,n}| + |\widetilde{\mathcal{Z}}_{ij,n} - \mathcal{Z}_{ij}| = |\widetilde{\mathcal{Z}}_{ij,n} - \mathcal{Z}_{ij}| + \mathcal{O}_p(n^{-1}) $. Moreover, it follows from Lo\`eve's $ c_r $-inequality, \Cref{Ass_dC:Sampling}, the CS inequality, and \Cref{Ass_dC:Reg}\Cref{Ass_dC:Reg_Dominance} that
	\begin{equation}\label{eqn:zcheck_op1}
	\begin{split}
	\E[|\widetilde{\mathcal{Z}}_{ij,n} -  \mathcal{Z}_{ij}|^2] & \leq  \frac{3}{(n-1)^2}\E\Big[\Big(\sum_{k=1}^{n}(||\widetilde{Z}_{ik}|| - \E[(||\widetilde{Z}_{ik}||)\mid Z_i])\Big)^2\Big] \\
    &+  \frac{3}{(n-1)^2}\E\Big[\Big(\sum_{k=1}^{n}(||\widetilde{Z}_{kj}|| - \E[(||\widetilde{Z}_{kj}||)\mid Z_j])\Big)^2\Big] \\
	& + \frac{3}{n^2(n-1)^2}\E\Big[\Big(\sum_{k=1}^{n}\sum_{l=1}^{n}(||\widetilde{Z}_{kl}|| - \E[||\widetilde{Z}_{kl}||])\Big)^2\Big]\\
	& = \frac{3}{(n-1)^2} \sum_{k=1}^{n} \E[\mathrm{var}((||\widetilde{Z}_{ik}||)\mid Z_i)] + \frac{3}{(n-1)^2} \sum_{k=1}^{n} \E[\mathrm{var}((||\widetilde{Z}_{kj}||)\mid Z_j)] \\
	& + \frac{3}{n^2(n-1)^2} \sum_{k=1}^{n}\sum_{l=1}^{n} \mathrm{var}(||\widetilde{Z}_{kl}||) + \frac{6}{n^2(n-1)^2} \sum_{k=1}^{n}\sum_{l=1}^{n}\sum_{l'\not=l} \mathrm{cov}(||\widetilde{Z}_{kl}||,||\widetilde{Z}_{kl'}||)\\
	& \leq \frac{6(n-1)}{(n-1)^2}\E[||\widetilde{Z}||^2] + \frac{3n(n-1)}{n^2(n-1)^2}\E[||\widetilde{Z}||^2] + \frac{6n(n-1)(n-2)}{n^2(n-1)^2}\E[||\widetilde{Z}||^2]\\
	& = \mathcal{O}(n^{-1}).
	\end{split}
	\end{equation} The conclusion follows from Markov's inequality.
\end{proof}

\subsection{The Hessian matrix}

\begin{lemma}\label{Lem:Hessian_Derivation}
Suppose Assumptions \ref{Ass_dC:Reg}, \ref{Ass_dC:Sampling}, \ref{Ass_dC:AsymN}\Cref{Ass_dC:AsymN_DiffFU} hold, then the Hessian matrix is given by $ \mathcal{H} = 2\E\Big[f_{\widetilde{U} \mid \widetilde{\sigma}( [X,Z] )}(0)\mathcal{Z}\widetilde{X^g}'\widetilde{X^g}\Big] + \E\big[\operatorname{sgn}(\widetilde{U}) \mathcal{Z}\widetilde{X^{gg}}\big] $.
\end{lemma}
\begin{proof}
By the LIE and given that $\widetilde{U}(\theta) = \widetilde{U} - \widetilde{X^g}(\bar{\theta})(\theta-\theta_o) $ holds by the MVT and \Cref{Ass_dC:Reg}\Cref{Ass_dC:Reg_UdiffXMeas},
	\begin{align*}
	\mathcal{S}(\theta) &:= \frac{\partial Q(\theta)}{\partial \theta} =  \E\big[\big(1-2\E_{\widetilde{U} \mid \widetilde{\sigma}( [X,Z] )}[\indicator{ \widetilde{U}(\theta) \leq 0 }]\big) \mathcal{Z}\widetilde{X^g}(\theta)\big] \\
	&=  \E\big[(1-2\E_{\widetilde{U} \mid \widetilde{\sigma}( [X,Z] )}[\indicator{ \widetilde{U} \leq \widetilde{X^g}(\bar{\theta})(\theta-\theta_o)  }]) \mathcal{Z}\widetilde{X^g}(\theta)\big] \\
	&=  \E\big[\big(1-2F_{\widetilde{U} \mid \widetilde{\sigma}( [X,Z] )}( \widetilde{X^g}(\bar{\theta})(\theta-\theta_o) )\big) \mathcal{Z}\widetilde{X^g}(\theta)\big].
	\end{align*}
	
    Under the assumptions of \Cref{Lem:Verif_Intgrable_dSntheta}, the expectation and the derivative are exchangeable by the dominated convergence theorem. The expression for $\displaystyle \mathcal{H}(\theta) := \frac{\partial \mathcal{S}(\theta)}{\partial \theta'} $ becomes
	\begin{align*}
	\mathcal{H}(\theta)& = 2\E\Big[f_{\widetilde{U} \mid \widetilde{\sigma}( [X,Z] )}\big( \widetilde{X^g}(\bar{\theta})(\theta-\theta_o) \big) \mathcal{Z}\widetilde{X^g}(\theta)'\widetilde{X^g}(\bar{\theta})\Big] + \E\Big[\big(1-2F_{\widetilde{U} \mid \widetilde{\sigma}( [X,Z] )}\big( \widetilde{X^g}(\bar{\theta})(\theta-\theta_o) \big)\big) \mathcal{Z}\widetilde{X^{gg}}(\theta)\Big].
	\end{align*}
	Since $ \bar{\theta} $ satisfies $ ||\bar{\theta}-\theta_o|| \leq ||\theta-\theta_o|| $, evaluating $ \mathcal{H}(\theta) $ at $ \theta = \theta_o $ gives  
    \[
    \mathcal{H} = 2\E\big[ f_{\widetilde{U} \mid \widetilde{\sigma}( [X,Z] )}(0) \mathcal{Z}\widetilde{X^g}'\widetilde{X^g} \big] + \E\big[\operatorname{sgn}(\widetilde{U}) \mathcal{Z}\widetilde{X^{gg}}\big].
    \]
    
\end{proof}

The following result verifies the dominance condition used in the proof of \Cref{Lem:Hessian_Derivation}. Define \begin{equation*}\label{}
\begin{split}
\eta(\theta) &:= \Big[2f_{\widetilde{U} \mid \widetilde{\sigma}( [X,Z] )}\big( \widetilde{X^g}(\bar{\theta})(\theta-\theta_o) \big) \mathcal{Z}\widetilde{X^g}(\theta)'\widetilde{X^g}(\bar{\theta})\Big] + \Big[\Big(1-2F_{\widetilde{U} \mid \widetilde{\sigma}( [X,Z] )}\big( \widetilde{X^g}(\bar{\theta})(\theta-\theta_o) \big)\Big) \mathcal{Z}\widetilde{X^{gg}}(\theta)\Big]\\
& := \eta^A(\theta) + \eta^B(\theta).
\end{split}
\end{equation*} 

\begin{lemma}\label{Lem:Verif_Intgrable_dSntheta}
	Under Assumptions \ref{Ass_dC:Reg}\Cref{Ass_dC:Reg_Dominance} and  \ref{Ass_dC:AsymN}\Cref{Ass_dC:AsymN_DiffFU},
    \[
    \E\Big[\sup_{\theta \in \Theta }\Big|\Big|\eta^A(\theta)\Big|\Big|\Big] \leq 2f_oC^{1/2} \text{ and } \E\Big[\sup_{\theta \in \Theta }\Big|\Big|\eta^B(\theta)\Big|\Big|\Big] \leq C^{1/2}.
    \]
\end{lemma}

\begin{proof}[\textbf{Proof of \Cref{Lem:Verif_Intgrable_dSntheta}}]
	For any $ \theta \in \Theta $,
	\begin{equation*}
	\begin{split}
	||\eta^A(\theta)|| &= \big|\big|2\mathcal{Z}f_{\widetilde{U} \mid \widetilde{\sigma}( [X,Z] )}\big( \widetilde{X^g}(\bar{\theta})(\theta-\theta_o) \big) \widetilde{X^g}(\theta)'\widetilde{X^g}(\bar{\theta})\big|\big| \leq 2f_{\widetilde{U} \mid \widetilde{\sigma}( [X,Z] )}\big( \widetilde{X^g}(\bar{\theta})(\theta-\theta_o) \big)  ||\mathcal{Z}\widetilde{X^g}(\theta)'\widetilde{X^g}(\bar{\theta})|| \\
	&\leq 2f_{\widetilde{U} \mid \widetilde{\sigma}( [X,Z] )}\big( \widetilde{X^g}(\bar{\theta})(\theta-\theta_o) \big) \cdot\sup_{\theta \in \Theta}||\{|\mathcal{Z}| \vee 1\}\widetilde{X^g}(\theta)||^2\\
	&\text{ by the Schwarz inequality and }\\
	||\eta^B(\theta)|| &= \big|1-2F_{\widetilde{U} \mid \widetilde{\sigma}( [X,Z] )}\big( \widetilde{X^g}(\bar{\theta})(\theta-\theta_o) \big)\big| \times \Big|\Big|\mathcal{Z}\widetilde{X^{gg}}(\theta)\Big|\Big| \leq \Big|\Big|\mathcal{Z}\widetilde{X^{gg}}(\theta)\Big|\Big|
	\end{split}
	\end{equation*} noting that $ \Big|1-2F_{\widetilde{U} \mid \widetilde{\sigma}( [X,Z] )}(\cdot)\Big| \leq 1 $.
		
	From the foregoing, the CS inequality, the Lyapunov inequality, \Cref{Ass_dC:Reg}\Cref{Ass_dC:Reg_Dominance}, and \Cref{Ass_dC:AsymN}\Cref{Ass_dC:AsymN_DiffFU},
	\begin{equation*}
	\begin{split}
	\E[\sup_{\theta \in \Theta }||\eta^A(\theta)||] &\leq 2\big(\E\big[\big(\sup_{\theta \in \Theta }f_{\widetilde{U} \mid \widetilde{\sigma}( [X,Z] )}\big( \widetilde{X^g}(\bar{\theta})(\theta-\theta_o) \big)\big)^4\big]\big)^{1/4} \big(\E\big[\sup_{\theta \in \Theta } \big|\big|\{|\mathcal{Z}| \vee 1\}\widetilde{X^g}(\theta)\big|\big|^4\big]\big)^{1/2}\\
	& \leq 2f_o^{1/4}C^{1/2} \text{ and }\\
	\E[\sup_{\theta \in \Theta }||\eta^B(\theta)||] & \leq \big(\E\big[\sup_{\theta \in \Theta }\big|\big|\mathcal{Z}\widetilde{X^{gg}}(\theta)\big|\big|^2\big]\big)^{1/2} \leq C^{1/2}.
	\end{split}
	\end{equation*}
\end{proof}

\subsection{Consistency of the covariance matrix estimator}
\begin{lemma}\label{Lem:Conv_op1_H_Terms}
			Let the conditions of \Cref{Theorem:Consistency} hold. Then, in addition to \Cref{Ass_dC:AsymN,Ass_dC:Bandwith_Consistent}, \( \displaystyle ||\widehat{\mathcal{H}}_{1n} - \mathcal{H}_{1n}|| = o_p(1). \)
		\end{lemma}
	
		\begin{proof}[\textbf{Proof}]
        The following expression is useful in subsequent analyses. For any positive $ \epsilon_1, \epsilon_2 $ in a neighbourhood of zero, 
\begin{equation}\label{eqn:Cond_Exp_fu}
		 \begin{split}
		 \E_{\widetilde{U} \mid \widetilde{\sigma}( [X,Z] )}[\indicator{ |\widetilde{U}|\leq \epsilon_1 }]/(2\epsilon_2) &= \frac{F_{\widetilde{U} \mid \widetilde{\sigma}( [X,Z] )}(\epsilon_1) - F_{\widetilde{U} \mid \widetilde{\sigma}( [X,Z] )}(-\epsilon_1)}{2\epsilon_2}\\
		 &= \frac{F_{\widetilde{U} \mid \widetilde{\sigma}( [X,Z] )}(\epsilon_1) - \big(F_{\widetilde{U} \mid \widetilde{\sigma}( [X,Z] )}(\epsilon_1) - f_{\widetilde{U} \mid \widetilde{\sigma}( [X,Z] )}((1-2\lambda)\epsilon_1)(2\epsilon_1)\big)}{2\epsilon_2}\\ 
		 &= (\epsilon_1/\epsilon_2)f_{\widetilde{U} \mid \widetilde{\sigma}( [X,Z] )}((1-2\lambda)\epsilon_1)
		 \end{split}
		\end{equation} for some $ \lambda \in (0,1) $ by \Cref{Ass_dC:AsymN}\Cref{Ass_dC:AsymN_DiffFU} and the MVT (taken about $ \epsilon_1 $).
        
        By the triangle inequality, $ ||\widehat{\mathcal{H}}_{1n} - \mathcal{H}_{1n}||\leq \frac{c_n}{\hat{c}_n}(A_{n,0} + A_{n,1} + A_{n,2} + A_{n,3} + A_{n,4}) $ where				
		\begin{align*}
			A_{n,0} &:= \frac{1}{n(n-1)c_n}\sum_{i=1}^{n}\sum_{j\neq i}^{n} \Big\{|\mathcal{Z}_{ij,n} - \mathcal{Z}_{ij}|\times \indicator{ |\widetilde{U}_{ij}(\widehat{\theta}_n)| \leq \hat{c}_n } \times \big\lVert \widetilde{X^g}_{ij}(\widehat{\theta}_n)'\widetilde{X^g}_{ij}(\widehat{\theta}_n) \big\rVert\Big\};\\
			A_{n,1} &:=  \frac{1}{n(n-1)c_n}\sum_{i=1}^{n}\sum_{j\neq i}^{n} \Big\{\big| \indicator{ |\widetilde{U}_{ij}(\widehat{\theta}_n)| \leq \hat{c}_n } - \indicator{ |\widetilde{U}_{ij}| \leq \hat{c}_n }\big| \times \big\lVert \mathcal{Z}_{ij} \widetilde{X^g}_{ij}(\widehat{\theta}_n)'\widetilde{X^g}_{ij}(\widehat{\theta}_n) \big\rVert \Big\};\\
			A_{n,2} &:= \frac{1}{n(n-1)c_n}\sum_{i=1}^{n}\sum_{j\neq i}^{n} \Big\{\indicator{ |\widetilde{U}_{ij}| \leq \hat{c}_n } \times \big\lVert \mathcal{Z}_{ij}[ \widetilde{X^g}_{ij}(\widehat{\theta}_n)'\widetilde{X^g}_{ij}(\widehat{\theta}_n) - \widetilde{X}_{ij}^{g\prime}\widetilde{X^g}_{ij}]\big\rVert \Big\};\\
			A_{n,3} &:= \frac{1}{n(n-1)c_n}\sum_{i=1}^{n}\sum_{j\neq i}^{n} \Big\{ \big| \indicator{ |\widetilde{U}_{ij}| \leq \hat{c}_n } - \indicator{ |\widetilde{U}_{ij}| \leq c_n } \big| \times \big\lVert \mathcal{Z}_{ij} \widetilde{X}_{ij}^{g\prime}\widetilde{X^g}_{ij} \big\rVert \Big\}; \text{ and }\\
            A_{n,4} &:= \frac{1}{n(n-1)c_n}\sum_{i=1}^{n}\sum_{j\neq i}^{n} \Big\{\indicator{ |\widetilde{U}_{ij}| \leq c_n } \times \big\lVert \mathcal{Z}_{ij} \widetilde{X}_{ij}^{g\prime}\widetilde{X^g}_{ij} \big\rVert \Big\} \times \Big| 1 - \frac{\hat{c}_n}{c_n} \Big|.
		\end{align*}
        
		The verification of the elements of $ [A_{n,0}, A_{n,1}, A_{n,2}, A_{n,3},A_{n,4}] $ proceeds in the following. 
		
		$ A_{n,0} $:\\
		By the Schwarz inequality, the Cauchy-Schwarz (CS) inequality, and the identical distribution of the data \Cref{Ass_dC:Sampling},
		\begin{equation}\label{eqn:An0_consistency}
		\begin{split}
		\E[A_{n,0}]  &= \frac{1}{n(n-1)c_n}\sum_{i=1}^{n}\sum_{j\neq i}^{n} \E\big[|\mathcal{Z}_{ij,n} - \mathcal{Z}_{ij}|\times \indicator{ |\widetilde{U}_{ij}(\widehat{\theta}_n)| \leq \hat{c}_n } \times || \widetilde{X^g}_{ij}(\widehat{\theta}_n)'\widetilde{X^g}_{ij}(\widehat{\theta}_n)||\big]\\
		& \leq \frac{(\E[\sup_{\theta \in \Theta}|| \widetilde{X^g}(\theta)||^4])^{1/2}}{n(n-1)c_n}\sum_{i=1}^{n}\sum_{j\neq i}^{n} (\E[(\mathcal{Z}_{ij,n} - \mathcal{Z}_{ij})^2])^{1/2}.
		\end{split}
		\end{equation} Under the assumptions of \Cref{Lem:Zijn_Zij_op1}, \Cref{Ass_dC:Reg}\Cref{Ass_dC:Reg_Dominance}, and \Cref{Ass_dC:Bandwith_Consistent}, it follows that $ A_{n,0} = \mathcal{O}_p((\sqrt{n}c_n)^{-1}) = o_p(1) $.
		
		$ A_{n,1} $:\\
        Let $ \widetilde{\Delta}_X(\widehat{\theta}_n) := -\widetilde{X^g}_{ij}(\bar{\theta}_n)(\widehat{\theta}_n-\theta_o) $ for notational ease. Since \( | \indicator{A} - \indicator{B} | = \indicator{ (A\setminus B) \cup (B\setminus A) } \), the following equality holds.
        \begin{align*}
            \big| \indicator{&|\widetilde{U}+\Delta| \leq c}  -  \indicator{|\widetilde{U}| \leq c} \big|\\
            &= \indicator{c \wedge (c-\Delta) < \widetilde{U} \leq c \vee (c-\Delta) } + \indicator{-c \wedge -(c+\Delta) \leq \widetilde{U} < -c \vee -(c+\Delta) }.
        \end{align*}
        
		\noindent For $ \bar{\theta}_n $ that satisfies $||\bar{\theta}_n - \theta_o|| \leq ||\widehat{\theta}_n - \theta_o|| $, apply the above equality:
		\begin{equation*}\begin{split}
		 & \E_{\widetilde{U} \mid \widetilde{\sigma}( [X,Z] )}\Big[\big|\indicator{ |\widetilde{U}_{ij}(\widehat{\theta}_n)| \leq \hat{c}_n } - \indicator{ |\widetilde{U}_{ij}| \leq \hat{c}_n }\big|\Big] \\ 
		 =& \E_{\widetilde{U} \mid \widetilde{\sigma}( [X,Z] )}\Big[\big|\indicator{ |\widetilde{U}_{ij} + \widetilde{\Delta}_X(\widehat{\theta}_n)| \leq \hat{c}_n } - \indicator{ |\widetilde{U}_{ij}| \leq \hat{c}_n }\big|\Big] \\
		 =& \E_{\widetilde{U} \mid \widetilde{\sigma}( [X,Z] )}\big[\indicator{ \hat{c}_n \wedge (\hat{c}_n-\widetilde{\Delta}_X(\widehat{\theta}_n)) < \widetilde{U}_{ij} \leq \hat{c}_n \vee (\hat{c}_n - \widetilde{\Delta}_X(\widehat{\theta}_n)) }\big] \\
		 &+ \E_{\widetilde{U} \mid \widetilde{\sigma}( [X,Z] )}\big[\indicator{ -\hat{c}_n \wedge -(\hat{c}_n + \widetilde{\Delta}_X(\widehat{\theta}_n)) \leq \widetilde{U}_{ij} < -\hat{c}_n \vee -(\hat{c}_n + \widetilde{\Delta}_X(\widehat{\theta}_n)) }\big]\\
		 =&: \E_{\widetilde{U} \mid \widetilde{\sigma}( [X,Z] )}[\widetilde{I}_{ij}^{(1)}] + \E_{\widetilde{U} \mid \widetilde{\sigma}( [X,Z] )}[\widetilde{I}_{ij}^{(2)}]
		\end{split}
		\end{equation*} by \Cref{Ass_dC:Reg}\Cref{Ass_dC:Reg_UdiffXMeas} and the MVT. By \Cref{Ass_dC:AsymN}\Cref{Ass_dC:AsymN_DiffFU} and the MVT for some $ \lambda_1, \lambda_2 \in (0,1) $,
		\begin{equation*}\label{}
		\begin{split}
		\E_{\widetilde{U} \mid \widetilde{\sigma}( [X,Z] )}[\widetilde{I}_{ij}^{(1)}]  &= \Big| F_{\widetilde{U} \mid \widetilde{\sigma}( [X,Z] )}(\hat{c}_n - \widetilde{\Delta}_X(\widehat{\theta}_n)) - F_{\widetilde{U} \mid \widetilde{\sigma}( [X,Z] )}(\hat{c}_n)\Big| \\ 
		& = f_{\widetilde{U} \mid \widetilde{\sigma}( [X,Z] )}\big(\hat{c}_n - \lambda_1\widetilde{\Delta}_X(\widehat{\theta}_n)\big)\big|\widetilde{\Delta}_X(\widehat{\theta}_n)\big| \qquad \text{ and }\\
		\E_{\widetilde{U} \mid \widetilde{\sigma}( [X,Z] )}[\widetilde{I}_{ij}^{(2)}] & = \Big|F_{\widetilde{U} \mid \widetilde{\sigma}( [X,Z] )}(-\hat{c}_n) - F_{\widetilde{U} \mid \widetilde{\sigma}( [X,Z] )}(-\hat{c}_n - \widetilde{\Delta}_X(\widehat{\theta}_n))\Big|\\
		& = f_{\widetilde{U} \mid \widetilde{\sigma}( [X,Z] )}\big(-\hat{c}_n - \lambda_2\widetilde{\Delta}_X(\widehat{\theta}_n)\big)\big|\widetilde{\Delta}_X(\widehat{\theta}_n)\big|.
		\end{split}
		\end{equation*}
		 Since $ f_{\widetilde{U} \mid \widetilde{\sigma}( [X,Z] )}(\cdot)  \leq f_o^{1/4}$ a.s. by \Cref{Ass_dC:AsymN}\Cref{Ass_dC:AsymN_DiffFU} and $ |\widetilde{\Delta}_X(\widehat{\theta}_n)| = \mathcal{O}_p(n^{-1/2}) $ by \Cref{Ass_dC:Reg}\Cref{Ass_dC:Reg_Dominance} cum  \Cref{Theorem:Normality}, $ c_n^{-1}\widetilde{\Delta}_X(\widehat{\theta}_n) = o_p(1) $ by \Cref{Ass_dC:Bandwith_Consistent}. It follows that $ \frac{\E_{\widetilde{U} \mid \widetilde{\sigma}( [X,Z] )}[\widetilde{I}_{ij}^{(1)}] + \E_{\widetilde{U} \mid \widetilde{\sigma}( [X,Z] )}[\widetilde{I}_{ij}^{(2)}]}{2c_n} = o_p(1) $. 

		From the foregoing, the LIE, the CS inequality, the Lyapunov inequality, \Cref{Ass_dC:Reg}\Cref{Ass_dC:Reg_Dominance}, and the identical sampling of the data (\Cref{Ass_dC:Sampling})
		\begin{equation*}
		\begin{split}
		&\E[A_{n,1}]  = \frac{1}{n(n-1)c_n}\sum_{i=1}^{n}\sum_{j\neq i}^{n} \E\Big[\E_{\widetilde{U} \mid \widetilde{\sigma}( [X,Z] )}[|\indicator{ |\widetilde{U}_{ij}(\widehat{\theta}_n)| \leq \hat{c}_n }-\indicator{ |\widetilde{U}_{ij}| \leq \hat{c}_n }|] \times ||\mathcal{Z}_{ij} \widetilde{X^g}_{ij}(\widehat{\theta}_n)'\widetilde{X^g}_{ij}(\widehat{\theta}_n)||\Big]\\
		& \leq \frac{2}{n(n-1)}\sum_{i=1}^{n}\sum_{j\neq i}^{n} \Big\{\E\Big[\Big(\frac{\E_{\widetilde{U} \mid \widetilde{\sigma}( [X,Z] )}[\widetilde{I}_{ij}^{(1)}] + \E_{\widetilde{U} \mid \widetilde{\sigma}( [X,Z] )}[\widetilde{I}_{ij}^{(2)}]}{2c_n}\Big)^2\Big] \times \E\big[||\{|\mathcal{Z}_{ij}| \vee 1\}\widetilde{X^g}_{ij}(\widehat{\theta}_n)||^4\big]\Big\}^{1/2} \\
            & \leq 2\sup_{\theta \in \Theta} \big(\E\big[||\{|\mathcal{Z}| \vee 1\}\widetilde{X^g}(\theta)||^4\big]\big)^{1/2} \times \frac{1}{n(n-1)}\sum_{i=1}^{n}\sum_{j\neq i}^{n} \Big\{\E\Big[\Big(\frac{\E_{\widetilde{U} \mid \widetilde{\sigma}( [X,Z] )}[\widetilde{I}_{ij}^{(1)}] + \E_{\widetilde{U} \mid \widetilde{\sigma}( [X,Z] )}[\widetilde{I}_{ij}^{(2)}]}{2c_n}\Big)^2\Big] \Big\}^{1/2}.
		\end{split}
		\end{equation*} $ A_{n,1}= o_p(1) $ thanks to the Markov inequality.
		
		$ A_{n,2} $:\\
		First, by \Cref{Ass_dC:AsymN}\Cref{Ass_dC:AsymN_DiffFU}, \eqref{eqn:Cond_Exp_fu}, and the MVT, 
		\begin{equation*}\label{eqn:num_F_deriv}
		    \E_{\widetilde{U} \mid \widetilde{\sigma}( [X,Z] )}[\indicator{ |\widetilde{U}_{ij}| \leq \hat{c}_n }]/(2c_n) = f_{\widetilde{U} \mid \widetilde{\sigma}( [X,Z] )}(\lambda\hat{c}_n)(\hat{c}_n/c_n)
		\end{equation*}
		 for some $ \lambda \in (0,1) $. It follows from the LIE, \Cref{Ass_dC:AsymN}\Cref{Ass_dC:AsymN_DiffFU}, \Cref{Ass_dC:Bandwith_Consistent}, the CS inequality, the continuity of the Jacobian (\Cref{Ass_dC:Reg}\Cref{Ass_dC:Reg_UdiffXMeas}), the continuous mapping theorem (CMT), and the consistency of the MDep (\Cref{Theorem:Consistency}) that
		\begin{equation*}
		\begin{split}
		\E[A_{n,2}] &= \frac{1}{n(n-1)c_n}\sum_{i=1}^{n}\sum_{j\neq i}^{n} \E\Big[\E_{\widetilde{U} \mid \widetilde{\sigma}( [X,Z] )} \big[\indicator{ |\widetilde{U}_{ij}| \leq \hat{c}_n }\big] \times \big\lVert \mathcal{Z}_{ij} \big[ \widetilde{X^g}_{ij}(\widehat{\theta}_n)'\widetilde{X^g}_{ij}(\widehat{\theta}_n) - \widetilde{X}_{ij}^{g\prime}\widetilde{X^g}_{ij}\big] \big\rVert \Big]\\
		&\leq \frac{2}{n^2}\sum_{i=1}^{n}\sum_{j\neq i}^{n} \Big\{\Big(\E\big[\big((\hat{c}_n/c_n)f_{\widetilde{U} \mid \widetilde{\sigma}( [X,Z] )}(\lambda \hat{c}_n)\big)^2\big]\Big)^{1/2} \Big(\E\big[\mathcal{Z}_{ij}^2\big\lVert \widetilde{X^g}_{ij}(\widehat{\theta}_n)'\widetilde{X^g}_{ij}(\widehat{\theta}_n) - \widetilde{X}_{ij}^{g\prime}\widetilde{X^g}_{ij} \big\rVert^2\big]\Big)^{1/2}\Big\} \\
		&\leq 4f_o^{1/4}\times o(1) = o(1)
		\end{split}
		\end{equation*} noting in particular that $ \rho(\theta):= \E\big[\mathcal{Z}^2|| \widetilde{X^g}(\theta)'\widetilde{X^g}(\theta) - \widetilde{X^g}(\theta_o)'\widetilde{X^g}(\theta_o)||^2\big] $ under \Cref{Ass_dC:Reg}\Cref{Ass_dC:Reg_UdiffXMeas} is continuous in $\theta$. $ A_{n,2}= o_p(1) $ thanks to the Markov inequality.
		
		$ A_{n,3} $:\\
        Using the equality 
        \begin{align*}
            \big| \indicator{&|\widetilde{U}| \leq \hat{c}}  -  \indicator{|\widetilde{U}| \leq c} \big| = \indicator{(\hat{c} \wedge c) < \widetilde{U} \leq (\hat{c} \vee c) } + \indicator{-(\hat{c} \vee c) \leq \widetilde{U} < -(\hat{c} \wedge c) }
        \end{align*}
		\begin{equation*}
		\begin{split}
		& \E_{\widetilde{U} \mid \widetilde{\sigma}( [X,Z] )}\big[|\indicator{ |\widetilde{U}_{ij}| \leq \hat{c}_n }-\indicator{ |\widetilde{U}_{ij}| \leq c_n }|\big]/(2c_n)\\ 
        &= \frac{\E_{\widetilde{U} \mid \widetilde{\sigma}( [X,Z] )}\Big[\indicator{ (\hat{c}_n \wedge c_n) < \widetilde{U}_{ij} \leq (\hat{c}_n \vee c_n) } + \indicator{ -(\hat{c}_n \vee c_n) \leq \widetilde{U}_{ij} < - (\hat{c}_n \wedge c_n) }\Big]}{2c_n}\\
		&= \frac{F_{\widetilde{U} \mid \widetilde{\sigma}( [X,Z] )} \big( \hat{c}_n \vee c_n \big) - F_{\widetilde{U} \mid \widetilde{\sigma}( [X,Z] )}\big( \hat{c}_n \wedge c_n \big)}{2c_n} + \frac{F_{\widetilde{U} \mid \widetilde{\sigma}( [X,Z] )} \big(-(\hat{c}_n \wedge c_n) \big) - F_{\widetilde{U} \mid \widetilde{\sigma}( [X,Z] )}\big(-(\hat{c}_n \vee c_n)\big)}{2c_n}\\
		 & = 0.5f_{\widetilde{U} \mid \widetilde{\sigma}( [X,Z] )}(\bar{c}_n)\Big(\frac{(\hat{c}_n \vee c_n)-(\hat{c}_n \wedge c_n)}{c_n} \Big) - 0.5f_{\widetilde{U} \mid \widetilde{\sigma}( [X,Z] )}(-\underline{c}_n)\Big(\frac{(\hat{c}_n \wedge c_n) - (\hat{c}_n \vee c_n)}{c_n}\Big)\\
         & = 0.5f_{\widetilde{U} \mid \widetilde{\sigma}( [X,Z] )}(\bar{c}_n)\Big|\frac{ \hat{c}_n }{c_n} -1 \Big| + 0.5f_{\widetilde{U} \mid \widetilde{\sigma}( [X,Z] )}(-\underline{c}_n)\Big|\frac{ \hat{c}_n }{c_n} - 1\Big| = o_p(1)
		\end{split}
		\end{equation*} for some intermediate values $ \bar{c}_n, \underline{c}_n \in \big( (\hat{c}_n \wedge c_n), \; (\hat{c}_n \vee c_n) \big) $ by \Cref{Ass_dC:AsymN}\Cref{Ass_dC:AsymN_DiffFU}, \Cref{Ass_dC:Bandwith_Consistent}, \eqref{eqn:Cond_Exp_fu}, and the MVT. Apply the LIE and \Cref{Ass_dC:Reg}\ref{Ass_dC:Reg_Dominance} to conclude that $ A_{n,3} = o_p(1) $.

        $ A_{n,4} $:\\
        By \eqref{eqn:Cond_Exp_fu} and \Cref{Ass_dC:AsymN}\Cref{Ass_dC:AsymN_DiffFU}, $\E_{\widetilde{U} \mid \widetilde{\sigma}( [X,Z] )}[\indicator{ |\widetilde{U}_{ij}| \leq c_n }]/(2c_n) = f_{\widetilde{U} \mid \widetilde{\sigma}( [X,Z] )}(\tilde{\lambda} c_n) \leq f_o^{1/4} $ for some $\tilde{\lambda}\in (-1,1)$. It follows, thanks to the LIE, the Schwarz inequality, the Lyapunov inequality, and \Cref{Ass_dC:Reg}\Cref{Ass_dC:Reg_Dominance} that,
        \begin{align*}
            \frac{1}{n(n-1)c_n}\sum_{i=1}^{n}\sum_{j\neq i}^{n} &\E \Big[\indicator{ |\widetilde{U}_{ij}| \leq c_n }\times||\mathcal{Z}_{ij} \widetilde{X}_{ij}^{g\prime}\widetilde{X^g}_{ij}||\Big]\\ 
            &\leq  2f_o^{1/4}\frac{1}{n(n-1)}\sum_{i=1}^{n}\sum_{j\neq i}^{n} \E \Big[||\mathcal{Z}_{ij} \widetilde{X}_{ij}^{g\prime}\widetilde{X^g}_{ij}||\Big]\\
            &\leq  2f_o^{1/4}\frac{1}{n(n-1)}\sum_{i=1}^{n}\sum_{j\neq i}^{n} \E \big[||\{|\mathcal{Z}_{ij}| \vee 1\} \widetilde{X^g}_{ij}||^2\big]\\
            &\leq 2f_o^{1/4}\frac{1}{n(n-1)}\sum_{i=1}^{n}\sum_{j\neq i}^{n} \big(\E \big[||\{|\mathcal{Z}_{ij}| \vee 1\} \widetilde{X^g}_{ij}||^4\big]\big)^{1/2}\\
            &\leq 2(f_oC)^{1/2}.
        \end{align*}
        \noindent In addition to \Cref{Ass_dC:Bandwith_Consistent}, $ A_{n,4} = o_p(1) $ thanks to the Markov inequality.

        Combining all parts above concludes the proof.
        
		\end{proof}
		
        The next result shows that $ \mathcal{H}_{1n} - \mathcal{H}_1 $ converges to zero in quadratic mean. 
		
		\begin{lemma}\label{Lem:Conv_Quadratic_Mean_H}
			Under Assumption  \ref{Ass_dC:Reg}(\Cref{Ass_dC:Reg_Dominance}), \Cref{Ass_dC:Sampling}, \ref{Ass_dC:AsymN}\Cref{Ass_dC:AsymN_DiffFU}, and  \ref{Ass_dC:Bandwith_Consistent}, $ \mathcal{H}_{1n} - \mathcal{H}_1 $ converges to zero in quadratic mean.
		\end{lemma}
		
		\begin{proof}[\textbf{Proof}]
			$ \Big|\E_{\widetilde{U} \mid \widetilde{\sigma}( [X,Z] )}\big[\indicator{ |\widetilde{U}_{ij}| \leq c_n }\big]/(2c_n) - f_{\widetilde{U} \mid \widetilde{\sigma}( [X,Z] )}(0) \Big|  = \big|f_{\widetilde{U} \mid \widetilde{\sigma}( [X,Z] )}(\tilde{\lambda} c_n)-f_{\widetilde{U} \mid \widetilde{\sigma}( [X,Z] )}(0)\big| = o_p(1) $ a.s. for some $ \tilde{\lambda} \in (-1,1) $ by \eqref{eqn:Cond_Exp_fu}, \Cref{Ass_dC:AsymN}\Cref{Ass_dC:AsymN_DiffFU}, the MVT, \Cref{Ass_dC:Bandwith_Consistent}, and the CMT. In addition to \Cref{Ass_dC:Reg}\Cref{Ass_dC:Reg_Dominance} and the CS inequality, this implies 
			\begin{equation*}\label{}
			\begin{split}
			&||\E[\mathcal{H}_{1n}] - \mathcal{H}_1|| \leq\\
			& 2\big(\E[\big(\E_{\widetilde{U} \mid \widetilde{\sigma}( [X,Z] )}[\indicator{ |\widetilde{U}_{ij}| \leq c_n }]/(2c_n) - f_{\widetilde{U} \mid \widetilde{\sigma}( [X,Z] )}(0)\big)^2]\big)^{1/2} \big(\E[||\{|\mathcal{Z}_{ij}| \vee 1\} \widetilde{X^g}_{ij}||^4]\big)^{1/2} \\
			&= o(1).
			\end{split}
			\end{equation*} 
			
			Let $ \uptau_1 $ and $ \uptau_2 $ be two $ p_X \times 1 $ vectors with $ ||\uptau_1|| = ||\uptau_2|| = 1 $, then
			\begin{flalign*}
			&\mathrm{var}(\uptau_1'\mathcal{H}_{1n}\uptau_2)\\ 
			&= 
			\frac{1}{n^2(n-1)^2c_n^2}\sum_{i=1}^{n}\sum_{j\neq i}^{n}\sum_{i'=1}^{n}\sum_{j'=1}^{n} \mathrm{cov}\Big(\{\indicator{ |\widetilde{U}_{ij}| \leq c_n }\mathcal{Z}_{ij} \uptau_1'\widetilde{X}_{ij}^{g\prime}\widetilde{X^g}_{ij}\uptau_2\},\{\indicator{ |\widetilde{U}_{i'j'}| \leq c_n }\mathcal{Z}_{i'j'} \uptau_1'\widetilde{X}_{i'j'}^{g\prime}\widetilde{X^g}_{i'j'}\uptau_2\}
			\Big)\\
			&= 
			\frac{1}{n^2(n-1)^2c_n^2}\sum_{i=1}^{n}\sum_{j\neq i}^{n} \mathrm{var}\Big(\indicator{ |\widetilde{U}_{ij}| \leq c_n }\mathcal{Z}_{ij} \uptau_1'\widetilde{X}_{ij}^{g\prime}\widetilde{X^g}_{ij}\uptau_2\Big)\\
			& + \frac{2}{n^2(n-1)^2c_n^2}\sum_{i=1}^{n}\sum_{j\neq i}^{n}\sum_{i'\neq i}^{n} \mathrm{cov}\Big(\{\indicator{ |\widetilde{U}_{ij}| \leq c_n }\mathcal{Z}_{ij} \uptau_1'\widetilde{X}_{ij}^{g\prime}\widetilde{X^g}_{ij}\uptau_2\},\{\indicator{ |\widetilde{U}_{i'j}| \leq c_n }\mathcal{Z}_{i'j} \uptau_1'\widetilde{X}_{i'j}^{g\prime}\widetilde{X^g}_{i'j}\uptau_2\}\Big)\\
			&\leq
			\frac{1}{n^2(n-1)^2c_n^2}\sum_{i=1}^{n}\sum_{j\neq i}^{n} \mathrm{var}\Big(\indicator{ |\widetilde{U}_{ij}| \leq c_n }\mathcal{Z}_{ij} \uptau_1'\widetilde{X}_{ij}^{g\prime}\widetilde{X^g}_{ij}\uptau_2\Big)\\
			& + \frac{2}{n^2(n-1)^2c_n^2}\sum_{i=1}^{n}\sum_{j\neq i}^{n}\sum_{i'\neq i}^{n} \Big(\mathrm{var}\big(\indicator{ |\widetilde{U}_{ij}| \leq c_n } \uptau_1'\mathcal{Z}_{ij}\widetilde{X}_{ij}^{g\prime}\widetilde{X^g}_{ij}\uptau_2\big)\cdot\mathrm{var}\big(\indicator{ |\widetilde{U}_{i'j}| \leq c_n } \uptau_1'\mathcal{Z}_{i'j}\widetilde{X}_{i'j}^{g\prime}\widetilde{X^g}_{i'j}\uptau_2\big)
			\Big)^{1/2}\\
			&\leq \frac{1}{n^2(n-1)^2c_n^2}\sum_{i=1}^{n}\sum_{j\neq i}^{n} \E[||\{|\mathcal{Z}_{ij}| \vee 1\}\widetilde{X^g}_{ij}||^4]\\
			& + \frac{2}{n^2(n-1)^2c_n^2}\sum_{i=1}^{n}\sum_{j\neq i}^{n}\sum_{i'\neq i}^{n} \Big(\E[||\{|\mathcal{Z}_{ij}| \vee 1\}\widetilde{X^g}_{ij}||^4]\cdot \E[||\{|\mathcal{Z}_{i'j}| \vee 1\}\widetilde{X^g}_{i'j}||^4]
			\Big)^{1/2}\\
			& \leq \frac{C}{n(n-1)c_n^2} + \frac{2C}{nc_n^2}.
			\end{flalign*} The second equality follows from \Cref{Ass_dC:Sampling}, the first inequality follows from the CS inequality, and the second inequality follows from Jensen's inequality. The second inequality holds because \begin{align*}
				\mathrm{var}(\uptau_1'M\uptau_2) &\leq \E[(\uptau_1'M\uptau_2)^2] = \E[(\mathrm{vec}(\uptau_1'M\uptau_2))^2] = \E[\mathrm{vec}(M)'  (\uptau_2'\otimes\uptau_1')'(\uptau_2'\otimes\uptau_1')\mathrm{vec}(M)]\\ 
				&\leq \E[||\mathrm{vec}(M)||^2\cdot||\uptau_1'\otimes\uptau_2'||^2] = \E[||\mathrm{vec}(M)||^2\cdot||\uptau_1||^2\cdot||\uptau_2||^2] = \E[||M||^2]
			\end{align*}for a matrix-valued random variable $ M $, and $ ||\uptau_1'\otimes\uptau_2'|| = ||\uptau_1||\cdot||\uptau_2|| $ by \textcite[Fact 9.7.27]{bernstein2009matrix}.
			Thanks to Assumptions \ref{Ass_dC:Reg}\Cref{Ass_dC:Reg_Dominance} and \ref{Ass_dC:Bandwith_Consistent}, $ \mathrm{var}(\uptau_1'\mathcal{H}_n\uptau_2) \leq 3C/(nc_n^2) = o(1) $, and the assertion is proved as claimed.
		\end{proof} 

\begin{lemma}\label{lem:Hess_2_consistency}
    Let the conditions of \Cref{Theorem:Consistency} hold. Then, in addition to \Cref{Ass_dC:Sampling}, (a) \( \displaystyle \lVert \widehat{\mathcal{H}}_{2n} - \mathcal{H}_{2n} \rVert = o_p(1) \) and (b) \( \lVert\mathcal{H}_{2n} - \mathcal{H}_{2} \rVert = o_p(1). \)
\end{lemma}
\begin{proof}
\textbf{Part (a):}

Consider the following decomposition
    \begin{align*}
        \widehat{\mathcal{H}}_{2n} - \mathcal{H}_{2n} =& \frac{1}{n(n-1)}\sum_{i=1}^{n}\sum_{j\neq i}^{n} \Big\{\sgn\big(\widetilde{U}_{ij}(\widehat{\theta}_n)\big) \mathcal{Z}_{ij,n}\widetilde{X^{gg}_{ij}}\big( \widehat{\theta}_n \big)  - \sgn (\widetilde{U}_{ij}) \mathcal{Z}_{ij}\widetilde{X^{gg}_{ij}}\big( \theta_o \big) \Big\} \\
        =& \frac{1}{n(n-1)}\sum_{i=1}^{n}\sum_{j\neq i}^{n} \big( \mathcal{Z}_{ij,n} - \mathcal{Z}_{ij}\big) \sgn \big(\widetilde{U}_{ij}(\widehat{\theta}_n)\big) \widetilde{X^{gg}_{ij}}\big( \widehat{\theta}_n \big) \\
        +& \frac{1}{n(n-1)}\sum_{i=1}^{n}\sum_{j\neq i}^{n} \Big(\sgn\big(\widetilde{U}_{ij}(\widehat{\theta}_n)\big) - \sgn (\widetilde{U}_{ij}) \Big) \mathcal{Z}_{ij}\widetilde{X^{gg}_{ij}}\big( \widehat{\theta}_n \big) \\
        +& \frac{1}{n(n-1)}\sum_{i=1}^{n}\sum_{j\neq i}^{n} \sgn (\widetilde{U}_{ij}) \mathcal{Z}_{ij}\Big(\widetilde{X^{gg}_{ij}}\big( \widehat{\theta}_n \big)  - \widetilde{X^{gg}_{ij}}\big( \theta_o \big) \Big)\\
        =&: H_{1n} + H_{2n} + H_{3n}.
    \end{align*} 
    
    First, \( \displaystyle \sup_{\theta\in\Theta} \big| \sgn\big(\widetilde{U}_{ij}(\widehat{\theta}_n)\big) \big| \leq 1 \). Using arguments as applied to \eqref{eqn:An0_consistency} analogously, in addition to the dominance conditions of \Cref{Ass_dC:Reg}\ref{Ass_dC:Reg_Dominance}, deduce that \( H_{1n} = o_p(1). \) Second, observe that \( \sgn\big(\widetilde{U}_{ij}(\widehat{\theta}_n)\big) - \sgn (\widetilde{U}_{ij}) = -2\big( \indicator{ \widetilde{U}(\widehat{\theta}_n) \leq 0 } - \indicator{ \widetilde{U} \leq 0 }  \big) \). Using arguments analogous to those used in \eqref{eqn:cons_indicator}, conclude, in addition to the dominance conditions of \Cref{Ass_dC:Reg}\ref{Ass_dC:Reg_Dominance} that \( H_{2n} = o_p(1). \) Third, by the twice continuous differentiability of $ U(\theta) $ under \Cref{Ass_dC:Reg}\ref{Ass_dC:Reg_UdiffXMeas}, conclude by the CMT and the consistency of the MDep, namely \Cref{Theorem:Consistency}, that $H_{3n}=o_p(1)$. This completes the proof of part (a).

    \textbf{Part (b):} Under the dominance conditions of \Cref{Ass_dC:Reg}\ref{Ass_dC:Reg_Dominance}, \( \displaystyle \E\Big[ \sup_{\theta\in \Theta} \Big\lVert \sgn (\widetilde{U}) \mathcal{Z}\widetilde{X^{gg}}\big( \theta \big)  \Big\rVert \Big] \leq \E\Big[ \sup_{\theta\in \Theta} \Big\lVert \mathcal{Z}\widetilde{X^{gg}} ( \theta )  \Big\rVert \Big] \leq C^{1/2}. \) In addition to the sampling condition of \Cref{Ass_dC:Sampling}, conclude that $ \mathcal{H}_{2n} \xrightarrow{a.s.} \mathcal{H}_2 $ thanks to Hoeffding's strong law of large numbers for U-statistics \citep{hoeffding1961strong}.
    
\end{proof}

\subsection{Stochastic equi-continuity}
The following lemma verifies the stochastic equicontinuity condition used in the proof of \Cref{Theorem:Normality}.
\begin{lemma}\label{Lem:Verif_AssStochEqui}
	Under Assumptions~\ref{Ass_dC:Reg}\ref{Ass_dC:Reg_UdiffXMeas},~\ref{Ass_dC:Reg}\ref{Ass_dC:Reg_Dominance}, and~\ref{Ass_dC:AsymN}, $\displaystyle \sup_{\theta \in \Theta_o } \frac{||v_n(\theta) - v_n(\theta_o)||}{1+\sqrt{n}||\mathcal{S}(\theta)||} =o_p(1) $ in some open neighbourhood $ \Theta_o $ of $ \theta_o $.
\end{lemma}

\begin{proof}[\textbf{Proof}]
	The proof proceeds by verifying the conditions of \textcite[Lemma 2]{honore1994pairwise}. Recall $ \psi(W_i,W_j;\theta) := \mathcal{Z}\big(1-2\indicator{ \widetilde{U}(\theta) \leq 0 }\big) \widetilde{X^g}(\theta)' $. 
	
	First, from \Cref{Ass_dC:Reg}\Cref{Ass_dC:Reg_UdiffXMeas}, $ \widetilde{U}(\theta) $ and $ \widetilde{X^g}(\theta) $ are measurable in $ [U,U^\dagger,X,X^\dagger] $ for all $ \theta \in \Theta $. It follows that for any $ \theta_1,\theta_2 $ in an open neighbourhood $ \Theta_o \subset \Theta $ containing $ \theta_o $ (\Cref{Ass_dC:AsymN}\Cref{Ass_dC:AsymN_IntTheta}), $\displaystyle \sup_{||\theta_1 - \theta_2||<d} \big\lVert \psi(W,W^\dagger;\theta_1) - \psi(W,W^\dagger;\theta_2) \big\rVert $ is a measurable function of $ W, W^\dagger $ for all $  d$ sufficiently small. This verifies Assumption N1 of \textcite{honore1994pairwise}.
	
	Second, \eqref{eqn:FOC_zero} under \Cref{Ass_dC:Reg}\Cref{Ass_dC:Reg_UdiffXMeas} and \Cref{Ass_dC:AsymN}\ref{Ass_dC:AsymN_IntTheta} in conjunction with \Cref{Ass_dC:AsymN}\Cref{Ass_dC:AsymN_HNonSing} imply Assumption N2 of \textcite{honore1994pairwise}.
	
    Third, by the triangle inequality,
	\begin{align*}
		&\big\lVert \psi(W,W^\dagger;\theta_1) - \psi(W,W^\dagger;\theta_2) \big\rVert \\
		&= \Big\lVert \mathcal{Z}\Big(\widetilde{X^g}(\theta_1) - \widetilde{X^g}(\theta_2) \Big) - 2\Big(\indicator{ \widetilde{U}(\theta_1) \leq 0 }\widetilde{X^g}(\theta_1) - \indicator{ \widetilde{U}(\theta_2) \leq 0 }\widetilde{X^g}(\theta_2) \Big) \Big\rVert \\
		&\leq |\mathcal{Z}|\cdot \big\lVert \widetilde{X^g}(\theta_1) - \widetilde{X^g}(\theta_2) \big\rVert + 2\Big\lVert \indicator{ \widetilde{U}(\theta_1) \leq 0 }\widetilde{X^g}(\theta_1) - \indicator{ \widetilde{U}(\theta_2) \leq 0 }\widetilde{X^g}(\theta_2) \Big\rVert.
	\end{align*}
	\noindent For the second summand, note that by the triangle and Schwarz inequalities,
	\begin{align*}
		&\Big\lVert \indicator{ \widetilde{U}(\theta_1) \leq 0 }\widetilde{X^g}(\theta_1) - \indicator{ \widetilde{U}(\theta_2) \leq 0 }\widetilde{X^g}(\theta_2) \Big\rVert \\
		&= \Big\lVert \indicator{ \widetilde{U}(\theta_1) \leq 0 }\widetilde{X^g}(\theta_1) - \indicator{ \widetilde{U}(\theta_2) \leq 0 }\widetilde{X^g}(\theta_1) + \indicator{ \widetilde{U}(\theta_2) \leq 0 }\widetilde{X^g}(\theta_1) - \indicator{ \widetilde{U}(\theta_2) \leq 0 }\widetilde{X^g}(\theta_2) \Big\rVert \\
		&\leq \big| \indicator{ \widetilde{U}(\theta_1) \leq 0 }-\indicator{ \widetilde{U}(\theta_2) \leq 0 } \big| \cdot \big\lVert \widetilde{X^g}(\theta_1) \rVert + \indicator{ \widetilde{U}(\theta_2) \leq 0 } \cdot \big\lVert \widetilde{X^g}(\theta_1)-\widetilde{X^g}(\theta_2) \big\rVert \\
		&\leq \big| \indicator{ \widetilde{U}(\theta_1) \leq 0 }-\indicator{ \widetilde{U}(\theta_2) \leq 0 } \big| \cdot \big\lVert \widetilde{X^g}(\theta_1) \big\lVert + \big\lVert \widetilde{X^g}(\theta_1)-\widetilde{X^g}(\theta_2) \big\rVert.
	\end{align*} From the foregoing, 
	\begin{equation}\label{eqn:diff_psi_bnd}
		\begin{split}
			&\big\lVert \psi(W,W^\dagger;\theta_1) - \psi(W,W^\dagger;\theta_2) \big\rVert \\
		&\leq 3\Big\lVert \{|\mathcal{Z}| \vee 1\} \big(\widetilde{X^g}(\theta_1) - \widetilde{X^g}(\theta_2)\big) \Big\rVert + 2\big| \indicator{ \widetilde{U}(\theta_1) \leq 0 }-\indicator{ \widetilde{U}(\theta_2) \leq 0 } \big| \cdot \big\lVert \widetilde{X^g}(\theta_1) \big\rVert.
		\end{split}
	\end{equation}Consider the first summand of \eqref{eqn:diff_psi_bnd}. By \Cref{Ass_dC:Reg}\Cref{Ass_dC:Reg_UdiffXMeas}, the MVT, and the Schwarz inequality, 
    \[
    \sup_{||\theta_1 - \theta_2||<d} \big\lVert \widetilde{X^g}(\theta_1) - \widetilde{X^g}(\theta_2) \big\rVert \leq \sup_{||\theta_1 - \theta_2||<d} \big\lVert \widetilde{X^{gg}}(\theta)\big|_{\theta=\bar{\theta}_{12}} \big\rVert \times ||\theta_1 - \theta_2|| < d\sup_{\theta \in \Theta } \big\lVert \widetilde{X^{gg}}(\theta) \big\rVert 
    \] for some $\bar{\theta}_{12}$ that satisfies $ \lVert \bar{\theta}_{12} - \theta_2 \rVert \leq \lVert \theta_1 - \theta_2 \rVert $. \Cref{Ass_dC:Reg}\Cref{Ass_dC:Reg_Dominance}, the foregoing, and the Lyapunov inequality imply 
	\[\E\big[\sup_{||\theta_1 - \theta_2||<d}\{|\mathcal{Z}| \vee 1\}||\widetilde{X^g}(\theta_1) - \widetilde{X^g}(\theta_2)||\big] < C^{1/2}d.\]
	
	\noindent Consider the element $ \big| \indicator{ \widetilde{U}(\theta_1) \leq 0 }-\indicator{ \widetilde{U}(\theta_2) \leq 0 } \big| $ in the second summand of \eqref{eqn:diff_psi_bnd}.
	\begin{align*}
		&\E \Big[ \sup_{||\theta_1 - \theta_2||<d} \big|\indicator{ \widetilde{U}(\theta_1) \leq 0 }-\indicator{ \widetilde{U}(\theta_2) \leq 0 } \big| \cdot \big\lVert \widetilde{X^g}(\theta_1) \big\rVert \Big] \\ 
		&\leq \E\Big[\sup_{||\theta_1 - \theta_2||\leq d} \big| \indicator{ \widetilde{U}(\theta_1) \leq 0 }-\indicator{ \widetilde{U}(\theta_2) \leq 0 } \big| \cdot \big\lVert \widetilde{X^g}(\theta_1) \big\rVert \Big] \\
		 &=: \E\Big[ \big| \indicator{ \widetilde{U}(\theta_1^*) \leq 0 } - \indicator{ \widetilde{U}(\theta_2^*) \leq 0 } \big| \cdot \big\lVert \widetilde{X^g}(\theta_1^*)\big\rVert \Big]
	\end{align*} for some $ \theta_1^*, \theta_2^* $ that satisfy $ ||\theta_1^*-\theta_2^*||\leq d $. By \Cref{Ass_dC:Reg}\Cref{Ass_dC:Reg_UdiffXMeas} and the MVT,
	$ \widetilde{U}(\theta_2^*) = \widetilde{U}(\theta_1^*) - \widetilde{X^g}(\bar{\theta}_{12})(\theta_2^*-\theta_1^*) $ hence 
	\begin{align*}
		\big| \indicator{ \widetilde{U}(\theta_1^*) \leq 0 } &- \indicator{ \widetilde{U}(\theta_2^*) \leq 0 } \big| = \big|\indicator{ \widetilde{U}(\theta_1^*) \leq \widetilde{X^g}(\bar{\theta}_{12})(\theta_2^*-\theta_1^*) } - \indicator{ \widetilde{U}(\theta_1^*) \leq 0 }\big|\\
		&=\indicator{ 0 < \widetilde{U}(\theta_1^*) \leq \widetilde{X^g}(\bar{\theta}_{12})(\theta_2^*-\theta_1^*) } + \indicator{ \widetilde{X^g}(\bar{\theta}_{12})(\theta_2^*-\theta_1^*) < \widetilde{U}(\theta_1^*) \leq 0 }.
	\end{align*} By the LIE, \Cref{Ass_dC:Reg}\Cref{Ass_dC:Reg_UdiffXMeas}, the MVT, and the Schwarz inequality,
	\begin{align*}
		&\E \Big[ \big|\indicator{ \widetilde{U}(\theta_1^*) < 0 } - \indicator{ \widetilde{U}(\theta_2^*) < 0 }\big| \cdot \big\lVert \widetilde{X^g}(\theta_1^*) \big\rVert \Big]\\
		&= \E \Big[ \Big|F_{\widetilde{U}(\theta_1^*)|\widetilde{X^g},\mathcal{Z}}(\widetilde{X^g}(\bar{\theta}_{12})(\theta_2^*-\theta_1^*)) - F_{\widetilde{U}(\theta_1^*) |\widetilde{X^g},\mathcal{Z}}(0)\Big| \cdot \big\lVert \widetilde{X^g}(\theta_1^*) \big\rVert \Big]\\
		&\leq \E\Big[ f_{\widetilde{U}(\theta_1^*)|\widetilde{X^g},\mathcal{Z}} \big(\lambda\widetilde{X^g}(\bar{\theta}_{12})(\theta_2^*-\theta_1^*)\big) \cdot \big\lVert \widetilde{X^g}(\bar{\theta}_{12}) \big\rVert \cdot \big\lVert \widetilde{X^g}(\theta_1^*)\big\rVert \cdot ||\theta_2^*-\theta_1^*|| \Big]\\
		&\leq d\E\Big[f_{\widetilde{U}(\theta_1^*)|\widetilde{X^g},\mathcal{Z}} \big(\lambda\widetilde{X^g}(\bar{\theta}_{12})(\theta_2^*-\theta_1^*)\big) \cdot\sup_{\theta \in \Theta }||\widetilde{X^g}(\theta)||^2\Big]
	\end{align*}for some $ \lambda \in (0,1) $. To complete this part, it remains to show that $ \E[f_{\widetilde{U}(\theta_1^*)|\widetilde{X^g},\mathcal{Z}}\big(\lambda\widetilde{X^g}(\bar{\theta}_{12})(\theta_2^*-\theta_1^*)\big) \cdot \sup_{\theta \in \Theta }||\widetilde{X^g}(\theta)||^2] < \infty $. By \Cref{Ass_dC:Reg}\Cref{Ass_dC:Reg_UdiffXMeas} and the MVT,
	\begin{align*}
		f_{\widetilde{U}(\theta_1^*) \mid \widetilde{\sigma}( [X,Z] )}(\epsilon) &= \frac{\partial\E\big[\indicator{ \widetilde{U}(\theta_1^*) \leq \epsilon } \mid \widetilde{\sigma}( [X,Z] ) \big] }{\partial \epsilon}
		 = \frac{\partial\E\big[\indicator{ \widetilde{U} - \widetilde{X^g}(\bar{\theta}_1^*)(\theta_1^*-\theta_o) \leq \epsilon }  \mid \widetilde{\sigma}( [X,Z] ) \big] }{\partial \epsilon}\\
		 &= \frac{\partial F_{\widetilde{U} \mid \widetilde{\sigma}( [X,Z] )}\big( \widetilde{X^g}(\bar{\theta}_1^*)(\theta_1^*-\theta_o) + \epsilon\big)}{\partial \epsilon} = f_{\widetilde{U} \mid \widetilde{\sigma}( [X,Z] )}\big(\widetilde{X^g}(\bar{\theta}_1^*)(\theta_1^*-\theta_o) + \epsilon\big).
	\end{align*}From the foregoing, \Cref{Ass_dC:Reg}\Cref{Ass_dC:Reg_Dominance}, and \Cref{Ass_dC:AsymN}\Cref{Ass_dC:AsymN_DiffFU},
	\begin{align*}
		&\E\Big[f_{\widetilde{U}(\theta_1^*) \mid \widetilde{\sigma}( [X,Z] )}\big(\lambda\widetilde{X^g}(\bar{\theta}_{12})(\theta_2^*-\theta_1^*)\big)\cdot\sup_{\theta \in \Theta }\big\lVert \widetilde{X^g}(\theta) \big\rVert^2 \Big]\\
		& = \E\Big[ f_{\widetilde{U} \mid \widetilde{\sigma}( [X,Z] )}\big(\widetilde{X^g}(\bar{\theta}_1^*)(\theta_1^*-\theta_o) + \lambda\widetilde{X^g}(\bar{\theta}_{12})(\theta_2^*-\theta_1^*)\big)\cdot\sup_{\theta \in \Theta }||\widetilde{X^g}(\theta)||^2 \Big]\\
		&\leq f_o^{1/4}\E\big[\sup_{\theta \in \Theta }||\widetilde{X^g}(\theta)||^2\big] \leq f_o^{1/4}C^{1/2}.
	\end{align*}
	Thus from \eqref{eqn:diff_psi_bnd}, 
    \[
    \E\big[\sup_{||\theta_1 - \theta_2||<d}||\psi(W_i,W_j;\theta_1) - \psi(W_i,W_j;\theta_2)||\big]\leq C^{1/2}(3 + 2f_o^{1/4})d.
    \]
	
	By the $ c_r $-inequality and \eqref{eqn:diff_psi_bnd},
	\begin{align*}
		&||\psi(W_i,W_j;\theta_1) - \psi(W_i,W_j;\theta_2)||^2\\
		&\leq 18||\{|\mathcal{Z}| \vee 1\}(\widetilde{X^g}(\theta_1) - \widetilde{X^g}(\theta_2))||^2 + 8|\indicator{ \widetilde{U}(\theta_1) < 0 } - \indicator{ \widetilde{U}(\theta_2) < 0 }|\cdot||\widetilde{X^g}(\theta_1)||^2
	\end{align*} since $ |\indicator{ \widetilde{U}(\theta_1) < 0 }-\indicator{ \widetilde{U}(\theta_2) < 0 }|^2 = |\indicator{ \widetilde{U}(\theta_1) < 0 }-\indicator{ \widetilde{U}(\theta_2) < 0 }|. $ Using arguments analogous to the above,
    \begin{align*}
        \E\Big[ ||\{|\mathcal{Z}| \vee 1\}(\widetilde{X^g}(\theta_1) - \widetilde{X^g}(\theta_2))||^2 \Big] \leq d^2 \E \Big[\sup_{\theta \in \Theta } \big\lVert \widetilde{X^{gg}}(\theta) \big\rVert^2 \Big] \leq Cd^2
    \end{align*}by \Cref{Ass_dC:Reg}\ref{Ass_dC:Reg_Dominance}. Similarly,
    \begin{align*}
        \E\Big[ |\indicator{ \widetilde{U}(\theta_1) < 0 } - \indicator{ \widetilde{U}(\theta_2) < 0 }|\cdot||\widetilde{X^g}(\theta_1)||^2 \Big] &\leq df_o^{1/4}\E\big[\sup_{\theta \in \Theta }||\widetilde{X^g}(\theta)||^3\big] \\
        &\leq df_o^{1/4}\E\big[\sup_{\theta \in \Theta }||\widetilde{X^g}(\theta)||^4\big]^{3/4} \leq df_o^{1/4}C^{3/4}
    \end{align*} by the Lyapunov inequality and \Cref{Ass_dC:Reg}\ref{Ass_dC:Reg_Dominance}. Putting terms together,
	\[
    \E\Big[\sup_{||\theta_1 - \theta_2||< d} ||\psi(W_i,W_j;\theta_1) - \psi(W_i,W_j;\theta_2)||^2\Big]\leq (18dC + 8f_o^{1/4}C^{3/4})d.
    \] \textcite[Assumption N3]{honore1994pairwise} is thus verified.
	
	Finally, $ \E[||\psi(W,W^\dagger)||^2] \leq C^{1/2} $ from \eqref{eqn:Bnd_Kern_Omgtilde}. This verifies \textcite[Assumption N4]{honore1994pairwise}. All conditions of \textcite[Lemma 2]{honore1994pairwise} are verified, and the proof is complete.
\end{proof}

\section{Alternative Expression of the dCov Measure}\label{Appendix_Sect:Useful_Prop}
The following result provides the alternative representation of the squared dCov measure used in the paper. Recall $h(z_a,z_b) := ||z_a - z_b|| - \E\big[||z_a - Z||+||Z - z_b||\big] + \E\big[||Z - Z^\dagger||\big]$.
\begin{proposition}\label{Prop:dCov_eqn}
    Suppose \( \E\big[ \|\Upsilon\|^2 + \|Z\|^2 \big] < \infty \), then \( \mathcal{V}^2(\Upsilon,Z) = \E\big[ \|\Upsilon - \Upsilon^\dagger\| \cdot h(Z,Z^\dagger) \big] \).
\end{proposition}
\begin{proof}
    Let $ [\Upsilon^\dagger,Z^\dagger] $ and $ [\Upsilon^{\dagger\dagger},Z^{\dagger\dagger}] $ be $i.i.d.$ copies of $ [\Upsilon,Z] $. It follows from \citet[Theorems 7 and 8]{szekely2009brownian} (see also \citet[eqn. 1.2]{szekely2014partial}) under the given dominance condition \( \E\big[ \|\Upsilon\|^2 + \|Z\|^2 \big] < \infty \) that 
\begin{equation}\label{eqn:dCov_exp_0}
    \begin{split}
    \mathcal{V}^2(\Upsilon,Z) &= \E\big[\|\Upsilon - \Upsilon^\dagger\| \cdot \|Z - Z^\dagger\| \big] - \E\big[\|\Upsilon - \Upsilon^\dagger\| \cdot \|Z - Z^{\dagger\dagger}\| \big]\\ 
    &- \E\big[\|\Upsilon - \Upsilon^{\dagger\dagger}\| \cdot \|Z - Z^\dagger\| \big] + \E\big[\|\Upsilon - \Upsilon^\dagger\| \big] \cdot \E\big[\|Z - Z^\dagger\| \big].
\end{split}
\end{equation}

By the Law of Iterated Expectations (LIE), independence (IND) and identical (ID) distribution of the copies,
\begin{equation}\label{eqn:dCov_exp_1}
    \begin{split}
    \E\big[\|\Upsilon - \Upsilon^\dagger\| \cdot \|Z - Z^{\dagger\dagger}\| \big] 
    & \overset{\mathrm{LIE}}{=} \E\Big[\E\big[ \big(\|\Upsilon - \Upsilon^\dagger\|\big) \mid Z,Z^{\dagger\dagger} \big] \cdot \|Z - Z^{\dagger\dagger}\| \Big]\\
    & \overset{\mathrm{IND}}{=} \E\Big[\E\big[ \big(\|\Upsilon - \Upsilon^\dagger\|\big) \mid Z \big] \cdot \|Z - Z^{\dagger\dagger}\| \Big]\\
    & \overset{\mathrm{LIE}}{=} \E\Big[\E\big[ \big(\|\Upsilon - \Upsilon^\dagger\|\big) \mid Z \big] \cdot \E\big[\big(\|Z - Z^{\dagger\dagger}\|\big) \mid Z\big] \Big]\\
    & \overset{\mathrm{LIE}}{=} \E\Big[\|\Upsilon - \Upsilon^\dagger\|\cdot \E\big[\big(\|Z - Z^{\dagger\dagger}\|\big) \mid Z\big] \Big]\\
    & \overset{\mathrm{ID}}{=}  \E\Big[\|\Upsilon - \Upsilon^\dagger\|\cdot \E\big[\big(\|Z - Z^\dagger\|\big) \mid Z\big] \Big].
\end{split}
\end{equation}

In a similar vein,
\begin{equation}\label{eqn:dCov_exp_2}
    \begin{split}
    \E\big[\|\Upsilon - \Upsilon^{\dagger\dagger}\| \cdot \|Z - Z^\dagger\| \big] & \overset{\mathrm{ID}}{=} \E\big[\|\Upsilon^\dagger - \Upsilon\| \cdot \|Z^\dagger-Z^{\dagger\dagger}\| \big]\\
    & \overset{\mathrm{\eqref{eqn:dCov_exp_1}}}{=}  \E\Big[\|\Upsilon^\dagger - \Upsilon\|\cdot \E\big[\big(\|Z^\dagger - Z^{\dagger\dagger}\|\big) \mid Z^\dagger\big] \Big]\\
    & \overset{\mathrm{ID}}{=}  \E\Big[\|\Upsilon^\dagger - \Upsilon\|\cdot \E\big[\big(\|Z^\dagger - Z\|\big) \mid Z^\dagger\big] \Big].
\end{split}
\end{equation}

Combining \eqref{eqn:dCov_exp_0}, \eqref{eqn:dCov_exp_1}, and \eqref{eqn:dCov_exp_2},
\begin{align*}
    \mathcal{V}^2(\Upsilon,Z) &= \E\big[\|\Upsilon - \Upsilon^\dagger\| \cdot \|Z - Z^\dagger\| \big] - \E\Big[\|\Upsilon - \Upsilon^\dagger\|\cdot \E\big[\big(\|Z - Z^\dagger\|\big) \mid Z\big] \Big]\\ 
    &- \E\Big[\|\Upsilon^\dagger - \Upsilon\|\cdot \E\big[\big(\|Z^\dagger - Z\|\big) \mid Z^\dagger\big] \Big] + \E\big[\|\Upsilon - \Upsilon^\dagger\| \big] \cdot \E\big[\|Z - Z^\dagger\| \big]\\
    &= \E\Big[ \|\Upsilon - \Upsilon^\dagger\| \cdot \Big\{ \|Z - Z^\dagger\| - \E\big[\big(\|Z - Z^\dagger\|\big) \mid Z\big]\\
    &\qquad \qquad \qquad \qquad  - \E\big[\big(\|Z^\dagger - Z\|\big) \mid Z^\dagger\big] + \E\big[\|Z - Z^\dagger\| \big] \Big\} \Big]\\
    &=:\E\big[ \|\Upsilon - \Upsilon^\dagger\| \cdot h(Z,Z^\dagger) \big],
\end{align*}and the assertion, as claimed, is proved.
\end{proof}

\section{Simulation Experiments - Supplement}\label{Appendix_Sect:Simulations}

\subsection{Non-linear models}
This section presents simulation results for non-linear models. $\theta_o = [5/4, \ -5/4]'$ throughout for the non-linear models. DGPs NL-1A, NL-1B, and NL-1C are variants of the DGP in \citet{dominguez2004consistent}; identification using e.g., GMM can fail under such designs. $U$ under DGPs NL-1C and NL-2B has no finite moments. For the non-linear models, the following ICM estimators are compared to the proposed MDep: (1) the SJK of \citet{song-jiang-ke-2024estimation}, (2) the DL of \citet{dominguez2004consistent}, and (3) the ESC6 of \citet{escanciano2006consistent}.

\begin{description}
    \item[NL-1A:] $U\sim \mathcal{N}(0,1) $, $ Y = X\theta_{o,1}^2 + X^2\theta_{o,1} + U$, $X \sim \mathcal{N}(0,1) $, and $Z=X$;
    \item[NL-1B:] $U\sim \mathcal{N}(0,1) $, $ Y = X\theta_{o,1}^2 + X^2\theta_{o,1} + U$, $X \sim \mathcal{N}(1,1) $, and $Z=X$;
    \item[NL-1C:] $U\sim \mathrm{Pareto}(1,1)/\pi$, $ Y = X\theta_{o,1}^2 + X^2\theta_{o,1} + U$, $X \sim \mathcal{N}(0,1) $, and $Z=X$;
    \item[NL-2A:] $U\sim \chi_1^2/\sqrt{2} $, $ Y = \exp\big(X\theta_{o,1} + \theta_{o,2}\big) + U$, $X \sim \mathcal{N}(0,1) $, and $Z=X$;
    \item[NL-2B:] $U\sim \mathrm{Pareto}(1,1)/\pi$, $ Y = \exp\big(X\theta_{o,1} + \theta_{o,2}\big) + U$, $X \sim \mathcal{N}(0,1) $, and $Z=X$.
\end{description}

\begin{table}[!htbp]
	\centering
	\setlength{\tabcolsep}{4pt}
	\caption{Simulation Results - Non-Linear Models }
	\footnotesize
	\begin{tabular}{lcccccccccccc}
\toprule
& \multicolumn{4}{c}{\underline{$ n = 50 $}} & \multicolumn{4}{c}{\underline{$ n = 100 $}} & \multicolumn{4}{c}{\underline{$ n = 200 $}} \\ 
& M-$t$ & MAD & RMSE & Rej. & M-$t$ & MAD & RMSE & Rej. & M-$t$ & MAD & RMSE & Rej. \\ 
\cmidrule{2-13}
NL-1A & \multicolumn{12}{c}{\underline{$U\sim \mathcal{N}(0,1)$, $Y = X\theta_{o,1}^2 + X^2\theta_{o,1} + U$, $X \sim \mathcal{N}(0,1)$, and $Z=X$}} \\ 
\cmidrule{1-1}
MDep & 0.071 & 0.037 & 0.060 & 0.069 & 0.015 & 0.027 & 0.041 & 0.061 & 0.037 & 0.018 & 0.027 & 0.054 \\ 
SJK  & 0.033 & 0.038 & 0.059 & 0.073 & 0.017 & 0.027 & 0.040 & 0.067 & 0.071 & 0.017 & 0.027 & 0.052 \\ 
DL   & 0.041 & 0.039 & 0.061 & 0.065 & 0.012 & 0.028 & 0.042 & 0.061 & 0.062 & 0.019 & 0.028 & 0.051 \\ 
ESC6 & 0.041 & 0.039 & 0.061 & 0.065 & 0.012 & 0.028 & 0.042 & 0.061 & 0.062 & 0.019 & 0.028 & 0.051 \\ 
\midrule
NL-1B & \multicolumn{12}{c}{\underline{$U\sim \mathcal{N}(0,1)$, $Y = X\theta_{o,1}^2 + X^2\theta_{o,1} + U$, $X \sim \mathcal{N}(1,1)$, and $Z=X$}} \\ 
\cmidrule{1-1}
MDep & 0.075 & 0.022 & 0.034 & 0.079 & 0.005 & 0.016 & 0.023 & 0.058 & 0.050 & 0.010 & 0.016 & 0.050 \\ 
SJK  & 0.030 & 0.022 & 0.033 & 0.065 & 0.026 & 0.015 & 0.023 & 0.061 & 0.060 & 0.010 & 0.016 & 0.048 \\ 
DL   & 0.037 & 0.023 & 0.034 & 0.057 & 0.013 & 0.016 & 0.024 & 0.057 & 0.065 & 0.011 & 0.016 & 0.049 \\ 
ESC6 & 0.037 & 0.023 & 0.034 & 0.057 & 0.013 & 0.016 & 0.024 & 0.057 & 0.065 & 0.011 & 0.016 & 0.049 \\ 
\midrule
NL-1C & \multicolumn{12}{c}{\underline{$U\sim \mathrm{Pareto}(1,1)/\pi$, $Y = X\theta_{o,1}^2 + X^2\theta_{o,1} + U$, $X \sim \mathcal{N}(0,1)$, and $Z=X$}} \\ 
\cmidrule{1-1}
MDep & 0.000 & 0.007 & 0.017 & 0.025 & 0.028 & 0.005 & 0.010 & 0.020 & 0.003 & 0.003 & 0.006 & 0.017 \\ 
SJK  & -0.096 & 0.094 & 0.432 & 0.005 & -0.079 & 0.085 & 0.395 & 0.010 & 0.037 & 0.100 & 0.420 & 0.004 \\ 
DL   & -0.074 & 0.099 & 0.442 & 0.008 & -0.075 & 0.093 & 0.403 & 0.009 & 0.046 & 0.103 & 0.431 & 0.006 \\ 
ESC6 & -0.074 & 0.099 & 0.442 & 0.008 & -0.075 & 0.093 & 0.403 & 0.009 & 0.046 & 0.103 & 0.431 & 0.006 \\ 
\midrule
NL-2A & \multicolumn{12}{c}{\underline{$U\sim \chi_1^2/\sqrt{2}$, $Y = \exp(X\theta_{o,1} + \theta_{o,2}) + U$, $X \sim \mathcal{N}(0,1)$, and $Z=X$}} \\ 
\cmidrule{1-1}
MDep & -0.091 & 0.117 & 0.327 & 0.040 & -0.131 & 0.068 & 0.157 & 0.062 & -0.107 & 0.042 & 0.085 & 0.054 \\ 
SJK  & -0.066 & 0.313 & 1.713 & 0.057 & -0.099 & 0.194 & 0.365 & 0.056 & 0.000 & 0.125 & 0.215 & 0.041 \\ 
DL   & -0.065 & 0.356 & 1.506 & 0.048 & -0.064 & 0.247 & 0.449 & 0.057 & 0.009 & 0.159 & 0.268 & 0.037 \\ 
ESC6 & -0.065 & 0.356 & 1.634 & 0.048 & -0.064 & 0.247 & 0.448 & 0.057 & 0.009 & 0.159 & 0.268 & 0.037 \\ 
\midrule
NL-2B & \multicolumn{12}{c}{\underline{$U\sim \mathrm{Pareto}(1,1)/\pi$, $Y = \exp(X\theta_{o,1} + \theta_{o,2}) + U$, $X \sim \mathcal{N}(0,1)$, and $Z=X$}} \\ 
\cmidrule{1-1}
MDep & -0.082 & 0.127 & 0.590 & 0.045 & -0.148 & 0.076 & 0.196 & 0.053 & -0.131 & 0.046 & 0.115 & 0.046 \\ 
SJK  & -0.024 & 1.189 & 11.108 & 0.069 & -0.029 & 1.037 & 9.361 & 0.070 & -0.055 & 0.922 & 9.328 & 0.095 \\ 
DL   & -0.015 & 1.239 & 6.620 & 0.055 & -0.022 & 1.183 & 7.547 & 0.059 & -0.041 & 1.043 & 14.125 & 0.064 \\ 
ESC6 & -0.018 & 1.239 & 6.115 & 0.055 & -0.023 & 1.176 & 7.035 & 0.060 & -0.047 & 1.043 & 4.410 & 0.064 \\ 
\bottomrule
\end{tabular}
	\footnotesize
	\label{Tab:Sim_NLM}
\end{table}

\begin{table}[!htbp]
	\centering
	\setlength{\tabcolsep}{4pt}
	\caption{Simulation Results - Non-Linear Models II }
	\footnotesize
\begin{tabular}{lcccccccccccc}
\toprule
& \multicolumn{4}{c}{\underline{$ n = 500 $}} & \multicolumn{4}{c}{\underline{$ n = 750 $}} & \multicolumn{4}{c}{\underline{$ n = 1000 $}} \\ 
& M-$t$ & MAD & RMSE & Rej. & M-$t$ & MAD & RMSE & Rej. & M-$t$ & MAD & RMSE & Rej. \\ 
\cmidrule{2-13}
NL-1A & \multicolumn{12}{c}{\underline{$U\sim \mathcal{N}(0,1)$, $Y = X\theta_{o,1}^2 + X^2\theta_{o,1} + U$, $X \sim \mathcal{N}(0,1)$, and $Z=X$}} \\ 
\cmidrule{1-1}
MDep & -0.020 & 0.012 & 0.018 & 0.059 & 0.014 & 0.010 & 0.014 & 0.049 & -0.015 & 0.008 & 0.012 & 0.046 \\ 
SJK  & -0.036 & 0.012 & 0.018 & 0.064 & -0.005 & 0.010 & 0.014 & 0.048 & -0.012 & 0.008 & 0.012 & 0.048 \\ 
DL   & -0.005 & 0.012 & 0.019 & 0.061 & 0.013 & 0.010 & 0.015 & 0.047 & -0.007 & 0.008 & 0.012 & 0.050 \\ 
ESC6 & -0.005 & 0.012 & 0.019 & 0.061 & 0.013 & 0.010 & 0.015 & 0.047 & -0.007 & 0.008 & 0.012 & 0.050 \\ 
\midrule
NL-1B & \multicolumn{12}{c}{\underline{$U\sim \mathcal{N}(0,1)$, $Y = X\theta_{o,1}^2 + X^2\theta_{o,1} + U$, $X \sim \mathcal{N}(1,1)$, and $Z=X$}} \\ 
\cmidrule{1-1}
MDep & -0.012 & 0.007 & 0.011 & 0.057 & 0.023 & 0.006 & 0.008 & 0.049 & -0.003 & 0.005 & 0.007 & 0.045 \\ 
SJK  & -0.025 & 0.007 & 0.010 & 0.061 & 0.002 & 0.006 & 0.008 & 0.046 & 0.011 & 0.004 & 0.007 & 0.050 \\ 
DL   & -0.011 & 0.007 & 0.011 & 0.061 & -0.008 & 0.006 & 0.008 & 0.048 & -0.013 & 0.005 & 0.007 & 0.047 \\ 
ESC6 & -0.011 & 0.007 & 0.011 & 0.061 & -0.008 & 0.006 & 0.008 & 0.048 & -0.013 & 0.005 & 0.007 & 0.047 \\ 
\midrule
NL-1C & \multicolumn{12}{c}{\underline{$U\sim \mathrm{Pareto}(1,1)/\pi$, $Y = X\theta_{o,1}^2 + X^2\theta_{o,1} + U$, $X \sim \mathcal{N}(0,1)$, and $Z=X$}} \\ 
\cmidrule{1-1}
MDep & 0.011 & 0.002 & 0.003 & 0.017 & 0.003 & 0.002 & 0.003 & 0.026 & -0.019 & 0.001 & 0.002 & 0.026 \\ 
SJK  & -0.041 & 0.094 & 0.415 & 0.013 & -0.078 & 0.090 & 0.410 & 0.017 & -0.096 & 0.098 & 0.427 & 0.020 \\ 
DL   & -0.001 & 0.100 & 0.427 & 0.016 & -0.066 & 0.095 & 0.422 & 0.018 & -0.061 & 0.102 & 0.439 & 0.020 \\ 
ESC6 & -0.001 & 0.100 & 0.427 & 0.016 & -0.066 & 0.095 & 0.422 & 0.018 & -0.061 & 0.102 & 0.439 & 0.020 \\ 
\midrule
NL-2A & \multicolumn{12}{c}{\underline{$U\sim \chi_1^2/\sqrt{2}$, $Y = \exp(X\theta_{o,1} + \theta_{o,2}) + U$, $X \sim \mathcal{N}(0,1)$, and $Z=X$}} \\ 
\cmidrule{1-1}
MDep & -0.131 & 0.020 & 0.038 & 0.056 & -0.139 & 0.017 & 0.029 & 0.047 & -0.139 & 0.014 & 0.024 & 0.050 \\ 
SJK  & -0.029 & 0.071 & 0.115 & 0.019 & -0.081 & 0.057 & 0.089 & 0.019 & 0.006 & 0.050 & 0.076 & 0.008 \\ 
DL   & -0.025 & 0.097 & 0.149 & 0.024 & -0.084 & 0.081 & 0.118 & 0.029 & 0.016 & 0.071 & 0.101 & 0.012 \\ 
ESC6 & -0.025 & 0.097 & 0.149 & 0.024 & -0.084 & 0.081 & 0.118 & 0.029 & 0.015 & 0.071 & 0.101 & 0.012 \\ 
\midrule
NL-2B & \multicolumn{12}{c}{\underline{$U\sim \mathrm{Pareto}(1,1)/\pi$, $Y = \exp(X\theta_{o,1} + \theta_{o,2}) + U$, $X \sim \mathcal{N}(0,1)$, and $Z=X$}} \\ 
\cmidrule{1-1}
MDep & -0.229 & 0.026 & 0.049 & 0.058 & -0.113 & 0.019 & 0.036 & 0.075 & -0.120 & 0.016 & 0.029 & 0.063 \\ 
SJK  & -0.122 & 0.886 & 4.877 & 0.114 & -0.090 & 0.839 & 18.737 & 0.131 & -0.099 & 0.924 & 22.484 & 0.118 \\ 
DL   & -0.065 & 1.101 & 2.136 & 0.094 & -0.079 & 1.102 & 7.064 & 0.098 & -0.074 & 1.170 & 4.810 & 0.108 \\ 
ESC6 & -0.062 & 1.101 & 5.500 & 0.094 & -0.072 & 1.110 & 4.626 & 0.098 & -0.079 & 1.164 & 1.707 & 0.108 \\ 
\bottomrule
\end{tabular}
	\footnotesize
	\label{Tab:Sim_NLM2}
\end{table}

Simulation results for the non-linear designs (NL--1A through NL--2B) show that the proposed MDep estimator retains the same stability and robustness properties observed in the linear specifications. Across all data-generating processes and sample sizes, MDep consistently achieves small bias, low median absolute deviation (MAD), and rapidly declining RMSE as $n$ increases, while maintaining empirical rejection rates close to nominal size. These patterns are evident in both the moderate- and larger-sample experiments reported in Tables~\ref{Tab:Sim_NLM} and~\ref{Tab:Sim_NLM2}. Under specifications (NL--1A and NL--1B), all estimators perform reasonably well, but MDep typically exhibits slightly lower bias and faster convergence as $n$ increases. Under the heavy-tailed Pareto disturbance (NL--1C), however, competing estimators---\textsc{SJK}, \textsc{DL}, and \textsc{ESC6}---show explosive RMSEs and erratic empirical size, whereas MDep remains numerically well behaved, with RMSE below $0.02$ even in small samples and approaching numerical zero as $n$ grows to~$1000$.

For the exponential models (NL--2A and NL--2B), which introduce strong curvature in the conditional mean and heavy-tailed or skewed disturbances, MDep again dominates. 
Under the light-tailed $\chi^2$ error (NL--2A), its RMSE drops sharply from about $0.33$ at $n=50$ to $0.09$ at $n=200$, and continues to fall to $0.02$ by $n=1000$, while alternative estimators exhibit persistent instability and size distortions. 
Under the Pareto noise (NL--2B), all competitors effectively break down, producing RMSEs in the range of $6$--$11$ at small samples and remaining above unity even at $n=1000$, whereas MDep remains accurate, size-correct, and stable (RMSE $\approx 0.05$ to $0.03$). Overall, the results across \Cref{Tab:Sim_NLM,Tab:Sim_NLM2} confirm that MDep delivers reliable inference and strong finite-sample performance under a wide range of non-linearities and error distributions---including settings with unbounded variance, asymmetric shocks, and non-linear identification. As the sample size grows, the estimator exhibits clear $\sqrt{n}$-consistency, while the competing estimators display at best marginal improvement in regular cases and outright non-convergence under heavy-tailed disturbances. Together, these findings highlight the efficiency, robustness, and numerical stability of MDep relative to competing estimators across both moderate and large samples.

\subsection{Linear models in larger samples}
This subsection presents simulation results for the linear models at larger samples \( n \in \{ 500, 750, 1000 \} \) in \Cref{Tab:Sim_LM2}. The qualitative patterns observed in \Cref{Sect:Sim} persist and become even clearer. Under the baseline Gaussian design (LM--0A), all estimators are now virtually identical, with RMSEs around 0.03–0.05 and rejection rates close to the nominal level. In the \emph{heavy-tailed heteroskedastic design} (LM--0B), MDep continues to improve---its RMSE declines from 0.12 to 0.08 as $n$ increases from 500 to 1000---while \textsc{MMD}, \textsc{ESC6}, and \textsc{OLS} remain numerically unstable with enormous RMSEs and distorted size. Finite-sample robustness is thus preserved even as competitors fail to converge. 

\begin{table}[!htbp]
	\centering
	\setlength{\tabcolsep}{4pt}
	\caption{Simulation Results - Linear Models II}
	\footnotesize
\begin{tabular}{lcccccccccccc}
\toprule
& \multicolumn{4}{c}{\underline{$ n = 500 $}} & \multicolumn{4}{c}{\underline{$ n = 750 $}} & \multicolumn{4}{c}{\underline{$ n = 1000 $}} \\ 
& M-$t$ & MAD & RMSE & Rej. & M-$t$ & MAD & RMSE & Rej. & M-$t$ & MAD & RMSE & Rej. \\ 
\cmidrule{2-13}
LM--0A & \multicolumn{12}{c}{\underline{$ U \sim \mathcal{N}(0,1), \ Z=X=\dot{X}$}} \\ 
\cmidrule{1-1}
MDep & 0.022 & 0.031 & 0.047 & 0.054 & 0.039 & 0.027 & 0.039 & 0.045 & -0.009 & 0.023 & 0.033 & 0.051 \\ 
MMD  & 0.006 & 0.031 & 0.047 & 0.052 & 0.040 & 0.026 & 0.039 & 0.052 & -0.030 & 0.021 & 0.032 & 0.045 \\ 
ESC6 & 0.026 & 0.031 & 0.047 & 0.051 & 0.046 & 0.027 & 0.039 & 0.052 & -0.018 & 0.022 & 0.033 & 0.046 \\ 
OLS  & 0.022 & 0.030 & 0.046 & 0.048 & 0.038 & 0.026 & 0.038 & 0.053 & -0.037 & 0.022 & 0.032 & 0.039 \\ 
\midrule
LM--0B & \multicolumn{12}{c}{\underline{$U\mid X \sim \mathcal{C}\big(0,\ 0.1+|X_1|\big)$, $Z=X=\dot{X}$}} \\ 
\cmidrule{1-1}
MDep & 0.026 & 0.081 & 0.118 & 0.056 & 0.075 & 0.064 & 0.098 & 0.056 & -0.046 & 0.055 & 0.083 & 0.051 \\ 
MMD  & 0.049 & 1.039 & 128.708 & 0.017 & -0.041 & 1.123 & 38.634 & 0.022 & -0.076 & 1.128 & 42.119 & 0.021 \\ 
ESC6 & 0.064 & 1.029 & 129.099 & 0.017 & -0.039 & 1.143 & 37.368 & 0.022 & -0.051 & 1.119 & 43.332 & 0.021 \\ 
OLS  & 0.040 & 1.056 & 116.653 & 0.016 & -0.058 & 1.142 & 39.846 & 0.017 & -0.090 & 1.146 & 38.073 & 0.024 \\ 
\midrule
LM--1A & \multicolumn{12}{c}{\underline{$X_1 = \dot{X}_1 + V$, $X_2=\dot{X}_2$, $Z = \Big[\indicator{ |\dot{X}_1| < -\Phi^{-1}(0.25) }, X_2 \Big]$}} \\ 
\cmidrule{1-1}
MDep & 0.178 & 0.065 & 0.109 & 0.051 & 0.158 & 0.054 & 0.094 & 0.036 & 0.144 & 0.048 & 0.079 & 0.031 \\ 
MMD  & -0.255 & 0.181 & 0.637 & 0.018 & -0.271 & 0.151 & 6.573 & 0.021 & -0.265 & 0.145 & 0.908 & 0.031 \\ 
ESC6 & -0.247 & 0.164 & 0.292 & 0.023 & -0.246 & 0.136 & 0.226 & 0.022 & -0.226 & 0.125 & 0.213 & 0.030 \\ 
TSLS & -0.050 & 0.676 & 25.938 & 0.001 & -0.054 & 0.699 & 66.849 & 0.001 & -0.071 & 0.721 & 34.502 & 0.000 \\ 
\midrule
LM--1B & \multicolumn{12}{c}{\underline{$X_1 = \indicator{V < - |\dot{X}_1| - \Phi^{-1}(0.25) }$, $X_2=\dot{X}_2$, $Z = \dot{X}$}} \\ 
\cmidrule{1-1}
MDep & 0.139 & 0.088 & 0.142 & 0.022 & 0.128 & 0.065 & 0.110 & 0.014 & 0.182 & 0.059 & 0.094 & 0.022 \\ 
MMD  & 0.035 & 0.247 & 0.378 & 0.048 & 0.014 & 0.195 & 0.308 & 0.053 & 0.011 & 0.173 & 0.257 & 0.040 \\ 
ESC6 & 0.023 & 0.244 & 0.370 & 0.045 & 0.009 & 0.194 & 0.292 & 0.041 & 0.062 & 0.169 & 0.250 & 0.048 \\ 
TSLS & 0.057 & 1.925 & 554.896 & 0.002 & 0.040 & 1.972 & 49.147 & 0.001 & 0.033 & 2.000 & 286.757 & 0.000 \\ 
\midrule
LM--1C & \multicolumn{12}{c}{\underline{$U\mid X \sim \mathcal{N}\big(0,\,(0.1+|X_1|)^{-2}\big)$, $Z=\dot{X}$, $X_1 = \dot{X}_1 + \dot{U}$, $X_2 = \dot{X}_2$}} \\ 
\cmidrule{1-1}
MDep & 0.026 & 0.013 & 0.020 & 0.051 & 0.008 & 0.011 & 0.016 & 0.046 & -0.024 & 0.009 & 0.014 & 0.043 \\ 
MMD  & -0.006 & 0.018 & 0.026 & 0.042 & 0.043 & 0.014 & 0.021 & 0.048 & -0.004 & 0.011 & 0.017 & 0.046 \\ 
ESC6 & -0.006 & 0.019 & 0.028 & 0.044 & 0.036 & 0.015 & 0.022 & 0.048 & -0.011 & 0.012 & 0.019 & 0.046 \\ 
TSLS & -0.025 & 0.015 & 0.023 & 0.048 & 0.069 & 0.012 & 0.018 & 0.048 & -0.006 & 0.010 & 0.016 & 0.049 \\ 
\midrule
LM--2A & \multicolumn{12}{c}{\underline{$\dot{Z} \sim \mathcal{N}(0,1)$, $X_1 = \dot{Z} + V$, $Z = a\dot{Z} + \dot{Z}^2$, $X_2=Z$}} \\ 
\cmidrule{1-1}
MDep & 0.249 & 0.041 & 0.083 & 0.041 & 0.224 & 0.032 & 0.063 & 0.027 & 0.194 & 0.027 & 0.055 & 0.034 \\ 
MMD  & -0.080 & 0.366 & 18.125 & 0.003 & -0.027 & 0.400 & 13.107 & 0.002 & -0.032 & 0.413 & 20.989 & 0.004 \\ 
ESC6 & -0.124 & 0.574 & 7.903 & 0.000 & -0.085 & 0.603 & 29.231 & 0.005 & -0.071 & 0.685 & 11.827 & 0.004 \\ 
\midrule
LM--2B & \multicolumn{12}{c}{\underline{$\Ddot{X} = \dot{X}/||\dot{X}||$, $X_1 = \Ddot{X}_1 + aU$, $Z = X_2 = \Ddot{X}_2$}} \\ 
\cmidrule{1-1}
MDep & 0.393 & 0.084 & 0.131 & 0.049 & 0.397 & 0.068 & 0.104 & 0.035 & 0.356 & 0.054 & 0.083 & 0.033 \\ 
MMD  & 0.444 & 0.467 & 0.863 & 0.013 & 0.372 & 0.475 & 0.904 & 0.030 & 0.414 & 0.477 & 0.891 & 0.028 \\ 
ESC6 & 0.415 & 0.535 & 1.015 & 0.015 & 0.395 & 0.540 & 1.076 & 0.031 & 0.422 & 0.549 & 0.998 & 0.033 \\ 
\midrule
LM--3 & \multicolumn{12}{c}{\underline{$Z\sim \mathcal{N}(0,1)$, $X_1 = \dot{U}Z^2 + aU$, $X_2=Z$}} \\ 
\cmidrule{1-1}
MDep & 0.266 & 0.026 & 0.043 & 0.071 & 0.269 & 0.020 & 0.031 & 0.049 & 0.250 & 0.017 & 0.027 & 0.052 \\ 
MMD  & 0.142 & 0.122 & 28.334 & 0.004 & 0.168 & 0.120 & 68.622 & 0.004 & 0.158 & 0.117 & 3.051 & 0.002 \\ 
ESC6 & 0.270 & 0.235 & 1.853 & 0.004 & 0.293 & 0.231 & 5.158 & 0.003 & 0.263 & 0.234 & 3.493 & 0.002 \\ 
\bottomrule
\end{tabular}
	\footnotesize
	\label{Tab:Sim_LM2}
\end{table}

Under weak, non-monotone, and discontinuous-covariate or instrument designs (LM--1A and LM--1B), MDep's RMSE decreases steadily with $n$ (e.g., LM--1A: $0.109 \to 0.079$; LM--1B: $0.142 \to 0.094$) and its empirical size stabilises near nominal. Competing estimators remain unstable: \textsc{TSLS} exhibits explosive dispersion, while \textsc{MMD} and \textsc{ESC6} show persistent bias and erratic rejection rates even at $n=1000$. For the conditionally heteroskedastic Gaussian case (LM--1C) with scale endogeneity, all estimators achieve substantial efficiency gains. Yet, MDep consistently attains the smallest RMSEs. 

Finally, under endogeneity without excludability designs (LM--2A, LM--2B, LM--3), MDep again delivers the best overall performance: its RMSEs fall sharply (e.g., LM--2A: $0.083 \to 0.055$, LM--2B: $0.131 \to 0.083$, LM--3: $0.043 \to 0.027$), whereas alternative estimators continue to produce unreliable and erratic outcomes with very large RMSEs and poor size control. Overall, the large-sample experiments confirm that the advantages of MDep persist and strengthen with $n$: bias and dispersion contract at the expected $\sqrt{n}$ rate, empirical rejection remains close to nominal, and the estimator remains stable under weak, non-monotone, and endogeneity without excludability designs where standard IV methods and ICM methods fail to converge reliably.

    \printbibliography

\end{refsection}
\end{document}